\documentclass[journal]{IEEEtran}

\usepackage{cite}
\usepackage{booktabs}
\usepackage{multirow}
\usepackage{pifont}
\usepackage{amssymb}
\usepackage{cite}
\usepackage{subfigure}
\usepackage{xspace}
\usepackage[table]{xcolor}
\usepackage{graphicx}
\usepackage{makecell} 
\usepackage{longtable}
\usepackage{hyperref}
\usepackage{amsmath}
\usepackage{wasysym}
\usepackage{threeparttable}
\usepackage{pdflscape}
\usepackage{rotating}
\usepackage{enumitem}
\usepackage{tabularx}
\usepackage{ragged2e}

\hypersetup{
    linkcolor=blue,
    urlcolor=magenta,
    citecolor=blue,
}
\begin{document}

\title{RF Sensing Security and Malicious Exploitation: \\A Comprehensive Survey}

\author{Mingda~Han, 
        Huanqi~Yang,
        Wenhao~Li,
        Weitao~Xu,
        Xiuzhen~Cheng,
        Prasant~Mohapatra,
        and~Pengfei~Hu$^*$
\IEEEcompsocitemizethanks{\IEEEcompsocthanksitem 
Mingda Han, Wenhao Li, Xiuzhen Cheng, and Pengfei Hu  are with School of Computer Science and Technology, Shandong University, Qingdao, China. 
\IEEEcompsocthanksitem 
Huanqi Yang and Weitao Xu are with Department of Computer Science, City University of Hong Kong, Hong Kong SAR, China.
\IEEEcompsocthanksitem 
Prasant Mohapatra is with the Department of Computer Science at UC Davis, CA, USA.
\IEEEcompsocthanksitem  $^*$Corresponding Author.
 }
}

\markboth{Journal of \LaTeX\ Class Files,~Vol.~14, No.~8, August~2025}
{Shell \MakeLowercase{\textit{et al.}}: Bare Demo of IEEEtran.cls for IEEE Journals}

\maketitle

\begin{abstract}
Radio Frequency (RF) sensing technologies have experienced significant growth due to the widespread adoption of RF devices and the Internet of Things (IoT). These technologies enable numerous applications across healthcare, smart homes, industrial automation, and human-computer interaction.
However, the non-intrusive and ubiquitous nature of RF sensing—combined with its environmental sensitivity and data dependency—makes these systems inherently vulnerable not only as attack targets, but also as powerful attack vectors.
This survey presents a comprehensive analysis of RF sensing security, covering both system-level vulnerabilities—such as signal spoofing, adversarial perturbations, and model poisoning—and the misuse of sensing capabilities for attacks like cross-boundary surveillance, side-channel inference, and semantic privacy breaches. We propose unified threat models to structure these attack vectors and further conduct task-specific vulnerability assessments across key RF sensing applications, identifying their unique attack surfaces and risk profiles.
In addition, we systematically review defense strategies across system layers and threat-specific scenarios, incorporating both active and passive paradigms to provide a structured and practical view of protection mechanisms. Compared to prior surveys, our work distinguishes itself by offering a multi-dimensional classification framework based on task type, threat vector, and sensing modality, and by providing fine-grained, scenario-driven analysis that bridges theoretical models and real-world implications.
This survey aims to serve as a comprehensive reference for researchers and practitioners seeking to understand, evaluate, and secure the evolving landscape of RF sensing technologies.
\end{abstract}


\IEEEpeerreviewmaketitle

\section{Introduction}
\subsection{Background}
Radio frequency (RF) sensing technology has undergone rapid evolution, driven by advances in wireless technology and the rise of the Internet of Things (IoT). 
The proliferation of IoT devices and ubiquitous wireless infrastructure has expanded RF sensing from traditional radar applications to everyday environments. 
Modern RF sensing systems utilize radio signals (e.g., Wi-Fi, mmWave, LoRa, RFID, etc.) to sense targets and surroundings without the need for direct contact or line-of-sight (LoS) visibility.
This technologies have a wide range of applications in areas such as smart homes~\cite{iyer2018rf,ren2020liquid,ding2020wifi,yu2019rfid,santhalingam2020mmasl,li2022towards}, healthcare~\cite{an2021mars,zheng2022catch,shi2022mmbp} and security~\cite{jiang2024behaviors, lin2020revisiting,fang2023nowhere}. 
For instance, RF sensing can non-intrusively monitor human presence~\cite{xin2018freesense} and vital signs~\cite{shi2022mmbp} in healthcare settings or elder care, using reflections of RF signals to track movement, breathing, and even heartbeat. 
In smart homes, device-free RF sensing enables gesture recognition~\cite{yu2019rfid,santhalingam2020mmasl,li2022towards} and occupancy detection~\cite{hsu2023novel, zou2017freedetector,tang2020occupancy} for automation and energy saving, providing convenience without the privacy concerns of cameras. 
In the security domain, RF sensing-based systems can be used for intrusion detection~\cite{devoti2020pasid, jin2018whole, lin2020revisiting}, surveillance~\cite{han2024seeing,li2022recovering,kefayati2020wi2vi} and authentication~\cite{xu2022mask,yang2023xgait,han2024mmsign}. 
RF sensing has become a key component of modern smart environments, providing low-cost, ubiquitous and unobtrusive sensing capabilities.

The expanding applications of RF sensing also bring serious security and privacy concerns. 
If RF sensing systems lack security measures, they can be targeted by attackers or even exploited for malicious purposes, thus posing a significant threat to system security and personal privacy .
First, privacy leakage is one of the major challenges facing RF sensing technology. 
Since RF signals have the ability to penetrate obstacles, they may inadvertently expose sensitive information. For instance, studies have shown that by analyzing RF signals, an attacker can infer the occupant's behavior patterns~\cite{lu2022actlistener,zhang2025radsee,mei2024mmspyvr}, and even the content of indoor conversations~\cite{hu2023mmecho,wang2014we,feng2023mmeavesdropper}.
In the real world, attackers can deploy hidden RF sensors behind walls to monitor the habits of house occupants without the victims being aware of it. 
Such device-free remote monitoring poses a serious threat to personal privacy and safety — a thief, for instance, might identify an empty house by monitoring its Wi-Fi signals
These real-world examples highlight that RF sensing can become a malicious tool for privacy invasion if security measures are not taken.

In addition to privacy risks, RF sensing systems face security attacks that jeopardize system integrity and reliability. 
Attackers may attempt to disrupt sensing operations or manipulate sensing data for nefarious purposes. 
For instance, an attacker could send RF jamming signals to drown out legitimate signals, rendering RF sensors non-detectable and leading to the failure of critical applications such as autonomous driving~\cite{yan2016can,yeh2017security}. 
More sophisticated attacks involve signal forgery, where an attacker tricks the system into detecting non-existent targets by constructing fake signals~\cite{chen2023metawave,lazaro2022spoofing}.
In addition, as RF sensing and deep learning technologies continue to be integrated, adversarial machine learning attacks have become a new security threat. 
For example, an attacker can add undetectable interference to Wi-Fi channel state information (CSI) data to spoof a deep learning-based gesture recognition system~\cite{zhou2022wiadv,ozbulak2021investigating}. 
Similarly, introducing a label flipping attack during the training process can lead to the manipulation of the learning model of the RF sensing system, which can lead to erroneous judgments in the inference phase~\cite{singha2024securing}. These vulnerabilities indicate that in the absence of proper security protection, RF sensing systems can be manipulated or compromised by attackers, leading to data leakage, identity impersonation, or even posing a direct threat to user security. 
Ensuring security and privacy in RF sensing is therefore paramount for building trust and enabling safe, widespread use of this technology.

\begin{figure*}
    \centering
    \includegraphics[width=0.93\linewidth]{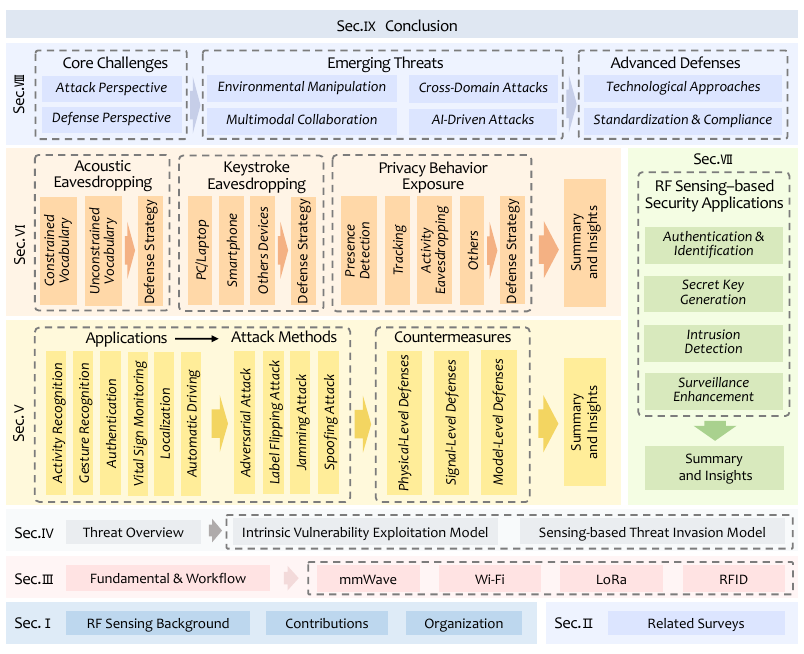}
    \caption{Organization of this survey}
    \label{fig:organization}
\end{figure*}

\subsection{Motivation and Contribution}
Despite the rapid development of RF sensing, the security and privacy implications have not yet been thoroughly examined from both theoretical and practical perspectives. 
While several recent works~\cite{liu2019wireless, zhang2023survey, venon2022millimeter, kong2024survey, wang2024multi,sikder2021survey,ma2019wifi,chen2023cross,geng2024survey} have surveyed RF sensing systems or focused on general IoT security, they often fall short in addressing the following critical aspects: 
(1) \emph{Formalized and unified threat modeling framework} that characterizes attacker capabilities, system exposure surfaces, and attack objectives in a modality-agnostic manner, enabling consistent evaluation across diverse tasks and system architectures;
(2) \emph{In-depth technical analysis} that covers emerging security threats involving RF sensing systems, including both direct attacks on sensing pipelines and malicious misuse of RF sensing capabilities to compromise user privacy;
(3) \emph{Task-specific vulnerability assessments} that systematically map and compare attack surfaces across key RF sensing applications;
(4) \emph{Multi-dimensional defense framework} that integrats hardware- and algorithm-level countermeasures, layered system design, and active–passive defense paradigms for comprehensive protection.
Furthermore, the convergence of deep learning and emerging RF infrastructures (e.g., 5G-A, 6G) is introducing novel security risks—ranging from cross-modal inference and semantic reconstruction to model-driven attacks—posing new challenges that existing frameworks are ill-equipped to address. 
These developments indicate the urgent need for a holistic, future-oriented security perspective tailored to the unique vulnerabilities and capabilities of RF sensing systems.

In light of these gaps, this survey aims to provide a detailed and technically grounded investigation of RF sensing security and privacy. 
Our main contributions are as follows:
\begin{itemize} 

\item \textbf{Unified Threat Modeling and Taxonomy:} We present a holistic threat model that captures both intrinsic vulnerabilities of RF sensing systems (e.g., adversarial attacks, signal spoofing, and data poisoning) and extrinsic threats (e.g., privacy breaches, side-channel exploitation). Our discussion bridges the gap between theoretical attack principles and real-world implications.

\item \textbf{Task-Specific Vulnerability Analysis with Case-Driven Insights:}
We conduct a fine-grained vulnerability analysis across key RF sensing tasks—such as HAR, gesture recognition, autonomous driving, eavesdropping—highlighting how different applications expose distinct attack surfaces. To illustrate practical relevance, we incorporate representative case studies with technical deep-dives and empirical findings.

\item \textbf{Multi-Dimensional Defensive Framework:} 
We present a comprehensive review of defense strategies across physical, signal, and model layers, and further distinguish between active and passive paradigms. Our framework spans robust waveform design, secure model training, cross-modal verification, and hardware-assisted countermeasures, offering a structured blueprint for resilient system design.

\item \textbf{Open Challenges and Forward-Looking Perspectives:} 
We identify emerging threats that extend beyond traditional attack surfaces, including infrastructure-level sensing misuse, AI-driven semantic inference, and multimodal coordination. 
In response, we systematically outline potential multi-layered defense strategies tailored to RF sensing, and emphasize the urgency of developing RF-specific security standards to guide the secure and compliant deployment of next-generation RF sensing systems.
\end{itemize}

\subsection{Organization of This Survey}
As shown in Fig.~\ref{fig:organization}, the remainder of this survey is organized as follows.
Section~\ref{sec:related} reviews related work and highlights the specific gaps addressed.  
Section~\ref{sec:fundamental} introduces RF sensing fundamentals, including system architectures and signal modalities.  
Section~\ref{sec:model} presents unified threat models and a taxonomy of security and privacy threats.  
Section~\ref{sec:attack1} analyzes attacks targeting sensing integrity, while Section~\ref{sec:attack2} examines RF sensing as a tool for privacy intrusion.  
Section~\ref{sec:security} explores how RF sensing can also be leveraged for positive security purposes.  
Section~\ref{sec:challenge} outlines key challenges and future directions.  
Section~\ref{sec:conclusion} concludes this survey.

\section{Related Work}
\label{sec:related}

RF sensing has recently gained attention not only as a non-intrusive tool for human activity recognition, but also as a potential threat vector for privacy leakage and malicious exploitation. A number of surveys have explored various aspects of RF sensing systems, such as the technologies employed (e.g., Wi-Fi, mmWave, LoRa), the application domains (e.g., smart homes, healthcare, autonomous driving), and sensing methodologies~\cite{liu2019wireless, venon2022millimeter, zhang2023survey, kong2024survey, wang2024multi}.

From a security perspective, the work by Sikder et al.~\cite{sikder2021survey} offers a foundational analysis of sensor-based threats to smart devices, primarily focusing on mobile and embedded sensors. Other surveys have touched on privacy concerns in specific modalities (e.g., Wi-Fi~\cite{liu2025survey,ma2019wifi}, mmWave~\cite{zhang2023survey}) or in cross-domain contexts~\cite{chen2023cross}, but often without a systematic treatment of RF sensing as both a target and vector of attacks.

The most relevant work to ours is the recent survey by Geng et al.~\cite{geng2024survey}, which introduces a role-based taxonomy—Victim, Weapon, and Shield—to categorize security issues in wireless sensing. While this perspective offers conceptual novelty and intuitive clarity, their discussion remains at a high level, focusing primarily on the roles of sensing signals, without formalizing unified threat models or systematically covering fine-grained sensing tasks, the spectrum of attack techniques, and multi-layered defense strategies.
In contrast, our work establishes a multi-dimensional security analysis framework along the axes of sensing modality, task type, and threat vector, providing a structured path from the physical principles of RF sensing to comprehensive attack chain modeling. 
We propose two unified threat models that rigorously characterize attacker capabilities and goals. 
Furthermore, we conduct task-specific vulnerability analyses across key applications—including HAR, indoor localization, identity authentication, autonomous driving, and acoustic eavesdropping—highlighting their unique threat surfaces and defense challenges.
On the defense side, we perform systematic, task-aware classification of countermeasures across different architectural layers and defensive paradigms, supported by illustrative case studies and tabular comparisons. This scenario-specific and layered organization enables readers to grasp the concrete types, severity, and characteristics of attacks across diverse RF sensing tasks, effectively filling a critical gap in current task-driven RF sensing security research by introducing a structured and comprehensive analytical dimension.

\begin{table*}[htbp]
\centering
\scriptsize
\caption{Comparison of Related Surveys}
\renewcommand{\arraystretch}{1.4}
\begin{tabularx}{\textwidth}{@{}l X X X *{5}{>{\centering\arraybackslash}p{1.1cm}}@{}}
\toprule

\textbf{Reference} & \textbf{Sensing Modality} & \textbf{Scenarios} & \textbf{Taxonomy} 
& \makecell{\textbf{Security}\\\textbf{\& Privacy}} 
& \makecell{\textbf{Threat}\\\textbf{Model}} 
& \makecell{\textbf{Attack}\\\textbf{Analysis}} 
& \makecell{\textbf{Task-level}\\\textbf{Analysis}} 
& \makecell{\textbf{Defense}\\\textbf{Methods}} \\
\midrule
Liu et al.~\cite{liu2019wireless} & Wi-Fi & Human sensing & Application  & \LEFTcircle & \Circle & \Circle  & \CIRCLE & \Circle \\ \hline
Sikder et al.~\cite{sikder2021survey} & Sensors & Internal sensors of smart devices & Attack path / sensor type  & \CIRCLE & \CIRCLE & \CIRCLE & \Circle & \CIRCLE \\ \hline
Zhang et al.~\cite{zhang2023survey} & mmWave & Human sensing  & Hardware platform / algorithm / sensing granularity  & \LEFTcircle & \Circle & \Circle & \LEFTcircle & \Circle \\ \hline
Venon et al.~\cite{venon2022millimeter} & mmWave & Autonomous driving & Application & \Circle & \Circle & \Circle & \CIRCLE & \Circle \\ \hline
Kong et al.~\cite{kong2024survey} & mmWave  & Autonomous driving, human sensing  & Algorithm / dataset / application & \Circle & \Circle & \Circle & \CIRCLE & \Circle \\  \hline
Wang et al.~\cite{wang2024multi} & mmWave + Vision / LiDAR / NIR / IMU & Autonomous driving & Application  & \LEFTcircle & \Circle & \Circle & \Circle & \LEFTcircle \\ \hline
Ma et al.~\cite{ma2019wifi} & Wi-Fi & Human sensing & Algorithm / task output  & \LEFTcircle & \Circle & \Circle & \CIRCLE & \Circle \\ \hline
Chen et al.~\cite{chen2023cross} & Wi-Fi & Cross-domain sensing generalization & Algorithm  & \Circle & \Circle & \Circle & \Circle & \Circle \\ \hline
Yao et al.~\cite{yao2024radar} & mmWave + Camera & Autonomous driving & Fusion framework / task  & \LEFTcircle & \Circle & \Circle & \CIRCLE & \Circle \\ \hline
Sun et al.~\cite{sun2022recent} & LoRa & Smart city, industrial IoT, agricultural monitoring & Protocol / Application  & \CIRCLE & \Circle & \Circle & \LEFTcircle & \CIRCLE \\ \hline
Geng et al.~\cite{geng2024survey} & Wi-Fi, Radar, LoRa, RFID, Bluetooth, EM, Acoustic & Human sensing, autonomous driving & Wireless signal role & \CIRCLE & \Circle & \LEFTcircle & \Circle & \CIRCLE \\ \hline
Liu et al.~\cite{liu2025survey} & Wi-Fi & Human sensing & Attack / defense method & \CIRCLE & \Circle & \LEFTcircle & \CIRCLE & \CIRCLE \\ \hline

\textbf{This survey} & Wi-Fi, Radar, LoRa, RFID, Bluetooth, EM, ZigBee & Human sensing, autonomous driving  & Attack principle / application / attack type & \CIRCLE & \CIRCLE & \CIRCLE & \CIRCLE & \CIRCLE \\
\bottomrule
\end{tabularx}
\begin{flushleft}
\scriptsize
\Circle~for Not covered, \LEFTcircle~for Partial discussion, \CIRCLE~for Comprehensive coverage.
\end{flushleft}
\label{tab:comparison} 
\end{table*}

\section{Fundamentals of RF Sensing}
\label{sec:fundamental}

\begin{figure*}
    \centering
    \includegraphics[width=0.98\linewidth]{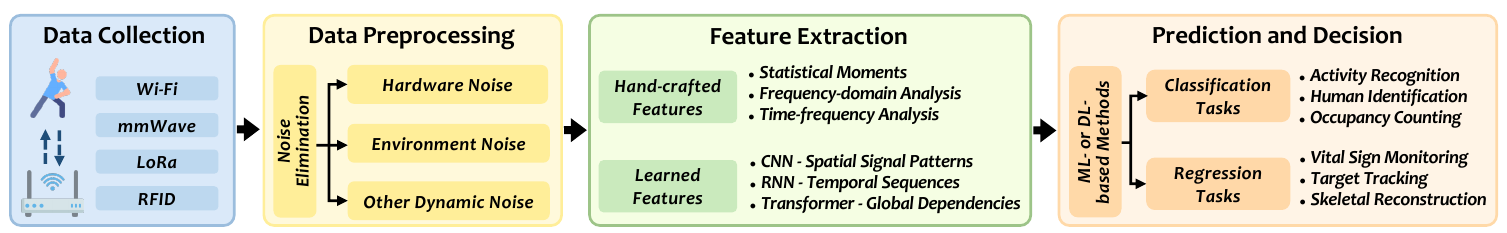}
    \caption{General Process of RF Sensing.}
    \label{fig:process}
\end{figure*}

\subsection{Fundamentals and Workflow of RF Sensing}
The fundamental principle of RF sensing technology is to achieve effective perception of objects and environments through the interaction of RF signals with the target and surrounding environment. 
Specifically, RF signals interact with nearby objects in the following key ways during propagation:

\begin{itemize}[leftmargin=*]
    \item Reflection. When an RF signal encounters an object, part of the signal is reflected back. By accurately measuring parameters such as the signal's strength, time delay, and angle of arrival, it is possible to infer the object's location, shape, and motion state. Reflection is the most common interaction in wireless sensing and is widely applied in fields such as indoor localization, motion tracking, and object detection.
    \item Diffraction. When an RF signal encounters an obstacle, it can diffract around the obstacle and continue to propagate. This property allows RF signals to propagate in complex environments, particularly in situations where the signal needs to pass through walls or navigate around obstacles, thereby supporting applications like through-wall sensing, target tracking, and detection of hidden objects.
    \item Refraction. When an RF signal passes through different media (e.g., walls, glass, liquids), its propagation characteristics (such as speed and phase) change. By analyzing these changes, it is possible to identify the properties of different materials, enabling applications like material identification and liquid detection.
\end{itemize}

Based on these physical phenomena of signal-environment interaction, RF sensing systems analyze various physical quantities such as time delay, frequency shift, signal strength, and phase changes to extract valuable information, establishing a mapping relationship with specific targets or tasks. 

The process of RF sensing typically consists of four main stages, as illustrated in Fig.~\ref{fig:process}:

\begin{itemize}[leftmargin=*]
\item \textbf{RF Data Collection.} In this stage, RF signals are transmitted and received using specialized hardware, such as software-defined radios (SDRs), Wi-Fi devices, radar systems, LoRa devices, or RFID readers. The collected raw RF data typically includes CSI, received signal strength indicator (RSSI), frequency shift, or phase information, depending on the sensing method used. The accuracy and robustness of the sensing system heavily rely on the quality and resolution of the acquired RF data.

\item \textbf{Data Preprocessing.} Raw RF signals are often affected by hardware noise, environmental interference, and hardware imperfections. Preprocessing techniques, such as denoising, filtering, and calibration, are applied to improve signal quality. Common preprocessing steps include bandpass filtering, principal component analysis (PCA) for noise reduction, and phase calibration to mitigate hardware-induced phase distortions. This step ensures that the subsequent analysis is based on clean and reliable data.  

\item \textbf{Feature Extraction.} The preprocessed RF data is transformed into meaningful representations through feature extraction. This step involves analyzing signal properties such as Doppler shift, time-of-flight, angle of arrival, signal variance, or spectral characteristics. Depending on the sensing task, hand-crafted features (e.g., statistical moments, wavelet transforms) or learned features (e.g., deep learning-based embeddings) are used to capture the most informative aspects of the signal.  

\item \textbf{Prediction.} The extracted features are fed into machine learning or deep learning models to perform classification or regression tasks. For example, classification models can be used for human activity recognition, gesture identification, or object detection, while regression models are applied to estimate continuous physiological or environmental variables such as breathing rate and heart rate. The final output of an RF sensing system varies based on the application but typically provides meaningful insights about the sensed environment.  
\end{itemize}

By integrating these four stages, RF sensing enables a wide range of applications, from smart home automation and healthcare monitoring to security surveillance and industrial automation.

\begin{table*}[htbp]
\centering
\setlength{\tabcolsep}{9pt}  
\caption{Comparison of Common RF Sensing Modalities.}
\label{tab:rf_modalities}
\renewcommand{\arraystretch}{1.3}
\begin{tabular}{@{}ccccc@{}}
\toprule
\textbf{ Dimension} & \textbf{mmWave} & \textbf{Wi-Fi} & \textbf{LoRa} & \textbf{RFID} \\
\bottomrule

Frequency & 24–81 GHz & 2.4/5 GHz & $<$ 1 GHz  & 860–960 MHz \\
Bandwidth & High ($>$ 1 GHz) & Medium (20–160 MHz) & Low ($<$ 500 kHz) & Low ($<$ 1 MHz) \\
Modulation Type & CW & OFDM & CSS & OOK \\
Signal & IF signal & RSSI / CSI & Reviced chirp signal & Backscattered signal \\
Resolution & High (cm-level) & Medium (cm-level) & Low (m-level) & Medium (cm/dm-level) \\
Sensing Range & Short ($<$ 10 m) & Indoor range ($<$ 30 m) & Long range ($>$ 100 m) & Short ($<$ 6 m) \\
Devices Required & mmWave radar & Commodity Wi-Fi AP / NIC & LoRa gateway + antennas & RFID reader + tag \\
Advantages & High accuracy & Low cost, widely available & Long range, low power & Low cost \\
Limitations & Poor penetration & Sensitive to multipath & Low resolution  & RFID reader required \\
\bottomrule
\end{tabular}
\vspace{0.5em}
\begin{tablenotes}
\footnotesize
\item \textbf{Note:} Reported values are typical or theoretical values. 
Actual performance may vary depending on hardware, environment, and signal processing techniques.
\end{tablenotes}
\end{table*}

\subsection{RF Sensing Modalities}
While a wide range of RF sensing technologies are currently under exploration, mmWave, Wi-Fi, LoRa, and RFID-based approaches have emerged as the most prevalent and well-established modalities, each offering unique advantages tailored to specific sensing applications and environmental conditions.  
Tab.~\ref{tab:rf_modalities} provides a comparative overview of their key characteristics, including frequency, bandwidth, signal type, sensing resolution, and deployment complexity.  
The following subsections offer detailed discussions on the sensing principles and practical considerations of each modality.

\subsubsection{\textbf{mmWave}}
mmWave radar can be classified into two main categories based on the signal transmission and reception methods: continuous wave (CW) radar and pulse radar. CW radar continuously emits electromagnetic waves and measures the changes in the reflected signal (such as frequency shifts). It primarily detects the relative velocity of a target using the Doppler effect. In contrast, pulse radar transmits a series of short, high-energy pulses and measures the time delay of the returned signal (i.e., the echo time) to calculate the target's range.
In current sensing research, CW radar, especially frequency modulated continuous wave (FMCW) radar, is widely used. 
FMCW radar is a specific type of CW radar that modulates the frequency of the transmitted signal to simultaneously measure both the target's range and velocity.

\begin{figure}
    \centering
    \includegraphics[width=0.92\linewidth]{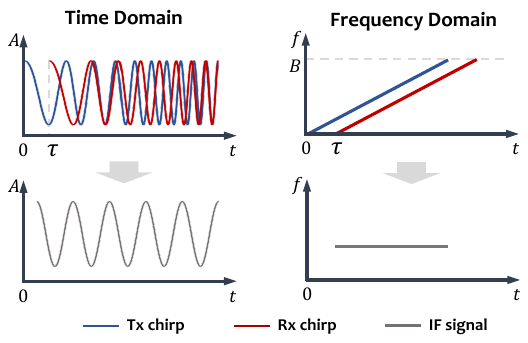}
    \caption{Chirp Signals.}
    \label{fig:chirp}
\end{figure}

The FMCW mmWave radar transmits FMCW signal, a.k.a, chirp, which is shown in Fig.~\ref{fig:chirp}.
The frequency of the chirp signal increases linearly with time $t$ and can be expressed as
\begin{equation}\label{equ:Chirp Frequency}
    f=f_0+St,
\end{equation}
where $f_0$ is the starting frequency and $S$ is the frequency modulation slope. Suppose the amplitude of the transmitted signal at time $t$ is $A$, then the transmitted sinusoidal FMCW signal $s_{_{T}}(t)$ can be expressed as
\begin{equation}\label{equ:Tx Signal}
    s_{_{T}}(t)=A \cos \left[2 \pi\left(f_0 t+\frac{St^2}{2}\right)\right].
\end{equation}
When the transmitted signal encounters an obstacle (e.g., the user's hand) at a distance $d$, the radar will receive a delayed version of the transmitted signal $s_{_{R}}(t)$, which can be expressed as
\begin{equation}\label{equ:Rx Signal}
    s_{_{R}}(t)=\alpha A \cos \left[2 \pi\left(f_0\left(t-\tau\right)+\frac{S(t-\tau)^2}{2}\right)\right],
\end{equation}
where $\alpha $ is the path loss, $\tau = 2d/c $ is the time delay, and $c$ is the speed of light.
Finally, the transmitted signal $s_{_{T}}(t)$ is mixed with the received signal $s_{_{R}}(t)$, and a low-pass filter is used to filter out the sum frequency components to obtain the IF signal:
\begin{equation}\label{equ:IF Signal}
    s_{_{IF}}(t)=LPF\{ s_{_{T}}(t)\cdot s_{_{R}}(t)\} = A_{_{IF}}\cos\left(2\pi f_{_{IF}}t+\phi_{_{IF}}\right),
\end{equation}
where $A_{_{IF}}$ is the amplitude of the IF signal, $f_{_{IF}} = S\tau = 2dS/c$ is known as the beat frequency, and $\phi_{_{IF}}$ is the phase.

The FMCW mmWave radar offers the capability to extract essential information regarding the range, velocity, and angle properties of the target. 
Specifically, the range information of the target is determined by applying the Fast Fourier Transform (Range FFT) to multiple sampling points along the fast time dimension of the IF signal.
Furthermore, the velocity information of the target is obtained by performing the Fast Fourier Transform (Doppler FFT) on multiple IF signals spanning the slow time dimension within a radar frame. 
Moreover, the angle information of the target is derived by subjecting the IF signals acquired from distinct receiving antennas to the Fast Fourier Transform (Angle FFT) operation. Collectively, these processes are commonly referred to as the 3D FFT, which responds to phase changes in different dimensions of the IF signal.

The combination of range, velocity, and angle information can be further organized into a spatial point cloud, representing the positions of detected targets in a 3D space.
The point cloud provides a spatial-temporal view of the sensing target and enables downstream tasks such as human tracking and 3D mesh.

\subsubsection{\textbf{Wi-Fi}}
Wi-Fi sensing leverages the existing Wi-Fi infrastructure to detect and interpret environmental changes based on variations in wireless signals. By analyzing how transmitted Wi-Fi signals interact with objects and people, Wi-Fi sensing enables applications such as motion detection, activity recognition, and human presence estimation. The two primary signal measurements utilized in Wi-Fi sensing are \textbf{RSSI} and \textbf{CSI}~\cite{ma2019wifi}.

\paragraph{RSSI-based Approach}
RSSI represents the power level of the received signal at a given receiver. It is a coarse-grained metric that provides an overall indication of signal strength but lacks fine-grained information about the multipath propagation effects. The RSSI of a received signal can be expressed as
\begin{equation}\label{equ:RSSI}
    \text{RSSI} = P_t + G_t + G_r - 10n \log_{10}(d) + X,
\end{equation}
where \( P_t \) is the transmitted power, \( G_t \) and \( G_r \) are the transmitter and receiver antenna gains, respectively, \( n \) is the path loss exponent (which depends on the environment), \( d \) is the distance between the transmitter and receiver, and \( X \) represents a random variable accounting for shadowing effects and small-scale fading~\cite{phillips2012survey}. 
Variations in RSSI occur due to environmental factors such as user movement, multipath effects, and signal obstructions.
While RSSI-based sensing is widely used due to its simplicity and compatibility with existing Wi-Fi devices, its sensitivity to noise and lack of spatial resolution limit its effectiveness in fine-grained sensing tasks.

\paragraph{CSI-based Approach}
Unlike RSSI, CSI provides detailed information about the Wi-Fi channel by characterizing the frequency response of each subcarrier in an Orthogonal Frequency Division Multiplexing (OFDM) system. CSI captures the impact of multipath propagation, enabling more precise and robust sensing. 
The CSI reported by commodity Wi-Fi NIC (e.g., Intel 5300~\cite{halperin2011tool} and Atheors 9580~\cite{xie2015precise}) can be expressed as
\begin{equation}\label{equ:CSI}
    H(f, t) = \sum_{i=1}^{N} \alpha_i e^{-j2\pi f \tau_i},
\end{equation}
where \( \alpha_i \) and \( \tau_i \) denote the amplitude attenuation and delay of the \( i \)-th multipath component, respectively, and \( N \) is the total number of multipath components. CSI is typically extracted from the physical layer (PHY) of Wi-Fi signals.
The extracted CSI data matrix for an \( N_t \)-transmit and \( N_r \)-receive antenna system over \( K \) subcarriers can be represented as
\begin{equation}
    \mathbf{H} = \begin{bmatrix}
        H_{1,1}(f_1) & H_{1,1}(f_2) & \cdots & H_{1,1}(f_K) \\
        H_{1,2}(f_1) & H_{1,2}(f_2) & \cdots & H_{1,2}(f_K) \\
        \vdots & \vdots & \ddots & \vdots \\
        H_{N_t,N_r}(f_1) & H_{N_t,N_r}(f_2) & \cdots & H_{N_t,N_r}(f_K)
    \end{bmatrix},
\end{equation}
where \( H_{i,j}(f_k) \) represents the CSI value between the \( i \)-th transmit and \( j \)-th receive antenna at the \( k \)-th subcarrier. 
Motion target or environment changes can cause CSI changes, by tracking CSI variations over time, Wi-Fi sensing systems can infer user motion, gestures, and even breathing patterns.

Similar to mmWave signal processing, Wi-Fi sensing employs various signal processing techniques to extract meaningful features from CSI data. These include analyzing amplitude and phase changes to detect movement patterns and object presence, applying time-frequency domain methods like Short-Time Fourier Transform (STFT) or Wavelet Transform (WT) to capture transient signal variations, and using dimensionality reduction techniques like PCA to reduce noise and keep key features.

\subsubsection{\textbf{LoRa}}
Long Range (LoRa) is a low-power wide-area network (LPWAN) communication technology based on Chirp Spread Spectrum (CSS) modulation. It is widely recognized for its long-range transmission and ultra-low power consumption, making it ideal for applications in environmental monitoring, smart agriculture, and smart cities. Recent studies have revealed that LoRa signals are not only effective for data communication but also possess long-range sensing capabilities, enabling applications such as target detection and human activity recognition.

LoRa employs Linear Frequency Modulation (LFM) chirp signals, where the frequency sweeps over a given bandwidth \( B \) within a symbol duration \( T \). The transmitted chirp signal can be expressed as:
\begin{equation}
    T_x(t) = e^{j(2\pi f_c t + \pi k t^2)},
\end{equation}
where \( f_c \) is the carrier frequency, and \( k = \frac{B}{T} \) is the chirp rate.

\begin{figure}
\centering
\subfigure[Multipath in LoRa Sensing.]{
\begin{minipage}[t]{0.48\linewidth}
\centering
\includegraphics[width=\linewidth]{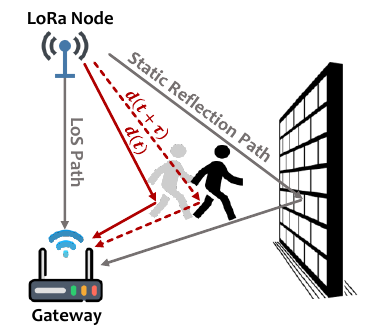}
\label{fig:loraMultipath}
\end{minipage}%
}
\subfigure[I/Q domain Representation.]{
\begin{minipage}[t]{0.48\linewidth}
\centering
\includegraphics[width=\linewidth]{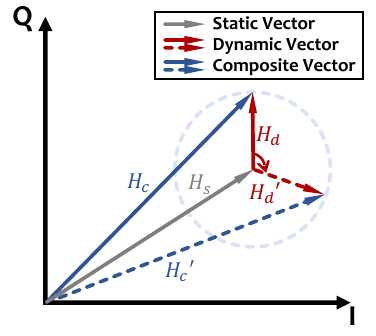}
\label{fig:loraIQ}
\end{minipage}%
}
\centering
\caption{LoRa Sensing Primer.}
\label{fig:lora}
\end{figure}

When the signal propagates in real-world environments, it experiences multipath propagation, where multiple reflections from objects lead to multiple delayed copies of the transmitted signal arriving at the receiver. 
Multipath effects can be categorized into static and dynamic components, as shown in Fig.~\ref{fig:lora}. The static multipath (\( H_s \)) is caused by reflections from stationary objects such as walls and furniture, introducing a constant phase shift:
\begin{equation}
    H_s = \sum_{i=1}^{N_s} \alpha_i e^{-j2\pi f_c \tau_i}.
\end{equation}
In contrast, dynamic multipath (\( H_d \)) results from moving objects, leading to time-varying path delays:
\begin{equation}
    H_d = \sum_{i=1}^{N_d} a_i(t) e^{-j2\pi f_c \tau_i(t)}.
\end{equation}

The total received signal, considering both static and dynamic multipath, can be written as:
\begin{equation}
    R_x(t) = e^{j(\pi k t^2 + \theta_c + \theta_s)} (H_s + H_d),
\end{equation}
where \( \theta_c \) and \( \theta_s \) represent carrier frequency offset and sampling frequency offset, respectively.
While \( H_d \) carries valuable motion-related information, \( H_s \) introduces unwanted phase distortions. 

To mitigate static multipath interference, signal ratio-based sensing method~\cite{zhang2020exploring} can be applied. This technique exploits the fact that static multipath signals remain consistent across multiple receiving antennas, whereas dynamic multipath signals vary due to motion-induced phase changes.
Taking the ratio of received signals from two antennas eliminates the static multipath term:
\begin{equation}
    R(t) = \frac{R_{x1}(t)}{R_{x2}(t)} = \frac{H_s + H_d^{(1)}}{H_s + H_d^{(2)}},
\end{equation}
where \( H_d^{(1)} \) and \( H_d^{(2)} \) represent the dynamic multipath components observed at the two antennas.
Since \( H_s \) is common in both signals, the ratio operation effectively cancels out its effect, leaving a function primarily dependent on the motion-induced variations \( H_d^{(1)} \) and \( H_d^{(2)} \). The extracted phase difference between the two antennas provides a more robust measure of motion:
\begin{equation}
    \Delta \theta_{\text{R}} = \angle R(t) = \angle H_d^{(1)} - \angle H_d^{(2)}.
\end{equation}
By tracking \(\Delta \theta_{\text{R}}\) over time, motion-related phase variations can be isolated while reducing interference from static multipath reflections. This method significantly improves motion detection accuracy and ensures robustness against environmental static clutter.

\subsubsection{\textbf{RFID}}
An RFID system primarily consists of two key components: the reader and the tag.
RFID sensing is based on the interaction between the reader and tag through backscattering, where the tag reflects the reader’s transmitted signal to convey information.

RFID systems can be broadly classified into two types based on the power source of the tags: passive RFID, where tags derive energy from the reader's transmitted signal; and active RFID, where tags possess an onboard power source, enabling long-range communication.
Among these, passive RFID is the most commonly used type due to its low cost, long lifespan, and ease of deployment. 
Passive tags operate without a battery, instead harvesting energy from the reader’s transmitted signal
Once powered, the tag modulates the incident RF signal through on-off keying (OOK), a simple yet effective technique where the tag alternates between reflecting and absorbing the RF signal to encode data.
The RFID reader transmits a continuous wave (CW) signal and simultaneously receives the backscattered signal from the tag. 
The backscattered signal $R(t)$ received at the reader can be expressed as:
\begin{equation}
    R(t)=A_{mod}b(t)e^{j(2\pi f_ct+\phi)},
\end{equation}
where $A_{mod}$ is the modulated amplitude, $b(t)$ is the binary modulation function (1 for reflection, 0 for absorption), $f_c$ is the carrier frequency, and $\phi$ is the phase shift introduced by the tag.
By decoding the tag’s response, the reader can extract physical layer information, which are essential for various sensing applications. 

In addition to the backscattered signal, the reader receives without-tag-reflection signals, which are simply delayed copies of the transmitted continuous wave and do not carry modulated information~\cite{wang2018modeling}.
Since without-tag-reflection signals do not contribute to data transmission, they are filtered out by the self-interference cancellation circuits of commercial RFID readers ~\cite{yang2015see}. 
As a result, only the tag-reflected backscattered signals are demodulated by the reader, providing sensing information, which can be further leveraged for localization, tracking, and activity recognition.

Like Wi-Fi sensing, RFID sesning commonly use the RSSI to infer object motion and presence. The RSSI value represents the power level of the received backscattered signal.
Variations in RSSI can indicate the movement of an RFID tag, making it useful for presence detection and coarse-grained localization. However, RSSI is highly sensitive to environmental noise and multipath interference, limiting its precision in fine-grained sensing tasks.

Compared to RSSI, phase information of RFID signals provides more fine-grained information. 
The received signal's phase can be expressed as
\begin{equation}\label{equ:Phase_RFID}
\phi = 2\pi \frac{2d}{\lambda} + \phi_0,
\end{equation}
where $\lambda$ is the signal wavelength, $d$ is the distance between the tag and reader, and $\phi_0$ is an initial phase offset. 
Small variations in $d$ cause periodic shifts in $\phi$, making phase-based sensing particularly effective for fine-grained motion tracking.
Since phase wraps around every $2\pi$, it is ambiguous for absolute distance estimation beyond one wavelength.
To resolve this ambiguity, phase unwrapping techniques can be applied by leveraging the continuity of phase changes over time. 

Similar to Wi-Fi and LoRa signals, RFID signals also experience multipath effects, where the transmitted signal reflects off surrounding objects before reaching the receiver.
Multipath propagation introduces both beneficial and detrimental effects in RFID-based sensing.
On one hand, multipath reflections encode additional spatial characteristics of the sensing target. By analyzing these reflections, RFID sensing systems can extract richer features about object movement, shape, and material properties. This enables applications such as gesture recognition,and activity monitoring.
On the other hand, multipath propagation also causes phase distortion and signal fading, leading to errors in distance estimation and localization. Signals arriving from different paths interfere with each other, creating phase noise and ambiguity that can degrade sensing accuracy. In dynamic environments, multipath interference becomes a major challenge in maintaining robust RFID sensing performance.

\section{Threat Models of RF Sensing}
\label{sec:model}
\subsection{Security and Privacy Threats of RF Sensing}
RF sensing systems, whether based on traditional signal processing with handcrafted feature extraction or deep learning models, face inherent security and privacy threats. 
These threats may arise from vulnerabilities within the system itself or from external adversaries exploiting RF sensing for malicious purposes.
Broadly, based on the target of the attack, these threats can be categorized into two main aspects: \textit{1) the intrinsic vulnerabilities and weaknesses of RF sensing systems, particularly when integrated with deep learning}, and \textit{2) the malicious exploitation of RF sensing as a tool for security and privacy invasions}.

\subsubsection{\textbf{Intrinsic Vulnerabilities of RF Sensing Systems}}
RF sensing systems, especially those empowered by deep learning, exhibit many vulnerabilities that adversaries can exploit. 
First, deep learning models introduce inherent security risks due to their black-box nature and sensitivity to training data. 
Unlike traditional algorithm-based RF sensing systems, deep learning-based systems rely heavily on data-driven feature extraction, making them prone to adversarial perturbations. By introducing imperceptible but structured noise into RF data, adversaries can manipulate system outputs, leading to misclassification in applications such as activity recognition~\cite{ambalkar2021adversarial,li2024practical}, gesture recognition~\cite{ozbulak2021investigating,zhou2022wiadv}, or authentication~\cite{cao2024security,huang2023phyfinatt}. 
Meanwhile, model inversion~\cite{fredrikson2015model,yang2019neural,yin2023ginver} attacks enable adversaries to reconstruct private information by exploiting the trained deep learning models, raising concerns about unintended data leakage.
Furthermore, RF sensing performance is often environment-dependent, leading to overfitting issues where a system trained in one setting fails to generalize to new conditions. This lack of robustness opens opportunities for attackers to exploit environmental changes or sensing-device variations to compromise the sensing process.

\subsubsection{\textbf{Exploiting RF Sensing for Security and Privacy Invasions}}
Beyond inherent vulnerabilities within RF sensing systems, RF sensing can also serve as an attack vector, enabling the extraction of sensitive information, tracking of target behaviors, and compromising security and privacy. 
Unlike traditional cybersecurity threats that primarily target software or network infrastructures, RF sensing possesses non-contact and remote sensing capabilities, allowing it to passively and covertly invade privacy without the target's awareness. 
This unique characteristic makes RF sensing-based threats more difficult to detect and defend against, posing significant security challenges.
A notable example is through-wall sensing eavesdropping, where adversaries exploit RF signals to detect the presence of a target~\cite{wang2017see,uysal2022new}, track the target~\cite{adib2013see,yang2015see}, and even recognize activities~\cite{lu2022actlistener,wang2018device} behind physical barriers. 
Furthermore, by analyzing fine-grained information from RF signals—such as keystroke behaviors or subtle vibrations caused by target-generated sounds—attackers can infer sensitive information, including passwords~\cite{ali2015keystroke,shen2021wipass} or private speaking~\cite{hu2023mmecho,basak2022mmspy}. This raises serious concerns regarding home security, confidential meetings, and personal privacy.
In addition, identity leakage via RF signatures poses a significant risk, as unique physiological and behavioral traits (e.g., gait patterns~\cite{yang2023xgait}, respiration~\cite{wang2022your}, and heartbeat~\cite{wang2022heartprint}) can be extracted from RF signals, enabling unauthorized tracking and biometric identification without user consent.

Based on the two types of threats mentioned above, we abstracted two threat models of RF sensing.
\textit{1) Intrinsic Vulnerability Exploitation Model (IVEM)}: This model describes attacks that exploit the inherent weaknesses of RF sensing systems, including adversarial perturbations, data poisoning, model inversion, and signal spoofing, leading to misclassification, authentication failures, and privacy leakage.
\textit{2) Sensing-based Threat Invasion Model (STIM)}: This model represents the misuse of RF sensing technology itself as an attack vector, enabling unauthorized surveillance, through-wall monitoring, keystroke inference, and identity tracking, leading to serious privacy invasions and covert intelligence gathering.

\subsection{Intrinsic Vulnerability Exploitation Model (IVEM)}
The IVEM captures attacks that leverage inherent weaknesses in RF sensing systems, particularly those integrating deep learning techniques. Adversaries exploit both physical and algorithmic vulnerabilities to compromise RF sensing system integrity and privacy.

\subsubsection{\textbf{Attacker Objectives}}
Depending on the objectives of the attacker, IVEM can be divided into two categories:
\begin{itemize}[leftmargin=*]
    \item Misclassification or Malfunction: Induces the RF sensing system to produce incorrect classification results (e.g., incorrect activity or gesture recognition) or faulty decision-making (e.g., failure to detect obstacles in autopilot systems).
    \item Privacy Leakage: Inferring sensitive user input data or private user attributes (e.g., height, weight, gender, etc.) from the output of the model's middle or output layer.
\end{itemize}

\subsubsection{\textbf{Problem Formalization}}
Let the RF sensing system be represented by a function:
\begin{equation}
    f_D: \mathcal{X} \rightarrow \mathcal{Y},
\end{equation}
where \( D \) is the training dataset, \( \mathcal{X} \) is the input space of RF signals, and \( \mathcal{Y} \) is the output space corresponding to the RF sensing system's predictions, such as classification labels or decision results.
Depending on the specific objective of the attacker, this problem can be formalized into the following two types:
\begin{itemize}[leftmargin=*]
     \item Misclassification or Malfunction Objective. The adversary seeks to introduce a perturbation \( \delta \in \Delta \) to an input \( x \) in order to cause the RF sensing system to misclassify or malfunction. The objective is to generate a perturbation \( \delta \) such that:
    \begin{equation}
        \exists \, \delta, \quad \|\delta\|_n < \epsilon \quad \text{and} \quad f_D(x + \delta) \neq f_D(x),
    \end{equation}
    where \( \|\delta\|_n\) is the \( n \)-norm of the perturbation, and \( \epsilon \) is a small threshold, ensuring that the perturbation is imperceptible or minimally invasive to the input data.

    In the case of data poisoning, the adversary may modify the training dataset \( D \) by injecting malicious samples that lead to incorrect learning, thereby corrupting the model's parameters and causing sensing system malfunctions.

    \item Privacy Leakage Objective. The adversary's objective in this case is to extract sensitive private information from the model’s outputs or intermediate representations. 
    This process can be formalized as an optimization problem over an inversion function \( h: \mathcal{Y} \rightarrow \mathcal{P} \), where \( \mathcal{P} \) denotes the space of private user attributes (e.g., height, gender), and \( \mathcal{Y} \) represents the output space of the RF sensing model. The optimization objective is given by:

    \begin{equation}
        \min_{h \in \mathcal{H}} \; \mathbb{E}_{(x, p) \sim \mathcal{D}} \left[ \mathcal{L} \left( h(f_D(x)), p \right) \right],
    \end{equation}
    
    where \( f_D(x) \) denotes the output or intermediate representation of the deployed RF sensing model for input \( x \), and \( \mathcal{L}(\cdot, \cdot) \) is a suitable loss function used to measure the discrepancy between the inferred private attribute \( \hat{p} = h(f_D(x)) \) and the true attribute \( p \). The goal is to learn an inversion function \( h \) that minimizes this discrepancy, thereby enabling unauthorized extraction of sensitive user information.
\end{itemize}

\subsubsection{\textbf{Attacker Capabilities}}
The attacker's abilities depend on the specific type of attack, as follows:
    \begin{itemize}[leftmargin=*]
        \item Adversarial Perturbation: The attacker has the ability to generate a perturbation \( \delta \) that subtly alters an input \( x \). This perturbation is typically small enough that it is imperceptible to human observers, but it is enough to fool the RF sensing system into making erroneous predictions.
        \item Signal Spoofing: The attacker has the ability to introduce synthetic RF signals that mimic legitimate patterns to deceive the RF sensing system.
        \item Data Poisoning: The attacker has the ability to inject malicious samples into the training set $D$, corrupting the learned parameters $\Theta$ of RF sensing model $f$.
        \item Model Inversion: The attacker can access intermediate or output layers of the RF sensing model to obtain the output of a target RF sensing model. 
    \end{itemize}
    
The first three capabilities focus on Objective 1 (\textit{Misclassification or Malfunction}), while the last one focuses on Objective 2 (\textit{Privacy Leakage}).

\subsection{Sensing-based Threat Invasion Model (STIM)}
The STIM abstracts scenarios where RF sensing technology is maliciously repurposed as an invasive tool for unauthorized surveillance and privacy breaches. Unlike IVEM, which exploits system-inherent vulnerabilities, STIM focuses on the covert misuse of RF sensing capabilities. 
\subsubsection{\textbf{Attacker Objectives}}
The primary objective of the STIM is to repurpose RF sensing system as a tool for covert surveillance and unauthorized privacy invasion. In this model, attackers exploit the non-contact, remote sensing capabilities of RF signals to gather sensitive information without the target's knowledge or consent.  

Specifically, RF sensing can be maliciously exploited by an attacker to cause both coarse-grained and fine-grained privacy breaches of a target. At a coarse-grained level, attackers can detect the presence of a target and monitor general movements or activities. At a finer level, attackers can exploit RF signals to infer sensitive actions, such as gait characteristics, keystroke patterns, and even the conversation.

\subsubsection{\textbf{Problem Formalization}}
Let the mapping from physical activities to RF signal be defined as
\begin{equation}
    g: \mathcal{A} \rightarrow \mathcal{S},
\end{equation}
where \( \mathcal{A} \) denotes the space of physical activity representations (e.g., user activities), and \( \mathcal{S} \) represents the RF signal space. 
The adversary aims to design an inversion function: 
\begin{equation}
    h: \mathcal{S} \rightarrow \mathcal{A},
\end{equation}
which can reconstruct or infer the physical activity \( a \) from the observed RF signal \( s = g(a) \). This objective can be formalized as the following optimization problem:
\begin{equation}
    \min_{h \in \mathcal{H}} \; \mathbb{E}_{a \sim \mathcal{A}} \left[ \mathcal{L}\left(h(g(a)), a \right) \right],
\end{equation}
where \( \mathcal{H} \) denotes the hypothesis space of inversion functions, and \( \mathcal{L}(\cdot, \cdot) \) is a suitable loss function that measures the discrepancy between the inferred activity \( \hat{a} = h(g(a)) \) and the true activity \( a \). The goal of the adversary is to minimize this reconstruction error, thereby enabling accurate recovery of sensitive user activities from RF signals.

\subsubsection{\textbf{Attacker Capabilities}}
The attacker's primary capability lies in receiving RF signals that contain information about the target. 
These signals can either be actively transmitted and received by the attacker or passively captured through existing wireless signals in the environment. 
The passive capture may include signals from the target device's communication, reflections or scattering from physical objects in the environment.

\begin{table*}[ht!]
\centering
\renewcommand{\arraystretch}{1.4} 
\setlength{\tabcolsep}{4pt} 
\caption{\textbf{Attacks to RF-sensing-based HAR Systems}}
\begin{tabular}{c c c cc cc cc c}
\bottomrule
\multirow{2}{*}{\textbf{Reference}} & \multirow{2}{*}{\textbf{Year}} & \multirow{2}{*}{\textbf{Sensing Source}} & \multicolumn{2}{c}{\textbf{Objective}} & \multicolumn{2}{c}{\textbf{Assumption}} & \multicolumn{2}{c}{\textbf{Domain}} & \multirow{2}{*}{\textbf{Key Features}} \\
\cmidrule(lr){4-5} \cmidrule(lr){6-7} \cmidrule(lr){8-9}
 & & & \textbf{Target} & \textbf{Untarget} & \textbf{White-box} & \textbf{Black-box} & \textbf{Digital} & \textbf{Physical}  & \\
\bottomrule

\cite{yang2020adversarial}    & 2020 &  \makecell{Wireless \\Doppler sensor} &      & \ding{51} & \ding{51} &         & \ding{51} &       & \makecell{The first adversarial attack to wireless \\  sensing-based HAR systems}  \\ \hline
\cite{ambalkar2021adversarial}& 2021 & Wi-Fi                   &      & \ding{51} & \ding{51} &         & \ding{51} &         & \makecell{The first adversarial attack to DNN-\\based HAR using Wi-Fi CSI} \\ \hline
\cite{xu2022wicam}            & 2022 & Wi-Fi                   &      & \ding{51} &          & \ding{51} & \ding{51} &          &  \makecell{Generalized for authentication, \\ gesture recognition task}  \\ \hline

\cite{liu2022physical,liu2024time} & 2022 & Wi-Fi                   & \ding{51} & \ding{51} &          & \ding{51} &         & \ding{51}   & Generalized for gesture recognition task \\  \hline
\cite{li2024practical}        & 2024 & Wi-Fi                   & \ding{51} & \ding{51} & \ding{51} & \ding{51} &         & \ding{51}  &  Generalized for authentication task  \\ \hline
\cite{huang2021wars}          & 2021 & Wi-Fi                   &      & \ding{51} &          & \ding{51} &         & \ding{51}   & \makecell{Cross-technology interference\\  using ZigBee signal }  \\ \hline
\cite{xie2023universal}       & 2024 & mmWave                  & \ding{51} & \ding{51} & \ding{51} & \ding{51} & \ding{51} &     &  \makecell{The first targeted adversarial attack\\ to mmWave-based HAR systems} \\  \hline
\cite{nallabolu2023emulation} & 2023 & \makecell{FMCW and \\  Doppler radar} & \ding{51} & \ding{51} &          & \ding{51} &         & \ding{51}  & \makecell{Generalized for human presence sensing\\ and health monitoring tasks} \\ \hline

\cite{singha2024securing} & 2024 & \makecell{mmWave} &   & \ding{51} &   \ding{51}  &  &  \ding{51} &   & \makecell{The first label flipping attack \\ to mmWave-based HAR} \\

\bottomrule
\end{tabular}
\label{tab: har}
\end{table*}

\section{Intrinsic Security Threats in RF Sensing Systems}
\label{sec:attack1}
In RF sensing systems, model integrity attacks seek to manipulate or compromise the accuracy and reliability of models, directly impacting their intended functionality. Based on our survey of current attack methods, we categorize these attacks into five key task-based areas, each reflecting the unique characteristics and vulnerabilities of specific RF Sensing applications. 
Each category—human activity recognition, gesture recognition, authentication, indoor localization, and autonomous vehicles—faces different risks and operational challenges due to its reliance on distinct signal processing techniques and application contexts. By examining these tasks individually, we can better understand the underlying principles of RF Sensing, the critical applications they support, and the specific ways in which compromised model integrity impacts their reliability and safety. The following sections provide a detailed analysis of attacks within each category, exploring both the nature of these systems and the potential consequences of integrity breaches.

\subsection{Human Activity Recognition}

HAR systems using RF signals, including Wi-Fi, mmWave, etc., enable non-intrusive monitoring by capturing subtle movements and activity patterns through signal processing. This method has become crucial in applications such as healthcare monitoring, security, and human-computer interaction due to its privacy-preserving, device-free nature. 
The reliance on machine learning models in these systems has raised concerns regarding their vulnerability to \textbf{adversarial attacks} and  \textbf{label flipping attack}.
These attacks can mislead HAR models and potentially cause significant consequences in real-world applications. 
Existing attacks against the RF sensing-based HAR system are shown in Table~\ref{tab: har}.

\subsubsection{\textbf{Adversarial Attacks}}
\begin{figure}
    \centering
    \includegraphics[width=0.98\linewidth]{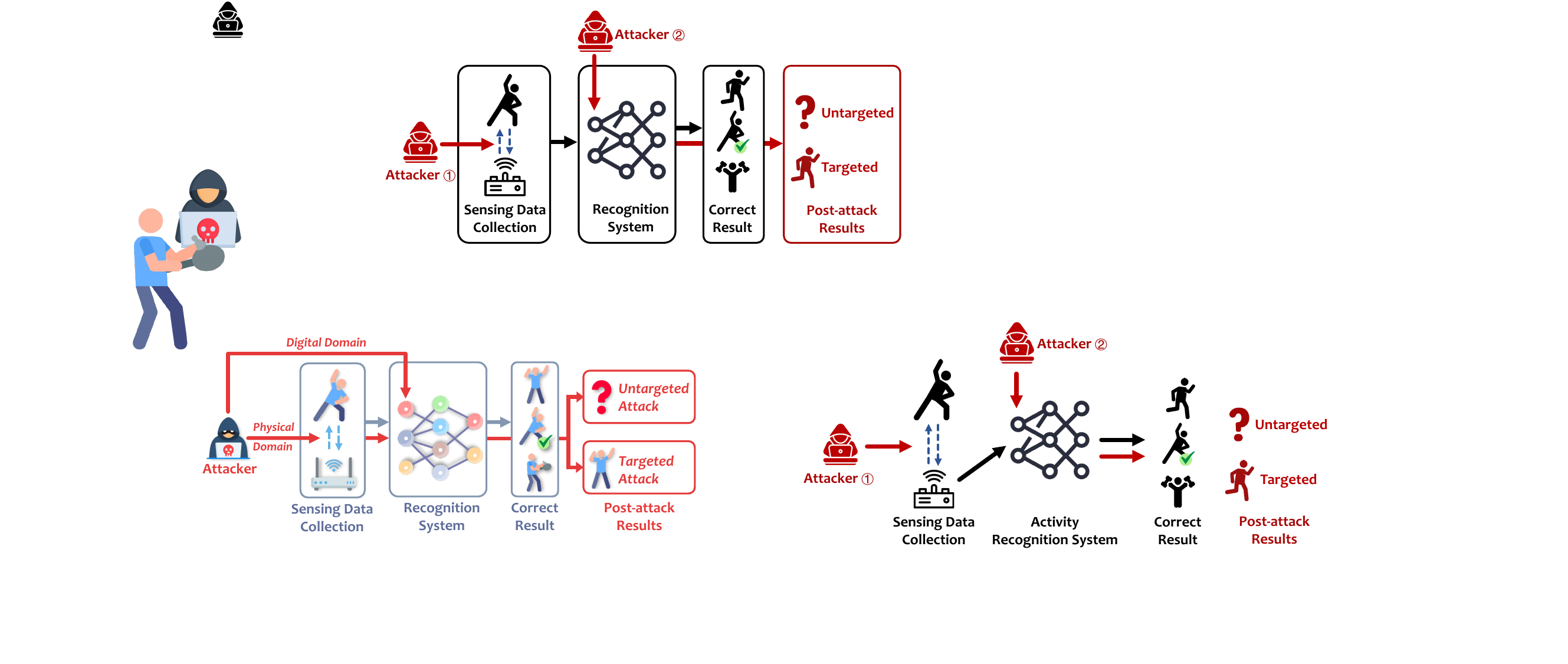}
    \caption{Adversarial attacks to RF sensing-based HAR systems.}
    \label{fig:HAR_attack}
\end{figure}

Adversarial attacks exploit the vulnerabilities of sensing systems by introducing subtle and often imperceptible perturbations to input data, leading to erroneous decision outcomes. Based on the attacker's intent, adversarial attacks can be categorized into targeted and untargeted attacks. Targeted attacks force the model to misclassify specific activities, while untargeted attacks aim to degrade the overall system performance. These attacks can occur in the digital domain (through direct modification of input data) or the physical domain (by manipulating the sensing process of the sensors), as illustrated in Fig.~\ref{fig:HAR_attack}.

Yang et al.~\cite{yang2020adversarial} were the first to investigate the vulnerability of wireless Doppler sensor-based human activity recognition system~\cite{zhao2016occupancy} to adversarial attacks. The authors applied three attack methods—Fast Gradient Sign Method (FGSM), Basic Iterative Method (BIM), and Carlini \& Wagner (C\&W)—on a system designed to recognize four types of activities: walking, lying on bed, turning over, and no signs. Experimental results indicate that minor perturbations in the input data (i.e., adversarial samples) can significantly degrade the classification performance of deep learning models, with accuracy reductions of up to 85\%. This study reveals potential security issues in Doppler-based HAR systems, laying the groundwork for future research on adversarial attacks targeting Wi-Fi and mmWave activity recognition systems.

Afterward, a significant amount of research began to focus on adversarial attacks against Wi-Fi sensing-based HAR systems.
Ambalkar et al.~\cite{ambalkar2021adversarial} first explore the vulnerability of deep neural network (DNN)-based human activity recognition systems that use Wi-Fi CSI to adversarial attacks. 
They investigated the impact of three white-box adversarial attacks—Fast Gradient Sign Method (FGSM), Projected Gradient Descent (PGD), and Momentum Iterative Method (MIM)—on an attention-based bidirectional long short-term memory (BiLSTM) activity classification model. Experimental results on a public Wi-Fi CSI dataset demonstrated that these adversarial attacks significantly reduced the classification accuracy of the DNN model.
Though the previous works considered the stealthy nature of adversarial samples, they did not consider the impact of adversarial samples on Wi-Fi communication.
WiCAM~\cite{xu2022wicam} builds upon adversarial attacks by defining imperceptibility, aiming to minimize the impact on Wi-Fi communication quality while maintaining effective attacks on Wi-Fi sensing systems. Additionally, the WiCAM employs a black-box attack strategy, using a ghost DNN to generate generalized adversarial perturbations that can be applied across various Wi-Fi sensing models. This approach enhances WiCAM's versatility and adaptability, making it suitable for a broader range of Wi-Fi sensing tasks.

Most prior works focused on adversarial attacks in the digital domain, which primarily operate by injecting perturbations directly into the input data of sensing systems, targeting the deep learning models responsible for activity classification. 
These methods focus on manipulating the data used by the system without altering the physical signals. In contrast, adversarial attacks in the physical domain involve directly overlaying adversarial perturbation signals on the received Wi-Fi sensing signals, thereby disrupting the CSI at the physical layer. 
Liu et al.~\cite{liu2022physical, liu2024time} are the first to implement a \textbf{physical-world} adversarial attack on Wi-Fi-based HAR systems, leveraging physical-layer jamming on Wi-Fi CSI signals rather than targeting deep learning models for HAR at the data level.
Their approach uses carefully timed jamming signals to induce packet loss in the CSI data, causing specific distortions that mislead the HAR system. This study introduces both untargeted (black-box) attacks and targeted (gray-box) attacks, allowing the system to misclassify a specific behavior as a designated alternative, thereby achieving a more precise and directed adversarial effect while minimizing the detectability of the interference.
Similarly, Li et al.~\cite{li2024practical} propose a practical physical adversarial attack method targeting deep learning-driven Wi-Fi sensing systems by embedding imperceptible perturbations in the preamble of Wi-Fi packets, thereby manipulating the CSI at the receiver side. This approach enables both targeted and untargeted attacks on Wi-Fi sensing-based HAR systems, under either white-box or black-box settings. The attack is designed to be stealthy, minimally impacting communication quality, making it difficult to detect, and demonstrating a high success rate.

While most adversarial samples are derived from the same type of Wi-Fi signals, it is worth noting that adversarial interference can also originate from other signal modalities.  For instance,
IS-WARS~\cite{huang2021wars} introduces the concept of cross-technology interference (CTI) by leveraging other ambient RF signals, such as ZigBee, to generate interference within Wi-Fi's 2.4~GHz band, thereby misleading human activity recognition systems. Compared to traditional interference methods, the CTI strategy used in IS-WARS is highly stealthy, with carefully adjusted signal strength and frequency that allows it to subtly disrupt Wi-Fi-based recognition systems without significantly impacting Wi-Fi communication quality.
However, due to the frequency band limitations of the interfering signals, this study only targets Wi-Fi sensing in the 2.4~GHz band and does not explore its effects on Wi-Fi sensing in the 5~GHz band.

While adversarial attacks have primarily targeted Wi-Fi-based HAR systems, recent research has extended these efforts to radar-based HAR systems. For instance, 
Xie et al.~\cite{xie2023universal} propose a universal targeted adversarial attack method tailored for mmWave sensing-based HAR systems. This approach enables both targeted and untargeted attacks in a black-box setting, leveraging knowledge distillation and GANs to enhance the attack's generalizability and stealthiness. The method effectively misclassifies activities into specified target classes, achieving over 90\% attack success rate in experiments. 
In addition to mmWave radar, Nallabolu et al.~\cite{nallabolu2023emulation} proposed a method to emulate or spoof 5.8~GHz ISM-band Doppler and FMCW radars using commercial off-the-shelf components. By injecting fake signals, attackers can mimic walking patterns, respiratory motion, and alter the perceived distance of static targets, all without requiring synchronization with the target radar. 

\begin{table*}[htbp]
\centering
\caption{Comparison of Label Flipping Attack and Adversarial Attack}
\renewcommand{\arraystretch}{1.4} 
\setlength{\tabcolsep}{5.3pt} 
\begin{tabular}{l c c }
\bottomrule
   & \textbf{Label Flipping Attack}                       & \textbf{Adversarial Attack}              \\ \bottomrule
\textbf{Operation Data}        & Training Data                                        & Testing Data                                    \\
\textbf{Modified Part}         & Labels (no changes to the data itself)               & Input features                                  \\
\textbf{Modification Scope}    & Explicit and visible label changes                   & Subtle and imperceptible feature perturbations  \\
\textbf{Attack Timing}         & Before model training                                & During model inference                          \\
\textbf{Impact}                & Long-term degradation of model performance           & Short-term misclassification of specific inputs \\ \bottomrule
\end{tabular}
\label{tab:attack_comparison}
\end{table*}

\subsubsection{\textbf{Label Flipping Attacks}}
The above-mentioned adversarial attacks are primarily focused on exploiting vulnerabilities in the inference process, another equally critical vector for compromising HAR systems lies in the training phase. Label flipping attacks, for example, target the integrity of training datasets by systematically or randomly altering activity labels, which misguides the learning process and fundamentally undermines the model's ability to generalize. Unlike adversarial examples that operate at the inference stage, label flipping attacks introduce disruptions during model development, posing long-term threats to the robustness and reliability of HAR systems. 
A comparison of the label flipping attack with the adversarial attack is shown in Table~\ref{tab:attack_comparison}.

Singha et al.~\cite{singha2024securing} first investigate the vulnerability of mmWave-based HAR systems to label flipping attacks and introduces three attack strategies: random attack, cross-trajectory attack, and inner-trajectory attack. Random attacks, which randomly alter labels, have the most significant impact on overall performance. Random attacks significantly impact overall performance by randomly altering labels, cross-trajectory attacks increase confusion by modifying labels to entirely different activities, and inner-trajectory attacks covertly disrupt contrastive learning by altering labels within similar trajectories. Experimental results demonstrate that these attacks significantly degrade classification accuracy, especially in contrastive learning models, highlighting the sensitivity of HAR systems to label flipping attacks and the critical need for robust defenses.

\subsection{Gesture Recognition}
Gesture recognition, particularly through RF sensing technologies like Wi-Fi and mmWave, detects subtle hand or body movements by analyzing disruptions in signal patterns. By capturing changes in phase and amplitude, these systems enable touchless control across applications such as smart homes, gaming, and automotive interfaces.
However, this fine-grained recognition comes with vulnerabilities. When gesture recognition models are attacked, they may misinterpret gestures or fail to respond accurately, resulting in unintended actions or posing security risks in interactive environments.

Ozbulak et al.~\cite{ozbulak2021investigating} were the first to propose adversarial attacks targeting CNN-based radar gesture recognition systems, revealing their significant vulnerability to adversarial examples. They demonstrated that these systems are not only susceptible to traditional white-box and black-box attacks but also designed an innovative Padding Attack. This attack successfully misleads model predictions by modifying the padding frames in the input data, without altering the key frames where gestures occur, showcasing for the first time how the specific structural characteristics of radar data can be exploited for stealthy adversarial attacks. Furthermore, the authors used Grad-CAM to deeply analyze the relationship between adversarial perturbations and critical model features, discovering that adversarial attacks tend to focus on the most significant features for model predictions. This finding establishes a novel link between adversarial optimization and model interpretability.

WiAdv~\cite{zhou2022wiadv} is the first to explore the feasibility of physical adversarial attacks against Wi-Fi-based gesture recognition systems. By leveraging signal synthesis techniques, WiAdv utilizes the dynamic multipath characteristics of Wi-Fi signals to simulate gesture motion features and generate adversarial signals. To address challenges posed by the non-continuous and non-differentiable processing modules in Wi-Fi signal pipelines, WiAdv introduces two attack strategies: Constant Attack and Greedy Attack. These strategies construct interference signals using constant and dynamic Doppler shifts, respectively, optimizing the temporal and frequency dimensions of the attack. Additionally, WiAdv enhances the robustness of adversarial signals in physical environments by adjusting signal power and reducing frequency jitter.

WiIntruder~\cite{cao2024security} introduces a universal black-box perturbation attack scheme targeting Wi-Fi sensing systems. By injecting perturbation signals into the physical space, WiIntruder can mislead deep learning-based Wi-Fi sensing applications, including not only the HAR systems but also systems like user authentication, respiratory monitoring, and indoor localization, causing them to produce incorrect recognition or prediction results. Additionally, the scheme leverages a Generative Adversarial Network (GAN) to generate diverse perturbation surrogates, enhancing attack stealthiness and effectively mitigating signal distortion issues in propagation.

It is worth noting that gesture recognition task is essentially a more refined form of HAR tasks focusing on intricate movements like the motion of a single finger or the rotation of a hand, rather than broader actions such as walking or running. Therefore, some of the current attack schemes against HAR systems~\cite{xu2022wicam, liu2022physical, liu2024time} are also effective against gesture recognition systems.

\begin{table*}[ht!]
\centering
\renewcommand{\arraystretch}{1.4} 
\setlength{\tabcolsep}{4pt} 
\caption{\textbf{Summary of Signal-level Attacks to Autonomous Vehicles}}
\begin{tabular}{c c c cc cc c c}
\toprule
\multirow{2}{*}{\textbf{Reference}} & \multirow{2}{*}{\textbf{Year}} & \multirow{2}{*}{\textbf{Attack Type}} & \multicolumn{2}{c}{\textbf{Target Type}} & \multicolumn{2}{c}{\textbf{Assumption}} & \multirow{2}{*}{\textbf{Attack Outcomes}} & \multirow{2}{*}{\textbf{Validation}} \\
\cmidrule(lr){4-5} \cmidrule(lr){6-7}
 &  &  & \textbf{Targeted} & \textbf{Untargeted} & \textbf{White-box} & \textbf{Black-box} &  &  \\
\toprule

\multirow{2}{*}{\cite{yan2016can}} & \multirow{2}{*}{2016} & Jamming &    & \ding{51} & &  \ding{51}  & \makecell{Disable obstacle  detection} & \multirow{2}{*}{Lab Simulation + Vehicle} \\ \cmidrule(lr){3-8}
 &  & Spoofing & \ding{51} &   & \ding{51} &   & \makecell{Fake targets} &  \\ \hline

\multirow{2}{*}{\cite{yeh2017security}} & \multirow{2}{*}{2017} & Jamming  &    & \ding{51} & &  \ding{51}  & \makecell{Disrupt radar  operation} & \multirow{2}{*}{ Lab Simulation} \\ \cmidrule(lr){3-8}
 &  & Spoofing & \ding{51} &   & \ding{51} &   & \makecell{Fake targets,  modify distance or speed} &  \\ \hline

\cite{miura2019low} & 2019 & Spoofing & \ding{51} &   & \ding{51} &   & \makecell{Distance estimation error} & Lab Simulation \\ \hline

\cite{komissarov2021spoofing} & 2021 & Spoofing & \ding{51} &   & \ding{51} &   & \makecell{Fake target distance  and speed data} & Lab Simulation \\ \hline

\cite{sun2021control} & 2021 & Spoofing & \ding{51} &   & \ding{51} &   & \makecell{Fake targets,  modify positions} & Lab Simulation + Vehicle \\ \hline

\cite{ordean2022millimeter} & 2022 & Spoofing & \ding{51} &   &   & \ding{51} & \makecell{Fake targets and hide real targets} & Lab Simulation \\ \hline

\cite{hunt2023madradar} & 2023 & Spoofing & \ding{51} &   &   & \ding{51} &\makecell{Fake targets, hide real targets, \\ and translate  target positions} & Lab Simulation \\






\bottomrule
\end{tabular}
\label{tab:av_attacks}
\end{table*}

\subsection{Authentication}
RF sensing-based \textbf{user authentication} leverages unique biometric and behavioral patterns through signal analysis for continuous, touch-free identity verification. In addition to user authentication, RF sensing is also utilized for \textbf{device authentication} by relying on physical-layer fingerprints (e.g., CSI) that uniquely identify devices. These approaches are widely applied in security systems, including restricted area access and device pairing. However, both user and device authentication are vulnerable to adversarial attacks, which can disrupt the integrity of the authentication process. For user authentication, attackers may mimic or spoof biometric patterns, while for device authentication, they may manipulate the surrounding environment to interfere with physical-layer fingerprints, allowing unauthorized access or causing legitimate devices to fail authentication.

Wi-Fi sensing-based user authentication systems are vulnerable to adversarial attacks. For instance, recent researches~\cite{xu2022wicam, cao2024security, li2024practical} focus on targeting user authentication systems based on biometric features such as gait, gestures, or breathing patterns. These methods introduce carefully crafted adversarial perturbations into Wi-Fi sensing signals, disrupting the decision-making process of deep learning models and causing them to produce incorrect user identity classifications. This leads to authentication failures for legitimate users or successful impersonation by attackers. 

In addition to user authentication, device authentication systems based on Wi-Fi CSI are also susceptible to significant security threats. The PhyFinAtt framework~\cite{huang2023phyfinatt} targets device authentication systems reliant on physical-layer fingerprints (PHY-layer information). By altering the physical environment (e.g., simulating human movement) to dynamically affect channel characteristics or manipulating wireless signal propagation paths, it disrupts the stability of device-specific PHY fingerprints, ultimately preventing legitimate devices from being authenticated. 
Unlike attacks on user authentication, PhyFinAtt exploits the high sensitivity of PHY fingerprints to environmental changes by externally interfering with signal characteristics (e.g., the amplitude and phase of CSI). This approach challenges the assumption of stability and uniqueness of PHY-layer fingerprints, revealing vulnerabilities in what are traditionally considered robust security features.

These two types of attacks highlight the potential security risks in Wi-Fi sensing systems for both user and device authentication scenarios. 
These studies not only expose the weaknesses in existing RF fingerprint-based authentication mechanisms but also underscore the necessity of enhancing systems to resist external interference. 
Improving the robustness of deep learning models and strengthening the resilience of PHY-layer fingerprints against environmental changes are crucial steps toward ensuring comprehensive security in Wi-Fi sensing applications.

\subsection{Vital Sign Monitoring}
RF sensing-based vital signs monitoring systems have been widely used in the fields of health monitoring, identity authentication, and security protection. 
However, these systems face the security threat of having their vital sign signals falsified or interfered with by attackers, which may lead to serious consequences. For example, in telemedicine scenarios, an attack may allow the system to misdiagnose abnormal conditions (e.g., sleep apnea, abnormal heart rhythms) of a patient that cannot be detected in a timely manner, affecting medical decisions. In addition, an attacker can forge the vital sign characteristics of a specific person to disable a contactless identification system based on cardiopulmonary exercise, resulting in the security risk of identity impersonation or illegal access.

Rodriguez et al.~\cite{rodriguez2021spoofing} investigates spoofing attacks on Doppler radar motion sensors using portable RF devices such as PIN diodes, analog phase shifters, and arbitrary signal generators. It proposes and validates two attack methods: BPSK modulation spoofing and analog phase shifter spoofing, successfully forging heartbeat and respiration signals. These attacks enable deception of vital sign monitoring and biometric authentication systems, highlighting potential security vulnerabilities in radar-based surveillance and identification technologies.
Ambalkar et al.~\cite{ambalkar2023adversarial} studied adversarial attacks on a Wi-Fi-based apnea detection system. The study evaluates three white-box adversarial attack methods: fast gradient sign method (FGSM), projected gradient descent (PGD), and momentum iteration method (MIM). Experimental results show that these attacks significantly reduce the classification accuracy of apnea detection models.

\subsection{Indoor Localization}
RF sensing-based indoor localization has gained significant attention in recent years due to its ability to determine precise positions without requiring additional infrastructure. Attacks on localization systems can introduce errors in positioning, leading to misdirected navigation, incorrect asset locations, or compromised emergency responses. 

Among various RF modalities, Wi-Fi has emerged as a core solution for indoor localization due to its low cost and widespread deployment. Wi-Fi-based localization systems, leveraging either RSSI or CSI, have proven effective in diverse scenarios. However, their vulnerability to adversarial attacks poses significant challenges to their security and reliability.
\textbf{Wi-Fi RSSI}-based localization systems, widely adopted for their simplicity and efficiency, have been shown to be highly susceptible to adversarial perturbations. For example, Patil et al.~\cite{patil2021adversarial} introduced three white-box attack methods (FGSM, PGD, and MIM) that apply small perturbations to RSSI data, significantly degrading the performance of deep learning models in both classification and regression tasks, exposing their inherent vulnerabilities. 
In contrast, \textbf{Wi-Fi CSI}-based localization systems, offering finer-grained channel characteristics, achieve higher accuracy in challenging environments. However, Wang et al.~\cite{wang2022adversarial} demonstrated that these systems also face threats from adversarial examples and proposed the AdvLoc system, which employs adversarial training and residual networks to enhance model robustness against first-order attacks. 
In addition, Liu et al.~\cite{liu2023exploring} presented RAFA, a hardware-based adversarial attack platform targeting CSI-driven systems. By injecting universal adversarial perturbations into the wireless channel, RAFA significantly increased localization errors in both white-box and black-box scenarios. 

Beyond Wi-Fi-based localization, \textbf{mmWave} has gained attention for its potential to enable high-precision localization in both indoor and outdoor environments. However, these systems are also vulnerable to sophisticated attacks.
Zhao et al.~\cite{zhao2023backdoor} investigated backdoor attacks on deep learning-based mmWave localization systems. By embedding triggers, such as one-pixel modifications or invisible random noise, into the training data, they demonstrated that these systems could produce significantly erroneous location predictions when the triggers were activated, while maintaining normal performance on benign inputs.

\subsection{Autonomous Vehicle}
Autonomous vehicles utilize mmWave radar to detect surroundings, navigate roads, and avoid obstacles by interpreting signal reflections and absorption patterns. This capability is vital for decision-making and ensuring passenger and public safety. 
Attacks on these models can manipulate the vehicle's perception of its environment, causing it to misinterpret obstacles, misread road signs, or alter routes, posing severe safety risks. 
Current research on automotive mmWave radar attacks is typically categorized into jamming attacks and spoofing attacks. 

\subsubsection{\textbf{Jamming Attacks}}
Jamming attacks are generally non-targeted and operate by transmitting high-power random noise or forged signals to the radar. This interference degrades the quality of the radar's received signals, rendering its sensing results unreliable or entirely ineffective, and ultimately preventing accurate detection of the surrounding environment. Such attacks are typically classified as black-box attacks because they do not require detailed knowledge of the radar's internal parameters. Instead, jamming attacks exploit the radar's operating frequency range to overwhelm its ability to process incoming signals. By relying on brute-force interference rather than precise synchronization or parameter matching, jamming attacks effectively disrupt radar functionality without needing to exploit specific system-level details. 

Yan et al.~\cite{yan2016can}  were the first to experimentally investigate the vulnerability of automotive mmWave radar to jamming attacks. Using a signal generator to produce interference signals in the 76-77~GHz band, they successfully rendered the mmWave radar on a Tesla Model S unable to detect obstacles, demonstrating the radar's sensitivity to jamming.
Furthermore, Yeh et al.~\cite{yeh2017security} corroborated these findings by simulating jamming attacks, transmitting strong noise signals within the same frequency band as automotive mmWave radar. This saturation of the radar receiver caused detection failures. Their study also highlighted that while mmWave radar is relatively resistant to interference in highly mobile environments due to its high directionality, mobile jamming devices—such as jammers mounted on trailing vehicles—can still have a persistent disruptive effect on its functionality.

\begin{figure*}
\centering
\subfigure[Backscatter tag~\cite{lazaro2022spoofing}]{
\begin{minipage}[t]{0.14\linewidth}
\centering
\includegraphics[width=\linewidth]{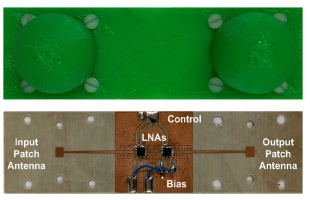}
\label{fig:tag1}
\vspace{-1cm}
\end{minipage}%
\vspace{-1cm}
}
\subfigure[Reflect array~\cite{vennam2023mmspoof}]{
\begin{minipage}[t]{0.14\linewidth}
\centering
\includegraphics[width=\linewidth]{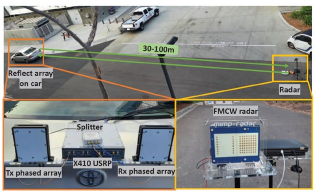}
\label{fig:tag2}
\vspace{-1cm}
\end{minipage}%
\vspace{-1cm}
}
\subfigure[Meta-material tag~\cite{chen2023metawave}]{
\begin{minipage}[t]{0.2\linewidth}
\centering
\includegraphics[width=\linewidth]{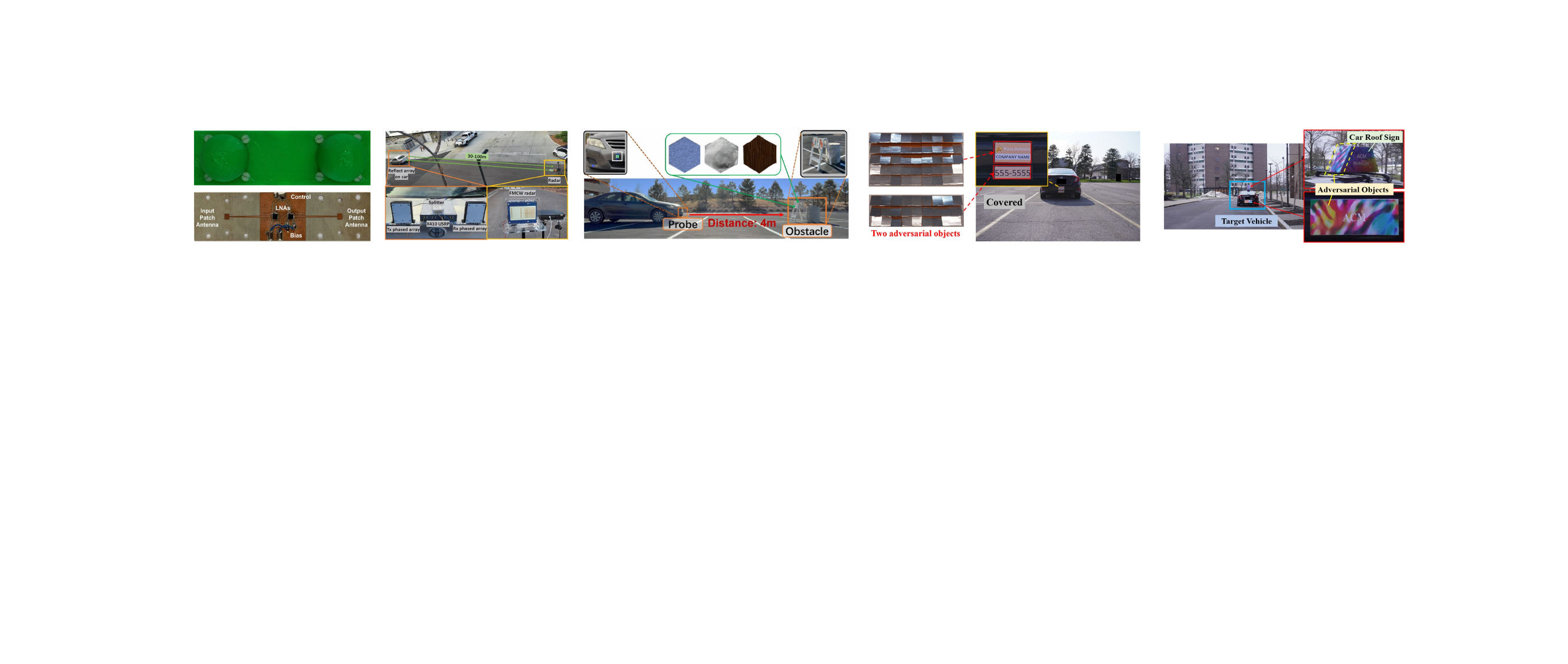}
\label{fig:tag3}
\vspace{-1cm}
\end{minipage}%
\vspace{-1cm}
}
\subfigure[Meta-material tag~\cite{zhu2023tilemask}]{
\begin{minipage}[t]{0.2\linewidth}
\centering
\includegraphics[width=\linewidth]{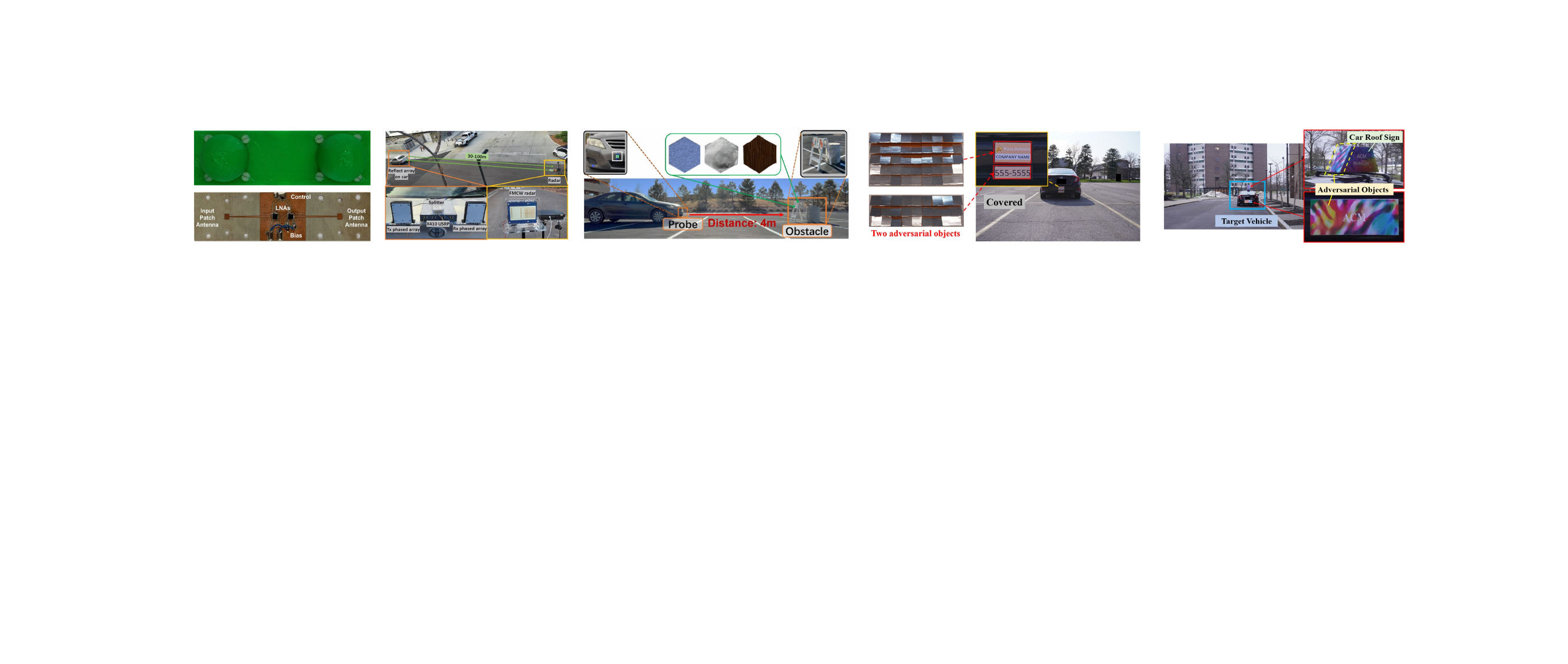}
\label{fig:tag4}
\vspace{-1cm}
\end{minipage}%
\vspace{-1cm}
}
\subfigure[Composite object~\cite{zhu2024malicious} ]{
\begin{minipage}[t]{0.2\linewidth}
\centering
\includegraphics[width=\linewidth]{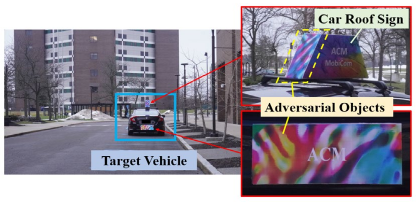}
\label{fig:tag5}
\vspace{-1cm}
\end{minipage}%
\vspace{-1cm}
}
\centering
\caption{Adversarial objects/tags leveraged in the object-level spoofing attack of autonomous vehicles.}
\label{fig:cycle}
\end{figure*}

\begin{table*}[htp!]
\centering
\renewcommand{\arraystretch}{1.5} 
\setlength{\tabcolsep}{5pt} 
\caption{\textbf{Comparison of Attack Objects in Autonomous Vehicles Radar Spoofing Attacks}}
\begin{tabular}{c c c c c c c}
\bottomrule
\textbf{Reference} & \textbf{Material Composition} & \textbf{Mechanism} & \textbf{Shape} & \textbf{Size} & \textbf{Attack Outcomes} & \textbf{Target Scene} \\
\bottomrule

\makecell{Backscatter \\Tag~\cite{lazaro2022spoofing}} & \makecell{Microstrip patch antenna, \\ LNA, microcontroller} & Semi-passive & Rectangular & 12cm × 3.7cm  & \makecell{Fake targets} & Static \\ \hline

Reflect Array~\cite{vennam2023mmspoof} & \makecell{Microstrip patch antenna, \\ LNA, mixer, metal foil} & Active & Rectangular & \makecell{10cm × 10cm, \\ 20cm × 20cm} & \makecell{Fake targets, manipulate \\ distance or speed} & Dynamic \\ \hline

\makecell{Meta-material \\Tag~\cite{chen2023metawave}} & \makecell{C-RAM LF, tin foil, \\ parallel copper wire grid} & Passive & Hexagonal & 6cm - 100cm & \makecell{Hide target, fake target, \\ manipulate speed or angle} &  Dynamic \\ \hline

\makecell{Meta-material \\Tag~\cite{zhu2023tilemask}} & Stainless steel foil & Passive & Square & 3cm × 3cm & \makecell{Hide target} & Dynamic \\ \hline

\makecell{Composite \\Object~\cite{zhu2024malicious}} & \makecell{Metal foil, stickers} & Passive & \makecell{Planar geometry,\\ curved surface} & 0.3m - 1m & \makecell{Disrupt multi-sensor fusion, \\ hide specific targets} & Dynamic \\

\bottomrule
\end{tabular}
\label{tab:attack_objects}
\end{table*}

\subsubsection{\textbf{Spoofing Attacks}}
Different from jamming attacks, spoofing attacks are typically targeted and involve generating precise forged echo signals to create false targets or manipulate the distance and velocity information of real targets. These attacks mislead the autonomous driving system's perception of the environment by injecting deliberately crafted data. For example, spoofing attacks can generate false targets to disrupt path planning or conceal real objects to create potential safety hazards. 
Fundamentally, spoofing attacks can be seen as a form of adversarial attack in the physical world, where carefully designed \textit{physical signals} or \textit{physical objects} act as adversarial perturbations to deceive the radar's sensing mechanism.
Unlike jamming attacks, spoofing requires a more detailed understanding of the radar’s signal processing mechanisms, making these attacks more sophisticated and tailored in their approach.

\textit{Signal-level spoofing attacks} target mmWave radar systems by manipulating or forging physical radar signals to mislead the radar's perception. These attacks operate directly at the signal level to create false targets, hide real targets, or manipulate key parameters like distance and velocity.

Yan et al.~\cite{yan2016can} systematically analyzed, for the first time, the security vulnerabilities of mmWave radars in autonomous vehicles. By transmitting physical adversarial signals to the target radar, the attack successfully misled it into detecting non-existent obstacles or ignoring real ones. These attacks could lead to incorrect vehicle decisions, such as triggering emergency braking or increasing collision risks. The feasibility of these attacks was validated in real-world experiments using a Tesla Model S. 
Yeh et al.~\cite{yeh2017security} further corroborated these vulnerabilities and theoretically investigated the impact of non-malicious signal interference (e.g., frequency overlap between radars of different vehicles) on radar sensing.

To reduce the cost of such attacks, Miura et al.~\cite{miura2019low} proposed a low-cost replica-based distance-spoofing attack that used inexpensive hardware, such as a replica radar chip and a microcontroller board, to target mmWave FMCW radars. By precisely synchronizing the replica radar signal with the target radar, attackers could manipulate the target radar’s distance measurements with a spoofed error margin of ±10 meters.

Komissarov et al.~\cite{komissarov2021spoofing} introduced a single-attack system targeting vehicle-mounted FMCW radars, capable of simultaneously spoofing both distance and velocity measurements. This was achieved by controlling signal delays and adjusting signal phases, ensuring that the spoofed measurements complied with physical laws, making them difficult to detect. 
Moving further, Sun et al.~\cite{sun2021control} proposed five realistic attack scenarios for mmWave radar spoofing, including emergency braking attacks, hard braking attacks, lane change attacks, multi-stage attacks, and cruise control spoofing. The feasibility of these attacks was experimentally validated on a Lincoln MKZ autonomous vehicle platform in real-world environments.

While these studies proposed effective spoofing techniques for mmWave radars, they relied heavily on precise knowledge of the radar’s internal parameters (e.g., chirp slope, frame period) or the ability to achieve signal synchronization with the target radar. Consequently, these attacks are largely classified as white-box attacks, requiring the attacker to possess detailed prior knowledge of the target radar system.
To overcome the limitations of white-box attacks, black-box frameworks have been developed to reduce reliance on target radar parameters. Ordean et al.~\cite{ordean2022millimeter} proposed a wireless asynchronous spoofing attack, demonstrating that successful attacks could be conducted without complete synchronization with the target radar. The study designed scenarios to inject virtual objects or remove real objects, and validated the effectiveness of the attacks in real-world environments using commercial off-the-shelf (COTS) hardware. This approach significantly lowered the technical barriers for carrying out such attacks while enhancing practical applicability.
Building on this, Hunr et al.~\cite{hunt2023madradar} introduced a black-box physical layer attack framework, MadRadar, capable of estimating the operating parameters of an unknown radar in real-time and executing complex attacks, including false object injection (False Positive), real object removal (False Negative), and object translation (Translation Attack). By leveraging software-defined radio (SDR), the MadRadar framework achieved high-precision attacks based solely on external observations of the radar signal, thereby bypassing the need for internal radar parameter knowledge.

\textit{Object-level spoofing attacks} exploit the vulnerabilities of mmWave radar systems by utilizing carefully designed physical objects, such as reflective surfaces or absorptive materials, to alter radar perception. These attacks manipulate the radar's response by leveraging the physical properties of objects, thereby affecting target detection.

Lazaro et al.~\cite{lazaro2022spoofing} introduces a low-cost spoofing attack scheme based on backscatter tags, utilizing a semi-passive modulated backscatter transponder. By altering the modulation frequency, the system generates false targets on the range-Doppler heatmap of FMCW radars, misleading the radar's object detection. Although experimental results in controlled laboratory settings have demonstrated the effectiveness of this method, its performance under dynamic real-world conditions remains unverified. Similarly, mmSpoof~\cite{vennam2023mmspoof} presents a spoofing approach for automotive mmWave radars using an active reflect array. This method eliminates the need for synchronization with the victim radar. By modulating the received radar signal and reflecting it back, it can fabricate arbitrary targets with specific distances and velocities on the radar’s range-Doppler map. The paper validates the feasibility of this scheme in dynamic environments with commercial automotive radars, achieving spoofing capabilities over a range exceeding 100 meters.
Both the semi-passive and active reflective devices incorporate key components such as transmitting and receiving antennas, Low-Noise Amplifiers (LNAs), and signal modulators. These components enable efficient reception, processing, and retransmission of radar signals. Through coordinated operation, the reflective devices manipulate existing radar signals without directly transmitting new ones, thereby creating deceptive target information to effectively spoof radar systems.

While the above semi-passive and active spoofing schemes achieve high precision and flexibility, their reliance on power amplification and complex modulation components increases deployment costs and limits their stealth. Addressing these limitations, fully passive solutions leverage the inherent properties of meta-materials to manipulate radar signals without requiring active circuitry. 
For instance, MetaWave~\cite{chen2023metawave} proposed a passive meta-material-based attack scheme, leveraging low-cost and easily accessible absorption, reflection, and polarization tags. By passively modulating mmWave signals, MetaWave enables precise manipulation of target distance, speed, or angle measurements. Combining simulation-optimized designs with deployment parameters, MetaWave achieves both ``disappearance attacks'' on obstacles and the generation of ``ghost targets'', significantly disrupting the environmental perception of mmWave radar systems. 
Similarly, TileMask~\cite{zhu2023tilemask} further reduces material costs and deployment complexity by introducing a passive reflection-based attack method. Using modular meta-material tags made of stainless steel foil fixed to the target surface, TileMask employs specific geometric designs and assembly methods to manipulate mmWave radar reflections, effectively hiding real targets. However, the proposed scheme targets deep neural network (DNN)-based radar object detection systems and may be ineffective for non-DNN-based radar object detection systems.

Previous attacks have predominantly targeted single sensors, (i.e. mmWave radar). However, autonomous vehicles rely heavily on multi-sensor fusion, which integrates data from multiple sources to ensure robust perception and decision-making.
Zhu et al.~\cite{zhu2024malicious} first investigated the vulnerability of multi-sensor fusion systems in autonomous vehicles, and proposed a composite adversarial object attack. This method combines materials such as metal foil and stickers to exploit their physical properties to interfere with the sensing capabilities of mmWave radar, LiDAR, and cameras. Metal foil manipulates radar signal reflections to weaken echo intensity, while adversarial patterns on stickers disrupt camera-based visual feature detection. For LiDAR, the attack leverages a combination of reflection and absorption to distort point cloud data. By simultaneously altering the input signals of these sensors, the method creates perception biases and disrupts the consistency of multi-sensor data fusion, leading to incorrect or conflicting environmental understanding at the fusion level.

\subsection{Countermeasures}
Existing defense strategies against RF sensing systems can be categorized into three main types: Physical-Level Defenses, Signal-Level Defenses, and Model-Level Defenses, as shown in Tab.~\ref{tab:defense1}.

\begin{table*}[htbp]
\centering
\caption{Summary of Defense Strategies Against RF Sensing Attacks}
\renewcommand{\arraystretch}{1.3} 
\setlength{\tabcolsep}{5.3pt} 
\begin{tabular}{l l l l}
\bottomrule
\textbf{} & \textbf{Method} & \textbf{Description} & \textbf{References} \\
\bottomrule


\multirow{2}{*}{\textbf{Physical-Level Defenses}} 
    & Sensor Enhancement & High-resolution radar. & \cite{vennam2023mmspoof} \\ 
    & & Multi-sensor fusion. & \cite{zhu2023tilemask, yan2016can, komissarov2021spoofing, hunt2023madradar, ordean2022millimeter} \\
\cmidrule(lr){2-4}
    & Environmental Interference & Passive reflectors disrupt radar echoes. & \cite{zhu2024malicious} \\ 
    & & Geofencing blocks interference. & \cite{liu2022physical} \\
\midrule
\multirow{3}{*}{\textbf{Signal-Level Defenses}} 
    & Parameter Randomization & Frequency hopping, chirp modulation & \cite{kunert2010d1, moon2022bluefmcw} \\ 
    & & prevent spoofing synchronization. & \cite{vennam2023mmspoof, sun2021control, komissarov2021spoofing, miura2019low, hunt2023madradar, ordean2022millimeter} \\
\cmidrule(lr){2-4}
    & Beamforming & Adjusts RF directionality to make & \cite{zhou2022wiadv} \\ 
    & & interference more difficult. & \\
\cmidrule(lr){2-4}
    & Protocol Enhancement & Encrypts Wi-Fi LTS fields to prevent & \cite{cao2024security} \\ 
    & & CSI-based perturbation attacks. & \\
\midrule
\multirow{2}{*}{\textbf{Model-Level Defenses}} 
    & Anomaly Detection & Detects spoofing via CSI, RSSI & \cite{liu2022physical, zhou2022wiadv} \\ 
    & & feature distribution analysis. & \cite{komissarov2021spoofing, ordean2022millimeter} \\
\cmidrule(lr){2-4}
    & Adversarial Training & Trains classifiers with attack samples & \cite{liu2022physical, cao2024security} \\ 
    & & to improve robustness. & \cite{zhu2023tilemask} \\
\bottomrule
\end{tabular}
\label{tab:defense1}
\end{table*}
\subsubsection{\textbf{Physical-Level Defenses}}
Physical-level defenses can enhance sensing systems' security through hardware enhancements or environmental interventions, making it more difficult for attackers to manipulate sensor inputs. 
\begin{itemize}[leftmargin=*]
    \item Sensor Enhancement. The approach aims to enhance the physical capabilities of the sensing system to reduce the impact of attacks. For instance, high-resolution imaging radars can improve target resolution, reduce the impact of false targets, and improve the detection of spoofing attacks ~\cite{vennam2023mmspoof}. In addition, multi-sensor fusion technology ~\cite{vennam2023mmspoof,zhu2023tilemask,yan2016can,komissarov2021spoofing,hunt2023madradar,ordean2022millimeter} can combine data from other sensors (e.g., RGB cameras, LiDAR, etc.) to enhance resistance to malicious attacks through cross-validation.
    \item Environmental Interference. Another defense strategy is to alter environmental conditions to disrupt the attacker's spoofing signals. For instance, passive reflectors or masks can alter radar echoes, thereby crippling an attacker's ability to manipulate the sensing system ~\cite{zhu2024malicious}. In addition, geofencing techniques ~\cite{liu2022physical} can limit an attacker's ability to interfere with legitimate RF signals through electromagnetic shielding or controlled signal propagation.
\end{itemize}

\subsubsection{\textbf{Signal-Level Defenses}}
Signal-level defenses can modify signal characteristics to make it more difficult for attackers to eavesdrop on or manipulate. These defenses primarily include parameter randomization, beamforming, and protocol enhancement.
\begin{itemize}[leftmargin=*]
    \item Parameter Randomization. Parameter randomization is one of the most common signal-level defense techniques. Techniques such as frequency hopping, phase randomization, and chirp modulation~\cite{kunert2010d1,moon2022bluefmcw,vennam2023mmspoof,sun2021control,komissarov2021spoofing,miura2019low,hunt2023madradar,ordean2022millimeter} make it difficult for attackers to predict and synchronize their spoofing signals with victim radar signals, thereby reducing the success rate of spoofing attacks.
    \item Beamforming. Beamforming techniques dynamically adjust the directionality of legitimate RF signals, increasing the difficulty of an attacker's interference~\cite{zhou2022wiadv}.
    \item Protocol Enhancement. Certain wireless protocols may contain inherent security vulnerabilities that attackers can exploit. For instance, the Long Training Sequence (LTS) field of the Wi-Fi protocol can be eavesdropped on, allowing attackers to infer CSI and launch perturbation attacks~\cite{cao2024security}. Encrypting or dynamically modifying the LTS field can prevent attackers from obtaining critical channel information.
\end{itemize}
\subsubsection{\textbf{Model-Level Defenses}}
Model-level defenses primarily rely on machine learning and artificial intelligence techniques to detect attacks and enhance model robustness. These defenses can be categorized into anomaly detection and adversarial training.
\begin{itemize}[leftmargin=*]
    \item Anomaly Detection. This approach trains anomaly detection models using specific RF signal characteristics to identify malicious spoofing signals. For instance, feature space anomaly detection identifies malicious attacks by training an anomaly detector by analyzing the CSI~\cite{liu2022physical, zhou2022wiadv} or RSSI~\cite{komissarov2021spoofing, ordean2022millimeter} feature distribution. However, these methods may have false alarms due to ambient noise.
    \item Adversarial Training. Adversarial training makes the classifier more robust against spoofing attacks by adding attack samples during model training ~\cite{liu2022physical,cao2024security,zhu2023tilemask }. However, this approach increases the computational overhead and may reduce the classification accuracy of normal samples.
\end{itemize}

While various defense strategies have been proposed to counter attacks targeted RF sensing systems, most existing studies primarily discuss potential countermeasures rather than providing detailed designs and comprehensive evaluations. 
Further research is needed to develop practical, systematically validated defense mechanisms that can be effectively deployed in real-world scenarios.

\subsection{Summary and Insights}
\subsubsection{\textbf{Summary}}
This section systematically reviews the model integrity threats faced by RF sensing systems across six representative application scenarios: HAR, gesture recognition, authentication, indoor localization, autonomous vehicles, and vital sign monitoring. 
With the growing adoption of RF sensing in contactless sensing tasks, the heavy reliance on deep learning models has exposed these systems to various forms of attacks during both the training and inference stages, including adversarial examples, physical-domain perturbations, label flipping, and backdoor attacks, etc.
Each task exhibits unique vulnerabilities and attack surfaces due to differences in signal processing techniques, sensing objectives, and application contexts. 
For example, HAR and gesture recognition are highly sensitive to subtle motion features and can be easily misled by fine-grained perturbations; 
authentication systems are susceptible to behavior imitation or channel manipulation; 
indoor localization and autonomous driving face multi-modal attacks from both the RF channel and physical environment; 
and vital sign monitoring can be compromised by forged physiological signals.

In addition, this section summarizes the primary defense strategies proposed to date, including physical-layer enhancement, signal-level randomization, and model-level robustness improvement. However, most existing studies adopt task-specific designs, lacking systematic deployment and real-world validation. Significant challenges remain in improving the generalizability and practicality of defense mechanisms for RF sensing systems.

\subsubsection{\textbf{Insights}}
The intrinsic security challenges of RF sensing systems lie in the fundamental contradiction between their dependence on data-driven models and the openness of the physical environment from which their sensing inputs originate. On the one hand, deep learning models endow RF sensing with powerful capabilities for recognition and classification. On the other hand, adversaries can exploit the black-box nature of these models, vulnerabilities in training data, and the manipulability of physical-layer signals to conduct highly precise and stealthy attacks—without needing physical access to the device or modification of communication protocols.

Case studies across six representative RF sensing tasks show that most of the current attacks have been migrated from the AI security domain, particularly adversarial examples, label-flipping, and backdoor attacks. 
These have been effectively adapted to RF-specific contexts such as HAR, gesture recognition, and authentication. 
At the same time, several emerging threats from AI security—which have not yet been widely explored in RF sensing—exhibit strong potential for future adaptation and systemic impact:
\begin{itemize}[leftmargin=*]
    \item Model extraction attacks: Adversaries query RF models with diverse input signals and analyze outputs to infer internal structures or decision boundaries, especially in edge-deployed black-box settings.
	\item Data inference attacks: By observing confidence scores or output behaviors, attackers may determine whether a sample was used during training or reconstruct private user attributes such as behavior patterns or physiological traits;
	\item Federated learning attacks: In distributed RF sensing scenarios, adversaries may upload malicious local updates to poison the global model or reverse-engineer gradient information, undermining system-wide integrity.
	\item Mask-aware attacks: These target lightweight, pruned models often used in edge devices by exploiting sparsity patterns to perform low-cost, structure-sensitive manipulations that evade detection.
\end{itemize}
These attacks indicate that the threat surface of RF sensing systems has extended well beyond input perturbations, now encompassing training procedures, signal propagation paths, model architectures, and even multi-device collaborative frameworks. 
This evolving landscape forms a closed-loop attack chain spanning algorithmic, physical, and system layers, demonstrating that RF sensing security is not merely a subset of AI safety, but a multi-dimensional challenge requiring coordinated protection across the physical–algorithm–application stack.

More critically, RF sensing systems are not solely AI-driven, they are deeply embedded within wireless communication infrastructures, governed by physical-layer interactions and protocol-level dynamics. 
Most research has focused on AI layer threats, while well-established attack strategies from the wireless communication layer remain largely overlooked. 
This gap reveals a crucial blind spot in current RF sensing security research. For example:
\begin{itemize}[leftmargin=*]
    \item CSI spoofing: Adversaries emulate the channel signature of a legitimate user or device, deceiving authentication or activity recognition systems into accepting a forged identity; this compromises user-level trust and enables impersonation or unauthorized access in device-free sensing environments.
    
    \item Relay attacks: Real-time signal forwarding misleads the system about the spatial location or motion trajectory of the target, compromising localization and behavior tracking; this severely undermines the spatial integrity of RF sensing systems and poses critical risks in applications such as indoor navigation or autonomous driving.
    
    \item Multipath manipulation: By shaping reflected paths, attackers can synthesize deceptive gesture or motion patterns, misleading HAR models; this results in false activity detection, reduced recognition accuracy, and can disrupt context-aware services relying on behavioral input.
    
    \item Synchronization spoofing: Disrupting frame synchronization introduces temporal misalignment, corrupting time-series or frequency-domain representations like spectrograms or CSI streams; this degrades signal quality and causes significant errors in model inference for time-sensitive tasks such as vital sign monitoring and autonomous driving.
\end{itemize}
These communication-layer attacks are inherently non-invasive, stealthy, and protocol-transparent, enabling adversaries to intervene at the signal source level and compromise the integrity of sensing pipelines. Systematically integrating these wireless-originated threats into RF sensing security research is vital to move beyond the current AI-centric paradigm, and toward the development of cross-layer, cross-domain security frameworks that better reflect the hybrid nature of RF systems.

Meanwhile, existing defense strategies remain largely fragmented, task-specific, and lack unified benchmarks or generalization capability. Most are tailored to specific attacks or application settings, offering limited adaptability across tasks and architectures. In the face of increasingly co-evolving offensive techniques, RF sensing urgently requires the development of transferable, verifiable, and deployable system-level defense architectures, ultimately facilitating a strategic transition from merely trustworthy sensing to trustworthy cognition.

\begin{figure}
    \centering
    \includegraphics[width=0.9\linewidth]{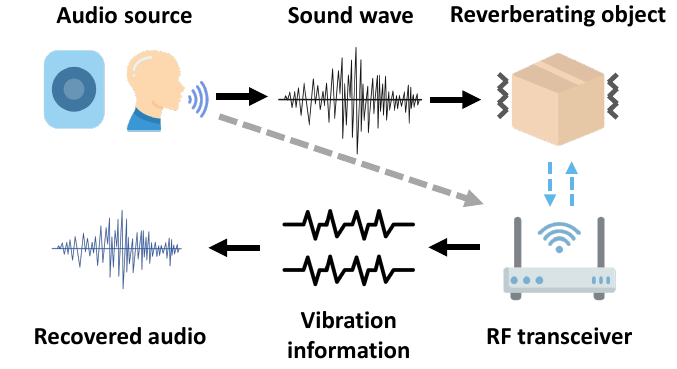}
    \caption{Acoustic eavesdropping via RF sensing.}
    \label{fig:acoustic-eve}
\end{figure}

\section{Exploiting RF Sensing for Privacy Invasion}
\label{sec:attack2}
\subsection{Acoustic Eavesdropping via RF Sensing}
The propagation characteristics of RF signals allow them to penetrate physical barriers such as walls and windows, making them a potential security threat in acoustic eavesdropping within private spaces. By capturing and analyzing RF signals, attackers can extract audio information from subtle vibrations in the environment, such as those caused by speakers, throats, or the surfaces of other objects, enabling them to eavesdrop on sensitive conversations or other acoustic information that should remain isolated.
This principle is illustrated in Fig.~\ref{fig:acoustic-eve}.

RF-based acoustic eavesdropping can be broadly divided into two categories: Constrained Vocabulary Eavesdropping (CVE) and Unconstrained Vocabulary Eavesdropping (UVE). 
CVE focuses on monitoring and analyzing a predefined, limited set of vocabulary, which typically includes digits, letters, and common keywords or commands. Attackers capture RF signal vibration patterns related to these specific words and train machine learning or deep learning models to recognize and reconstruct these words. To improve accuracy, CVE typically relies on large training datasets. Using this method, attackers may steal sensitive information such as numeric passwords, voice commands, and other authentication data. 
In contrast, UVE is not limited to a specific set of words and can capture and analyze any vocabulary or sentence in the environment. This enables UVE to reconstruct more complex and natural language content. Attackers can use advanced vibration extraction algorithms or machine learning techniques to reconstruct complete audio information from RF signals, thereby eavesdropping on and recovering private conversations, meeting records, phone calls, and other instances involving a wide range of language expressions. UVE is suitable for scenarios where arbitrary conversational content needs to be reconstructed and is typically more challenging than CVE.

\begin{table*}[htp!]
\centering
\begin{threeparttable}
\renewcommand{\arraystretch}{1.6} 
\setlength{\tabcolsep}{5.3pt} 
\caption{\textbf{Comparison of Acoustic Eavesdropping via RF Signals}}
\label{tab:acoustic}
\begin{tabular}{c c c c c c c c c c}
\bottomrule
\textbf{Attack Scheme} & \textbf{Year} & \textbf{Audio Source} & \makecell{\textbf{RF Signal Type} \\ \textbf{\& Hardware}}& \makecell{\textbf{Max.}\\ \textbf{Distance}\tnote{1}} &  \makecell{\textbf{Through-}\\ \textbf{wall}} & \makecell{\textbf{Training}\\ \textbf{Effort}\tnote{2}} &
\makecell{\textbf{Evaluation}\\ \textbf{Metrics}\tnote{3}}&
\makecell{\textbf{Audio }\\ \textbf{$f_s$}\tnote{4}}&
\makecell{\textbf{Eavesdropping}\\ \textbf{Type}}\\
\bottomrule

WiHear~\cite{wang2014we} & 2014 & \makecell{Human\\Subject} & \makecell{2.4GHz band \\ (USRP N210)}  &  - & \ding{51} & \CIRCLE & CA & - &  \makecell{Word\\Recognition}\\ \hline

WaveEar~\cite{xu2019waveear} & 2019 & \makecell{Human\\Subject} & \makecell{24GHz band\\(mmWave Radar)}  &  2m & \ding{55} & \RIGHTcircle & WER, MCD & 16kHz &  \makecell{Full\\Speech}\\ \hline

Wavesdropper~\cite{wang2022wavesdropper} & 2022 & \makecell{Human\\Subject} & \makecell{77GHz band \\ (mmWave Radar)}  &  5m & \ding{51} & \CIRCLE & CA & 0.4kHz &  \makecell{Word\\Recognition}\\ \hline

RF-Mic~\cite{chen2023rf} & 2023 & \makecell{Human\\Subject} & \makecell{920MHz \& 2.4GHz \\ (RFID\&USRP N210)}  &  2.6m & \ding{55} & \RIGHTcircle & CA, WER & 1.2kHz &  \makecell{Word\\Recognition}\\ \hline

mmMIC~\cite{fan2023mmmic} & 2023 & \makecell{Human\\Subject} & \makecell{60GHz band \\ (mmWave Radar)}  &  3m & \ding{55} & \CIRCLE & CA & 4kHz &  \makecell{Phonetic\\Recognition}\\ \hline

Wei et al.~\cite{wei2015acoustic} & 2015 & \makecell{Loudspeaker,\\Smartphone} & \makecell{2.4GHz band \\ (WARP SDR)}  &  4m & \ding{51} & \Circle & CA & 19.5kHz &  \makecell{Full\\Speech}\\ \hline

UWHear~\cite{wang2020uwhear} & 2020 & \makecell{Loudspeaker} & \makecell{10GHz band\\(IR-UWB Radar)}  &  9m & \ding{51} & \Circle & - & 1.2kHz &  \makecell{Tone\\Signal}\\ \hline

Tag-Bug~\cite{wang2021thru} & 2021 & \makecell{Loudspeaker} & \makecell{920MHz \& 2.4GHz \\ (RFID\&USRP N210)}  &  4m & \ding{51} & \RIGHTcircle & CA & 2kHz &  \makecell{Word\\Recognition}\\ \hline

MILLIEAR~\cite{hu2022milliear} & 2022 & \makecell{Loudspeaker} & \makecell{77GHz band \\ (mmWave Radar)}  &  5m & \ding{51} & \CIRCLE & MCD & 10kHz &  \makecell{Full\\Speech}\\ \hline

Feng et al.~\cite{feng2023mmeavesdropper} & 2023 & \makecell{Loudspeaker} & \makecell{77GHz band \\ (mmWave Radar)}  &  3m & \ding{55} & \RIGHTcircle & CA & 10kHz &  \makecell{Word\\Recognition}\\ \hline

mmPhone~\cite{wang2022mmphone} & 2022 & \makecell{Piezoelectric\\Film} & \makecell{77GHz band \\ (mmWave Radar)}  &  7m & \ding{51} & \CIRCLE & CA, STOI & 10.2kHz &  \makecell{Full\\Speech}\\ \hline

mmEcho~\cite{hu2023mmecho} & 2023 & \makecell{Everyday\\Object} & \makecell{77GHz band \\ (mmWave Radar)}  &  5m & \ding{51} & \Circle & WER, MCD & 10kHz &  \makecell{Full\\Speech}\\ \hline

VibSpeech~\cite{wang2024vibspeech} & 2024 & \makecell{Everyday\\Object} & \makecell{77GHz band \\ (mmWave Radar)}  &  5m & \ding{51} & \CIRCLE & MCD & 8kHz &  \makecell{Full\\Speech}\\ \hline

mmSpy~\cite{basak2022mmspy} & 2022 & \makecell{Smartphone} & \makecell{77GHz band \\ (mmWave Radar)}  &  1.82m & \ding{55} & \CIRCLE & CA & 8kHz &  \makecell{Word\\Recognition}\\ \hline

mmEve~\cite{wang2022mmeve} & 2022 & \makecell{Smartphone} & \makecell{77GHz band \\ (mmWave Radar)}  &  8m & \ding{55} & \CIRCLE & WER, STOI & 10.2kHz &  \makecell{Full\\Speech}\\ \hline

mmEar~\cite{xu2024mmear} & 2024 & \makecell{Headset} & \makecell{77GHz band \\ (mmWave Radar)}  &  2m & \ding{55} & \CIRCLE & STOI & 6kHz &  \makecell{Full\\Speech}\\ \hline

RFSpy~\cite{chen2024rfspy} & 2024 & \makecell{Headset} & \makecell{920MHz \& 2.4GHz \\ (RFID\&USRP N210)}  &  4.5m & \ding{51} & \CIRCLE & MCD, WER & 3kHz &  \makecell{Full\\Speech}\\ \hline

RF-Parrot~\cite{wang2024wireless} & 2024 & \makecell{Earphone\\Wires} & \makecell{USRP N210}  &  1m & \ding{51} & \CIRCLE & \makecell{MCD,\\WER, STOI} & 8kHz &  \makecell{Full\\Speech}\\ 

\bottomrule
\end{tabular}

\begin{tablenotes}
    \item[1] The distance between the victim and the RF receiver.
    \item[2]\Circle ~for Low (Training free), \RIGHTcircle  ~for Middle (Less than 10 samples for training), \CIRCLE  ~for High (More than 10 samples for training).
    \item[3] Objective metrics used when evaluating system performance: Classification Accuracy (CA), Mel-Cepstral Distortion (MCD), Word Error Rate (WER), Short-Time Objective Intelligibility (STOI).
    \item[4] Audio Sampling Frequency ($f_s$) is twice the maximum frequency response of the audio.
\end{tablenotes}

\end{threeparttable}

\end{table*}

\subsubsection{\textbf{Constrained Vocabulary Eavesdropping (CVE)}}

As a representative study in RF-based speech sensing, Wang et al.~\cite{wang2014we} proposed the WiHear system, which can monitor subtle mouth movements and infer spoken content without direct contact with the target. WiHear uses MIMO beamforming to precisely focus on the speaker’s mouth, capturing tiny RF reflections caused by speech. It then applies waveform analysis and machine learning to achieve lip reading based on wireless signals. 
Unlike inferring speech content by detecting mouth movements, Wavesdropper~\cite{wang2022wavesdropper} further utilizes subtle throat vibrations to directly capture acoustic information. This technique employs commercial mmWave radar combined with CEEMD dynamic clutter suppression and wavelet analysis techniques to extract clean speech signals, and uses a ResNet-based deep neural network model, WavesdropNet, for the retrieval of specific words. 
As a successor to human speech sensing technologies, RF-Mic~\cite{chen2023rf} proposes an eavesdropping approach that utilizes RFID tags attached to the bridge of eyeglasses. By capturing subtle facial speech dynamics, such as lip movements and bone conduction vibrations, RF-Mic effectively achieves the recognition of specific vocabulary.



Shifting the focus of attacks to eavesdropping on audio emitted by loudspeakers, UWHear~\cite{wang2020uwhear} proposed using sub-10~GHz band IR-UWB radar for non-contact acoustic eavesdropping. The system employs techniques such as phase noise correction, static clutter suppression, and vibrating object localization, without relying on machine learning. However, this study did not evaluate the system's performance in recovering human speech.
Similarly, Wang et al.~\cite{wang2021thru} proposed Tag-Bug, a through-wall eavesdropping system based on RFID tags that reconstructs audio by capturing vibrations from loudspeakers. This approach attaches RFID tags to various everyday objects to sense sub-millimeter-level vibrations and introduces a Conditional Generative Adversarial Network (CGAN) to compensate for incomplete frequency spectra.
mmEavesdropper~\cite{feng2023mmeavesdropper} further proposed a speech eavesdropping system based on commercial mmWave radar, utilizing beamforming and Chirp-Z transform for signal enhancement to improve audio recovery performance, and employing an encoder-decoder model for audio reconstruction.
To extend the target of eavesdropping to private phone conversations, mmSpy~\cite{basak2022mmspy} proposes a remote eavesdropping scheme based on commercial mmWave radar. The system captures subtle vibrations from earpiece to reconstruct the content of phone calls. By integrating phase noise correction, multipath suppression, and a deep learning model based on RCED, the system achieves speech recognition even in noisy environments.

Table~\ref{tab:acoustic} provides a detailed comparison of related works, showing that due to limited signal resolution, CVE is typically restricted to reconstructing low-sampling-rate audio. This low bandwidth leads researchers to focus more on detecting and analyzing specific keywords such as digits, letters, or common commands. To accomplish this task, systems often rely on machine learning, which introduces additional training effort.
As technology advances, eavesdropping mediums have evolved from Wi-Fi signals to IR-UWB radar, and now to mmWave radar. The continuous improvement in wireless signal sensing capability and granularity has made UVE increasingly feasible.

\subsubsection{\textbf{Unconstrained Vocabulary Eavesdropping (UVE)}}
Wei et al.~\cite{wei2015acoustic} proposed a Wi-Fi-based remote speech eavesdropping attack capable of recovering audio through walls. The study reveals that vibrations from the sound source can modulate nearby wireless signals, and by analyzing these modulated signals, an attacker can reconstruct complete speech information.
Xu et al.~\cite{xu2019waveear} further proposed a speech reconstruction technique based on mmWave radar, named WaveEar. The system utilizes mmWave signals to detect subtle skin vibrations near the throat, capturing vocal cord movement to reconstruct human speech. To enhance reconstruction quality, WaveEar incorporates a deep neural network model called Wave-voice Net. 
mmMIC~\cite{fan2023mmmic} proposes a multi-modal speech recognition system based on mmWave radar, applicable in multi-source and noisy environments. This work utilizes mmWave radar to simultaneously sense lip movements and vocal cord vibrations and introduces the TransFuser model with an attention mechanism to fuse the differences between vowels and consonants in vocal cord vibrations and lip movements.
Regarding loudspeaker eavesdropping, Hu et al.~\cite{hu2022milliear,hu2022towards} proposed the MILLIEAR system. This system employs commercial mmWave FMCW radar to achieve unconstrained vocabulary eavesdropping. By introducing the Virtual Sub-chirp technique, MILLIEAR improves the precision of vibration signal extraction and utilizes cGAN to enhance audio reconstruction quality.

To eliminate reliance on active audio sources, Wang et al.~\cite{wang2022mmphone,lin2023high} proposed mmPhone, a remote speech eavesdropping system based on mmWave radar and the piezoelectric effect. mmPhone utilizes PVDF piezoelectric film as a sensing material, which is pre-placed in the room to assist in eavesdropping. The system detects the subtle electromagnetic property changes of the film induced by sound waves via mmWave radar, enabling the extraction and reconstruction of audio signals.
mmEcho~\cite{hu2023mmecho,li2025acoustic} further explores the method of reconstructing audio from the vibrations of common objects near the audio source. This technique also employs FMCW radar for vibration sensing and introduces an intra-chirp-based micron-level vibration sensing scheme, enabling sound reconstruction from everyday objects such as tinfoil and chip bags. mmEcho relies on signal processing techniques, eliminating the demand for machine learning models, training data, or prior knowledge of the target audio.
Similarly, VibSpeech~\cite{wang2024vibspeech} also proposes an approach to reconstruct audio by capturing subtle surface vibrations on objects. It can recover speech with frequencies up to 8 kHz from a narrowband signal as limited as 500 Hz, covering both voiced and unvoiced phonemes. 

To investigate audio eavesdropping from smartphones, mmEve~\cite{wang2022mmeve} proposed a solution based on commercial mmWave radar, similar to mmSpy~\cite{basak2022mmspy}. 
The system identifies the side-channel correlation between the mmWave reflections from the phone's rear surface and the sound emitted by the earpiece to extract and reconstruct conversation content. To enhance performance, mmEve introduces a GAN-based denoising scheme and incorporates a motion-resilient mechanism. 
mmEar~\cite{xu2024mmear} extends the eavesdropping target to over-ear headphones and proposes a method that utilizes commercial mmWave radar to capture and reconstruct weak vibration signals. The system incorporates a pretraining-finetuning strategy based on cGAN to enhance its generalization performance across different headphone models and environments. 
RFSpy~\cite{chen2024rfspy} proposes an RFID-based eavesdropping system for online conversations, which secretly attaches an RFID tag to a headset to sense the vibration signals of the headset's speaker and microphone metal coils, thereby capturing and reconstructing conversation content. This technique propose a sound spectrogram reconstruction network with a phoneme-based pixel mapping mechanism to reconstruct out-of-vocabulary words.
Instead of eavesdropping on the vibrations of a headphone's surface, RF-Parrot\cite{yang2024rf,wang2024wireless} alternatively proposes a wired audio eavesdropping system specifically designed to intercept analog audio signals transmitted through earphone wires. The system embeds a miniature D-MOSFET reflector in the target earphone wire, whose reflection efficiency varies with the amplitude of the audio signal, thereby converting the voltage fluctuations of the audio signal into detectable RF signals.

In recent years, wireless acoustic eavesdropping has become more diverse and refined: initially, the focus was on capturing lip movements and throat vibrations;  later, the approach evolved to speaker-based eavesdropping, and most recently, it has targeted headphones.  At the same time, technological advances have improved signal sensing precision from the sub-millimeter to the micrometer level, enhancing the ability to capture target speech signals.  Moreover, the evaluation methods for eavesdropping systems have undergone a significant transformation.  Whereas in the past, evaluation relied solely on classification accuracy, current methods employ multi-dimensional metrics—such as MCD, STOI, and WER—thus providing a more comprehensive assessment.  These metrics not only quantify subtle distortions in speech signals but also better align with human perceptions of clarity and intelligibility, while directly reflecting the performance of downstream speech recognition systems.  This, in turn, offers specific and targeted feedback for system improvements, promoting dual advancements in both practicality and overall quality.

However, current wireless acoustic eavesdropping techniques remain limited in terms of attack range and wall penetration, thereby reducing their real-world threat. Future research should focus on extending the effective range of attacks—for example, by optimizing antenna designs, increasing signal power, and adopting more efficient signal processing algorithms to enhance the signal-to-noise ratio. Additionally, as eavesdropping scenarios diversify further, integrating multi-modal sensing technologies (such as visual, infrared, and ultrasonic methods) is expected to overcome current limitations and offer more comprehensive and precise acoustic reconstruction capabilities.

\subsubsection{\textbf{Countermeasures}}
While RF-based acoustic eavesdropping has demonstrated remarkable capabilities in capturing speech through vibrations, its rapid advancement has also prompted growing concerns about privacy and security. To mitigate this threat, researchers have proposed various countermeasures that aim to disrupt or obscure the signals used in eavesdropping.

Shaikhanov et al.~\cite{shaikhanov2024audio} proposed a defense mechanism against audio eavesdropping attacks that leverage mmWave radar to capture smartphone vibrations caused by speaker audio. Their method introduces an on-phone sub-terahertz metasurface designed to actively manipulate the phase of radar signals reflected from the phone's surface. By dynamically modulating the phase response with controlled voltage signals, the metasurface introduces intentional phase distortions that obscure the true audio signal. Moreover, this system can inject false audio information, effectively misleading the attacker with fabricated content.

In addition to hardware-based approaches, software-driven solutions have also been explored.
Chang et al.~\cite{chang2024eveguard} proposed EveGuard, a software-driven defense designed to protect voice privacy from vibration-based side-channel eavesdropping attacks. Unlike traditional hardware-based solutions, EveGuard introduces adversarial audio perturbations that interfere with eavesdropping sensors while maintaining natural speech quality for human listeners. It employs a Perturbation Generator Model (PGM), trained using a novel Eve-GAN framework that predicts eavesdropped audio from source speech, enabling robust perturbation generation across various attack scenarios. 


\subsection{Keystroke Eavesdropping via RF Sensing}
\begin{table*}[htp!]
\centering
\begin{threeparttable}

\setlength{\tabcolsep}{5.5pt} 
\caption{\textbf{Comparison of Keystroke Eavesdropping via RF Signals}}
\label{tab:keystroke}

\begin{tabular}{c c c c c c c c}
\toprule
\textbf{Attack Scheme} & \textbf{Year} & \textbf{Keyboard Type} & \makecell{\textbf{RF Signal Type} \\ \textbf{\& Hardware}}& \makecell{\textbf{Max. Attack}\\ \textbf{Distance}\tnote{1}} &  \makecell{\textbf{Through}\\\textbf{-wall}} & \makecell{\textbf{Training}\\ \textbf{Effort}\tnote{2}} &\makecell{\textbf{Attack}\\\textbf{Type}}\\
\toprule

Chen et al.~\cite{chen2015tracking} & 2015 & \makecell{Laptop keyboard \\ (QWERTY and numeric keypad)}&  \makecell{2.4GHz ISM band \\ (NI-based SDR)}  &  5m & \ding{55} & \RIGHTcircle & OKE \\ \hline

WiKey~\cite{ali2015keystroke,ali2017recognizing} & 2015 & \makecell{Laptop keyboard \\ (QWERTY and numeric keypad)} &  \makecell{Wi-Fi CSI \\ (Inter 5300 NIC)}  & 30cm   & \ding{55}  &   \CIRCLE & OKE \\ \hline

Fang et al.~\cite{fang2018no,yang2022wireless} & 2018 &  \makecell{Laptop keyboard \\(QWERTY and numeric keypad)}&  \makecell{ Wi-Fi CSI\\ (USRP X300) } &  50cm & \ding{51}  &  \Circle & OKE\\ \hline

SpiderMon~\cite{ling2020spidermon} & 2020 &   \makecell{Laptop keyboard \\(Numeric keypad)}&  \makecell{LTE CRS (USRP B210)}  & 15m & \ding{51}  &   \CIRCLE & OKE \\ \hline

WiPass~\cite{zhang2016privacy} & 2016 &  \makecell{Smartphone keyboard \\ (Graphical unlock pattern)} &  \makecell{Wi-Fi CSI \\ (Inter 5300 NIC)} &   1m &   \ding{55}  &   \RIGHTcircle & IKE\\ \hline

WindTalker~\cite{li2016csi,meng2020revealing} & 2016 & \makecell{Smartphone keyboard \\ (PIN keypad)} & \makecell{Wi-Fi CSI \\ (Inter 5300 NIC)}  &  1.6m &  \ding{55} & \RIGHTcircle  & IKE\\ \hline

Shen et al.~\cite{shen2021wipass} & 2021 &\makecell{Smartphone keyboard\\ (Numeric keypad) } &   \makecell{Wi-Fi CSI \\ (Inter 5300 NIC)} & 75cm  &  \ding{55} &    \CIRCLE & OKE\\ \hline

WINK~\cite{yang2022wink} & 2022 & \makecell{Smartphone keyboard\\ (PIN keypad)}  &   \makecell{ Wi-Fi CSI\\ (USRP X300) }   & 1.5m  &   \ding{51} & \Circle &  OKE \\  \hline

WiKI-Eve~\cite{hu2023password} & 2023 & \makecell{Smartphone keyboard \\ (Numeric keypad and QWERTY)}  & \makecell{Wi-Fi BFI \\ (Intel AX210 NIC)}   &  10m & \ding{51}  &  \CIRCLE &  IKE\\  \hline

MuKI-Fi~\cite{wang2024muki} & 2024 & \makecell{Smartphone keyboard \\ (Numeric keypad and QWERTY)}  & \makecell{Wi-Fi BFI \\ (Intel AX210 NIC)}   &  10m & \ding{51}  &  \CIRCLE & IKE\\  \hline

Periscope~\cite{jin2021periscope} & 2021 &  \makecell{Smartphone keyboard \\ (Numeric keypad)}   &  \makecell{Electromagnetic\\(EPS)}  &  90cm & \ding{55}  &   \Circle  & / \\   \hline

WiPOS~\cite{zhang2020wipos} & 2020 & \makecell{POS terminal keyboard \\(Numeric keypad)}  &   \makecell{Wi-Fi CSI \\ (Inter 5300 NIC)} &  23cm  & \ding{55}  &  \CIRCLE & OKE \\  \hline

SThief~\cite{chen2024silent} & 2024 & \makecell{POS terminal keyboard \\(Numeric keypad)} & \makecell{Wi-Fi BFI \\ (MacBook Pro built-in NIC)} & 1.5m & \ding{55} & \CIRCLE & IKE \\ \hline

VR-Spy~\cite{al2021vr} & 2021 &  \makecell{VR headset \\ (Virtual keyboard)} &   \makecell{Wi-Fi CSI \\ (Inter 5300 NIC)}   &  $\approx$~60cm &  \ding{55}  & \CIRCLE & OKE\\  \hline

mmSpyVR~\cite{mei2024mmspyvr} & 2024 &  \makecell{VR headset \\ (Virtual keyboard)} &   \makecell{mmWave \\ (TI IWR6843-ODS)}   &  8m &  \ding{51}  & \CIRCLE & / \\

\bottomrule
\end{tabular}

\begin{tablenotes}
    \item[1] The distance between the keyboard and the RF receiver.
    \item[2]\Circle ~for Low (Training free), \RIGHTcircle  ~for Middle (Less than 10 samples for training), \CIRCLE  ~for High (More than 10 samples for training).
\end{tablenotes}

\end{threeparttable}
\end{table*}

\begin{figure}
\centering
\subfigure[Out-of-band Keystroke Eavesdropping (OKE) Method]{
\begin{minipage}[t]{0.48\linewidth}
\centering
\includegraphics[width=\linewidth]{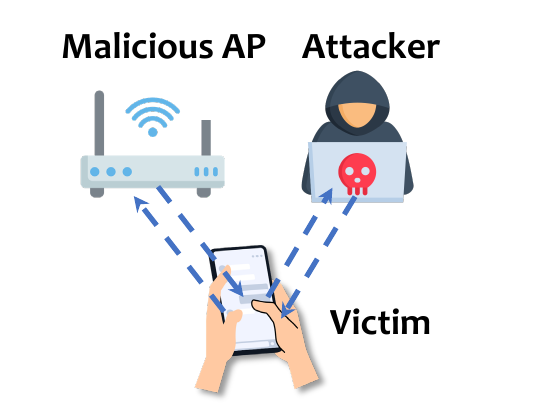}
\label{fig:OK}
\end{minipage}%
}
\subfigure[In-band Keystroke Eavesdropping (IKE) Method]{
\begin{minipage}[t]{0.48\linewidth}
\centering
\includegraphics[width=\linewidth]{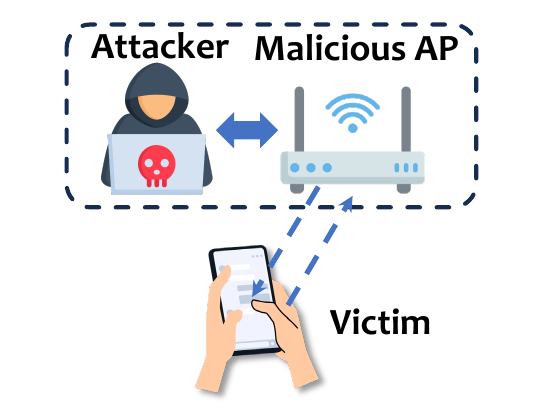}
\label{fig:IK}
\end{minipage}%
}
\centering
\caption{RF sensing-based keystroke eavesdropping methods}
\label{fig:keystroke}
\end{figure}

Leveraging RF signal's fine-grained sensing ability, attackers can accurately capture and analyze every keystroke a user makes on the keyboard. These techniques provide a potential means for attackers to listen in on sensitive information entered by users, such as passwords, personal data, and encryption keys, without having direct access to the user's device.

Currently, there are two primary methods for leveraging RF sensing to eavesdrop on keystrokes: Out-of-band Keystroke Eavesdropping (OKE) and In-band Keystroke Eavesdropping (IKE)~\cite{li2016csi}, as shown in Fig.~\ref{fig:keystroke}.
OKE involves deploying two Wi-Fi devices: a malicious access point (AP) and a receiver. The target device must be positioned between the AP and the receiver, with keystrokes inferred by analyzing the RF signal changes caused by multipath effects between the AP and the receiver. 
In contrast, IKE utilizes a single malicious AP that communicates directly with the target device. By analyzing the RF signal variations between the AP and the target device, IKI achieves keystroke eavesdropping with greater flexibility, making it particularly suitable for attacks in mobile scenarios.
Besides, keystroke eavesdropping can be categorized into three types: targeting physical keyboards on PCs or laptops, targeting virtual keyboards on smartphones or tablets (e.g., PIN keypads), and targeting other types of keyboards (e.g., POS terminals).
We summarize the RF sensing-based keystroke eavesdropping schemes in Table~\ref{tab:keystroke}.

\begin{table*}[!ht]
\centering
\renewcommand{\arraystretch}{1.4} 
\setlength{\tabcolsep}{5.3pt} 
\caption{Comparison between Beamforming Feedback Information (BFI) and Channel State Information (CSI)}
\begin{tabular}{l c c}
\toprule
   & \textbf{Beamforming Feedback Information (BFI)}  & \textbf{Channel State Information (CSI)}  \\ \bottomrule
\textbf{Hardware Requirement}  & No modification needed. & Requires specialized hardware and NIC modifications.                \\
\textbf{Information Type}     & Clear-text feedback for beamforming.             & Detailed channel state information.       \\
\textbf{Transmission Method}  & Exchanged automatically during communication.     & Requires specific protocols (e.g., 802.11n). \\
\textbf{Capture Difficulty}   & Easy to capture with any monitor mode device.     & Requires specialized hardware and tools.  \\ \bottomrule
\end{tabular}
\label{tab:bfi_vs_csi}
\end{table*}

\subsubsection{\textbf{PC or Laptop Keyboard}}
Chen et al.~\cite{chen2015tracking} first proposed to remotely detect keystrokes by analyzing changes in the phase and amplitude of RF signals. This approach utilizes multiple receiving antennas to capture RF signals and create imperfect phase and amplitude matching. These mismatches result in ``trough'' in the frequency spectrum. 
When a key is pressed, the movement of the user's fingers alters the signal propagation path, causing the position of these troughs to shift. By tracking these changes, the system can identify the specific keystroke being input by the user.
Building on this idea, WiKey~\cite{ali2015keystroke} proposed to utilize COTS Wi-Fi devices for the first time for contactless keystroke recognition. The scheme infers user keystrokes by capturing changes in the Wi-Fi CSI between the user's device and the access point. 
However, despite the success of these early keystroke eavesdropping schemes, they still suffer from certain limitations, such as the need for significant training data or model calibration. 
To address these challenges, Fang et al.~\cite{fang2018no} introduce an innovative training-agnostic keystroke inference attack. 
This attack utilizes the mapping relationship between Wi-Fi CSI changes and typed letters to infer keystrokes. 
Unlike early schemes that rely on large amounts of training data or model tuning, this approach quickly infers user input by analyzing real-time variations in the phase and amplitude of Wi-Fi signals without any prior training phase. 

In addition to Wi-Fi signals, other RF signals have also been explored for keystroke eavesdropping tasks. 
For instance, SpiderMon~\cite{ling2020spidermon} leverages the Cell-Specific Reference Signal (CRS) emitted by commercial LTE base stations as an illuminating source to passively monitor signal reflections near the target, enabling long-range (up to 15 meters) and non-line-of-sight (NLoS) keystroke eavesdropping. The system integrates directional antennas to enhance target signal reflections, employs Block Principal Component Analysis (Block PCA) for feature extraction, and utilizes a Hidden Markov Model (HMM) to infer continuous keystroke sequences.

\subsubsection{\textbf{Smartphone Keyboard}}
Besides laptop physical keyboards, the increasing prevalence of smart devices has led to more research targeting smartphone virtual keyboards, thus extending RF sensing-based keystroke eavesdropping attacks to mobile devices.
WiPass~\cite{zhang2016privacy} proposes leveraging the wireless hotspot functionality of smart devices to infer graphical unlock patterns by analyzing changes in Wi-Fi signals caused by finger movements during unlocking. However, this attack is limited by the requirement that the target device must have its hotspot feature enabled, and the attacker must connect to the target's hotspot to enable signal monitoring and CSI data collection. While effective in certain scenarios, these prerequisites make the attack a relatively narrow threat vector.

WindTalker~\cite{li2016csi} advances the above eavesdropping attack by targeting PIN passwords and utilizing public Wi-Fi networks. 
It operates in public Wi-Fi environments, capturing CSI data from nearby access points without requiring the target device to have its hotspot function enabled. 
Moreover, WindTalker employs more sophisticated signal processing techniques, such as PCA, to extract more precise patterns from noisy signals, enhancing its robustness and adaptability.
Additionally, it combines network traffic and CSI data to focus the inference only during the sensitive period when the victim enters their PIN or password. 
By enabling the eavesdropping of PINs, WindTalker significantly expands the attack surface, marking a major step forward in the scope and effectiveness of RF sensing-based side-channel attacks.
Similarly, Shen et al.~\cite{shen2021wipass} introduce the use of deep learning models, specifically 1-D Convolutional Neural Networks (CNNs), to automate feature extraction from CSI signals, improving the accuracy of keystroke recognition.

In prior work, a certain number of training samples are typically required to train the model or serve as a matching reference dataset, which increases the barrier to launching the attack.
WINK~\cite{yang2022wink} proposes a zero-training spatiotemporal analysis scheme to infer numerical keystrokes of the smartphone, rather than relying on traditional training methods (such as machine learning models). 
Specifically, WINK analyzes the spatial distribution of these disturbances (i.e., the location of the interference) and the temporal distribution (i.e., the time intervals between keystrokes) in the Wi-Fi signal. By extracting and analyzing these spatial and temporal features, WINK can efficiently and accurately infer the victim's input sequence, bypassing the need for training data or prior knowledge of the victim's typing behavior.

\begin{figure}
    \centering
    \includegraphics[width=0.6\linewidth]{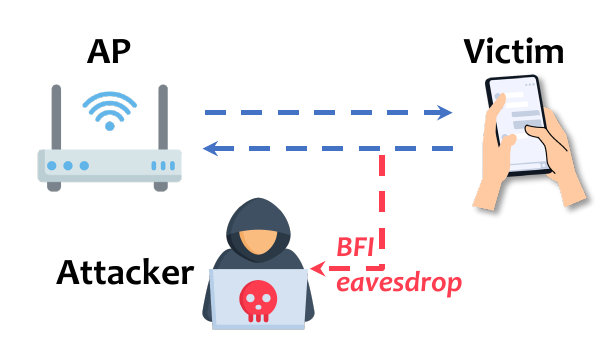}
    \caption{Wi-Fi BFI-based keystroke eavesdropping.}
    \label{fig:bfi-eve}
\end{figure}

Recent research has demonstrated the potential of Beamforming Feedback Information (BFI) for fine-grained sensing applications. BFI represents signal quality information exchanged between a Wi-Fi device and an AP during the beamforming process. Unlike CSI, which requires specialized hardware modifications for extraction, BFI is transmitted in clear text as part of routine communication between Wi-Fi devices and the AP. This accessibility enables any Wi-Fi device operating in monitor mode to capture BFI without the need for hardware hacking. A detailed comparison between BFI and CSI is provided in Table~\ref{tab:bfi_vs_csi}.
Building on this potential, WiKI-Eve~\cite{hu2023password} introduces a BFI-based keystroke eavesdropping method. 
Unlike the previous OKE method, this approach is referred to as overhearing in-band keystroke eavesdropping (o-IKE), eliminating the need to hack Wi-Fi hardware (i.e., AP), as illustrated in Fig.~\ref{fig:bfi-eve}.
By switching a Wi-Fi device into monitor mode, WiKI-Eve captures and analyzes BFI signals, which are then used to infer the user's keystrokes. This approach avoids the complexities of traditional attacks, which typically require compromising hardware or tricking the target into connecting to a malicious AP. 
To improve its adaptability across different environments, WiKI-Eve incorporates adversarial learning techniques, enabling the model to generalize effectively to new scenarios and devices, while maintaining high inference accuracy.
Furthermore, MuKI-Fi~\cite{wang2024muki} captures BFI variations from independent Wi-Fi links associated with different devices and leverages near-field domination effect to effectively eavesdrop on keystrokes from multiple targets.

While previous research on keystroke eavesdropping has focused on Wi-Fi-based techniques, such as using CSI and BFI to detect keystrokes, recent studies have explored keystroke eavesdropping using electromagnetic emissions affected by user interaction. 
For instance, Periscope~\cite{jin2021periscope} proposes to infer user input by exploiting human-coupled electromagnetic emissions from touchscreen devices. 
When a user's finger approaches or touches the screen, it creates a capacitive coupling between the finger and the touchscreen's electrode grid. This coupling causes variations in electromagnetic radiation, with the signal fluctuating in real time as the finger moves. By capturing and analyzing these electromagnetic signals, the attacker can infer the user's finger trajectory and eavesdrop on keystrokes. The prototype uses an Electric Potential Sensor (EPS) and an Arduino Nano board to capture electromagnetic signals. The EPS measures changes in electric potential, while the Arduino Nano collects and processes the signals to infer user input. 

It's worth noting that earlier research on eavesdropping keystrokes through electromagnetic leakage~\cite{vuagnoux2009compromising} primarily focused on signals generated by the keyboard's internal circuitry (such as clock signals and data transmission lines). These electromagnetic signals, which are emitted when a key is pressed, change with each keystroke and can be captured by an attacker to infer the typed keys. 
However, such studies largely neglect the influence of the user's fingers on the propagation characteristics of RF signals.
Consequently, electromagnetic leakage-based keystroke eavesdropping is fundamentally different from RF sensing techniques, as it relies on the passive capture of hardware-originated emissions rather than the dynamic interaction between human actions and RF signal propagation. Therefore, it falls outside the scope of RF sensing and is not the focus of this survey.

\subsubsection{\textbf{Other Types of Keyboards}}
In addition to traditional computer keyboards and smartphone virtual keyboards, other types of input devices are also vulnerable to eavesdropping. Zhang et al.~\cite{zhang2020wipos} proposed a password eavesdropping scheme, WiPOS, to reveal the security risks of password leakage when entering passwords on POS terminals in public Wi-Fi environments. The system collects wireless signals using two commercially available Wi-Fi devices and utilizes a keystroke segmentation algorithm, a Support Vector Machine (SVM) classifier, and a Global Alignment Kernel (GAK) technique to infer the user's entered password.
In contrast, SThief~\cite{chen2024silent} leverages Wi-Fi BFI to eavesdrop on keystrokes from POS terminals. It uses Maximum Ratio Combining (MRC) for signal enhancement and Connectionist Temporal Classification (CTC) for password inference, achieving higher accuracy and robustness across various scenarios.

Furthermore, with the rapid development of virtual reality (VR) technology, users frequently input sensitive information, such as passwords and search queries, through virtual keyboards in VR environments, making the security of virtual keystrokes increasingly important. Abdullah Al Arafat et al.~\cite{al2021vr} proposed VR-Spy, a side-channel attack method based on CSI for virtual keystroke recognition. By capturing the subtle Wi-Fi signal variations caused by hand movements associated with each virtual keystroke, attackers can eavesdrop on and infer the user's input, further exposing the risk of information leakage in virtual reality environments.
Beyond Wi-Fi signals, mmWave has gained prominence as a highly efficient approach for keystroke eavesdropping, owing to its exceptional resolution and strong penetration capabilities. For instance, mmSpyVR~\cite{mei2024mmspyvr} capitalizes on these advantages to enable precise eavesdropping on virtual keyboard inputs in VR environments, even when physical obstacles such as walls and doors are present.

\subsubsection{\textbf{Countermeasures}}
Existing defense strategies against RF keystroke eavesdropping attacks can be categorized into three main types: \textit{Access Control and Authentication}, \textit{Input Obfuscation}, and \textit{Signal Disruption and Shielding}, as shown in Tab.~\ref{tab:defense2}.

\begin{table}[t!]
\centering
\renewcommand{\arraystretch}{1.4} 

\caption{Summary of Defense Strategies Against RF Sensing-based Keystroke Eavesdropping}
\label{tab:defense2}
\resizebox{\linewidth}{!}{  
\begin{tabular}{>{\raggedright}p{2cm} p{3.3cm} p{3cm}} 
\bottomrule
\textbf{Defense Category} & \textbf{Method} & \textbf{Reference} \\
\bottomrule
\multirow{3}{*}{\shortstack[l]{\textbf{Access Control} \\ \textbf{\& Authentication}}}  
    & Traffic Encryption & \cite{hu2023password} \\ 
\cmidrule(lr){2-3}
    & Biometric Authentication & \cite{zhang2016privacy} \\ 
\cmidrule(lr){2-3}
    & Avoiding Untrusted Wi-Fi  & \cite{li2016csi} \\
\midrule
\multirow{3}{*}{\shortstack[l]{\textbf{Input} \\ \textbf{Obfuscation}}}  
    & Keystroke Encryption & \cite{fang2018no, yang2022wink} \\ 
\cmidrule(lr){2-3}
    & Typing Behavior Distortion & \cite{li2016csi, jin2021periscope} \\ 
\cmidrule(lr){2-3}
    & Keyboard Randomization & \cite{li2016csi, yang2022wink, hu2023password, jin2021periscope} \\
\midrule
\multirow{3}{*}{\shortstack[l]{\textbf{Signal Disruption} \\ \textbf{\& Shielding}}}  
    & Wireless Jamming & \cite{yang2022wireless, fang2018no, yang2022wink} \\ 
\cmidrule(lr){2-3}
    & Signal Obfuscation & \cite{hu2023password} \\ 
\cmidrule(lr){2-3}
    & Electromagnetic Shielding  & \cite{jin2021periscope} \\ 
\bottomrule
\end{tabular}
} 
\end{table}

\paragraph{\textbf{Secure Channel and Access Hardening}}
Secure channel and access hardening strategies aim to prevent adversaries from acquiring keystroke-related information by enhancing wireless communication security and minimizing user exposure to untrusted network environments.

\begin{itemize}[leftmargin=*]
    \item \textit{Traffic Encryption.} Encrypting Wi-Fi traffic hinders attackers from extracting sensitive metadata, such as beamforming feedback information (BFI), from transmitted packets~\cite{hu2023password}. This practice is widely adopted in enterprise and institutional networks to safeguard communication.
    \item \textit{Biometric Authentication.} Replacing conventional password input with biometric methods, such as fingerprint or facial recognition, eliminates keystroke generation altogether, thereby nullifying keystroke inference attacks~\cite{zhang2016privacy}.
    \item \textit{Avoiding Untrusted Wi-Fi Connections.} Avoiding connections to untrusted or public Wi-Fi networks mitigates the risk of channel state information (CSI) being intercepted and exploited for input inference~\cite{li2016csi}.
\end{itemize}

\paragraph{\textbf{Behavioral and Interface Obfuscation}}
Behavioral and interface obfuscation techniques aim to degrade the effectiveness of keystroke inference by modifying user input behaviors or disrupting the spatial mapping between keystrokes and interface elements, thereby undermining the reliability of signal-based eavesdropping.

\begin{itemize}[leftmargin=*]
    \item \textit{Keystroke Encryption.} Injecting random characters, decoy words, or dummy inputs during typing introduces semantic noise, significantly reducing the precision of keystroke classification and sequence reconstruction~\cite{fang2018no,yang2022wink}.
    \item \textit{Typing Behavior Distortion.} Altering temporal typing patterns—such as introducing irregular delays or varying typing speed—impairs the attacker’s ability to extract consistent signal features corresponding to key positions~\cite{li2016csi,jin2021periscope}.
    \item \textit{Keyboard Randomization.} Dynamically shuffling keyboard layouts disrupts the one-to-one mapping between physical keystrokes and input characters, effectively neutralizing inference models trained on fixed spatial arrangements~\cite{li2016csi,yang2022wink,hu2023password,jin2021periscope}.
\end{itemize}

\paragraph{\textbf{Physical-Layer Interference and Isolation}}
Physical-layer interference and isolation techniques aim to degrade the quality or availability of keystroke-induced RF signal features by disrupting signal propagation paths or suppressing electromagnetic emissions, thereby limiting the attacker’s ability to perform reliable inference.

\begin{itemize}[leftmargin=*]
    \item \textit{Wireless Jamming.} Introducing external noise through jamming devices masks the subtle signal variations caused by keystrokes, rendering CSI-based inference models ineffective~\cite{yang2022wireless,fang2018no,yang2022wink}.
    \item \textit{Signal Obfuscation.} Intelligent reflecting surfaces (IRS) and multiple-input multiple-output (MIMO) architectures can dynamically alter wireless propagation patterns, injecting controlled randomness into CSI measurements to confuse signal analysis~\cite{hu2023password}.
    \item \textit{Electromagnetic Shielding.} Applying electromagnetic interference (EMI) shielding materials to keyboards or touchscreens physically blocks electromagnetic leakage, eliminating the signal sources required for electromagnetic-based keystroke eavesdropping~\cite{jin2021periscope}.
\end{itemize}

Despite the effectiveness of these defense strategies, each approach has its limitations in terms of usability, implementation complexity, and hardware requirements. Future research should focus on developing adaptive and low-cost security mechanisms that provide robust protection without significantly compromising user experience. By integrating multiple defense strategies, a more resilient and comprehensive security framework can be established to counter emerging wireless keystroke eavesdropping threats.

\subsection{Privacy Behavior Exposure via RF Sensing}

RF sensing has been widely applied in the detection and recognition of human activities, showing great potential in areas such as smart homes, health monitoring, and elderly care. 
However, the widespread adoption of these technologies also introduces significant privacy risks. RF sensing can capture a large amount of personal activity data without direct contact with the user, which can reveal sensitive information such as an individual's daily routines, activity patterns, and even location, time, and behavior details. In the absence of proper authorization, the activity data generated by RF sensing may be maliciously intercepted or misused, leading to serious privacy leakage.
Furthermore, the wall-penetrating nature of RF sensing often leaves victims unaware of being monitored, exacerbating privacy concerns.

\subsubsection{\textbf{Through-wall Human Presence Detection}}
RF-based human presence detection poses significant privacy risks. It can enable covert monitoring, exposing sensitive information such as movement patterns and daily activities. Unauthorized access to or misuse of presence data may lead to malicious activities like stalking or burglary. 

Youssef et al.~\cite{youssef2007challenges} first introduce the concept of Device-free Passive (DfP), enabling the detection, tracking, and identification of targets by monitoring RF signal (i.e., Wi-Fi RSSI) variations, without requiring the target to carry any device or actively participate in the process.
Prior research~\cite{chetty2012through, banerjee2014violating, cushman2016exper, wang2017see, zhu2017r} has focused on leveraging variations in RF signal properties, particularly reflections, attenuation, and multipath effects, to detect the presence of \textit{moving individuals}. 
Human movement induces changes in the propagation paths, reflections, and signal strength of radio frequency waves, which can be captured through metrics such as RSS, CSI, or phase shifts. 
These signal dynamics, including temporal fluctuations in signal strength, inter-subcarrier correlations, and phase variations, serve as key features for analysis. 
By incorporating machine learning or statistical modeling techniques, such as PCA, SVM, and HMM, these features can be modeled to robustly distinguish between static environments and the presence of moving individuals, enabling precise human presence detection in complex scenarios.

In addition to detecting moving individuals, detecting the presence of \textit{stationary individuals} presents a unique challenge. Static individuals, unlike moving ones, do not cause significant variations in signal propagation paths, which makes them harder to detect. However, human presence can still be inferred through subtle, periodic changes in the wireless signal, such as those induced by breathing or small body movements. 
Wu et al.~\cite{wu2017non} utilize Wi-Fi CSI features to detect moving individuals and harness breathing-induced periodic signal variations to identify stationary individuals.
Further, Uysal et al.~\cite{uysal2022new} enable through-wall detection of stationary individuals by analyzing the minute signal changes caused by the rhythmic breathing movements of static individuals, using narrowband RF signals transmitted and received by a low-cost SDR module. 
In addition, Domenico et al.~\cite{domenico2018wifi} utilize the Doppler spectrum of Wi-Fi CSI to extract the subtle Doppler shift features caused by stationary individuals, enabling through-wall detection of stationary individuals. 
Similarly, Shen et al~\cite{shen2024attention} propose an attention-enhanced deep learning system that leverages spectral, spatial, and temporal features of Wi-Fi CSI to improve the accuracy of through-wall presence detection, addressing challenges in both static and dynamic scenarios.

\subsubsection{\textbf{Through-wall Human Tracking}}
Through-wall human tracking is a particularly concerning application due to its potential to infringe on personal privacy and security. By leveraging the wall-penetrating properties of RF signals, these systems can detect and track individuals without requiring direct line-of-sight access or the individual's awareness. This capability poses significant threats, as it enables the monitoring of individuals within private spaces, such as homes or offices, where they would typically expect a high degree of privacy. The ability to gather such detailed information covertly can be exploited for unauthorized surveillance, stalking, or other malicious activities.

Early approaches to through-wall tracking utilized narrow-band radars with antenna arrays~\cite{ram2008through, ram2008doppler}, which laid the groundwork for RF-based human tracking. 
Recently, Wi-Fi signals have been explored due to their widespread availability and potential for device-free tracking~\cite{qian2017widar, falcone2012localization, li2017indotrack, xie2019md, wu2021witraj, wang2022single}. 
By employing key features such as Doppler~\cite{falcone2012localization, li2017indotrack}, Angle of Arrival (AoA)~\cite{li2017indotrack, xie2019md,falcone2012localization}, Time of Flight (ToF)~\cite{xie2019md}, Doppler Frequency Shift (DFS)~\cite{wu2021witraj, wang2022single}, and Angle of Departure (AoD)~\cite{xie2019md}, these systems effectively address challenges like noisy signals, multipath interference, and hardware inconsistencies.  
Building on these advancements, Widar 2.0~\cite{qian2018widar2} and WiSen~\cite{jin2022wisen} achieve decimeter-level accuracy with single-link Wi-Fi tracking, simplifying deployment. Recent studies~\cite{tan2019multitrack, venkatnarayan2020leveraging} further demonstrate Wi-Fi's capability to track multiple individuals simultaneously.

Despite the promise shown by the above Wi-Fi signal-based tracking systems in light NLoS conditions, their through-wall tracking capabilities remain underdeveloped.
Wi-Vi~\cite{adib2013see} represents the first Wi-Fi signals-based through-wall tracking system by employing techniques such as MIMO interference nulling and inverse synthetic aperture radar (ISAR). This approach successfully detects and tracks the relative direction and angle of moving targets behind walls (e.g., hollow walls, wooden doors, and concrete walls up to 8 inches thick), distinguishing up to three individuals.
In addition to Wi-Fi signals, Yang et al.~\cite{yang2015see} introduced an innovative method utilizing COTS RFID tags affixed to walls as an antenna array. This setup captures reflections from moving individuals, facilitating accurate and effective through-wall human tracking. 
Meanwhile, Zhang et al.~\cite{zhang2020exploring} leverage LoRa signals for long-range through-wall sensing. It achieves human tracking and activity sensing at distances up to 30 meters and can penetrate up to two walls with high accuracy. 

In addition, mmWave~\cite{zeng2016human, cui2021high, li2023indoor, chen2023environment, chen2024mmtai} and UWB~\cite{li2022multistatic} signals have been widely used for human tracking.
However, the high frequency and short wavelength of mmWave limit its penetration capability, while UWB faces challenges such as multipath interference in complex environments, restricting its ability to covert through-wall tracking capability.

\subsubsection{\textbf{Through-wall Human Activity Eavesdropping}}
Through-wall human activity eavesdropping raises even greater privacy concerns by enabling detailed observation of individuals' actions and behaviors. Attackers can covertly identify gestures, postures, and interactions, increasing the risk of unauthorized surveillance and malicious exploitation, such as blackmail or manipulation.
Based on the granularity of the eavesdropped information, through-wall human activity eavesdropping can be broadly understood as encompassing two levels of intrusiveness:
\textit{1) Recognition-level Eavesdropping.} This involves recognizing general activities performed by individuals behind walls, such as standing, sitting, walking, or running. Advanced methods further enhance this capability by identifying finer details, such as specific gestures and subtle movements;
\textit{2) Reconstruction-level Eavesdropping.} This represents a more invasive approach, enabling the reconstruction of detailed human meshes, skeletal structures, or even video-like sequences of actions occurring behind walls, providing a highly precise depiction of movements and behaviors.

\paragraph{\textbf{Recognition-level Eavesdropping}}
Many RF sensing-based through-wall activity recognition systems focus on functional applications, such as smart home or health monitoring, but their techniques inherently enable malicious eavesdropping. 
While not explicitly addressing adversarial use, these methods can be repurposed for privacy-invasive scenarios. 

Building on prior research~\cite{adib2013see,pu2013whole,adib20143d} that relied on specialized hardware for device-free localization and activity recognition, Wang et al.~\cite{wang2014eyes} propose E-eyes, a system for device-free, location-based activity recognition in home environments using CSI from COTS Wi-Fi devices. E-eyes can distinguish between in-place activities (e.g., cooking, watching TV) and walking activities with only a single Wi-Fi access point and a few connected Wi-Fi devices.
Follow-up studies~\cite{zhu2016wisefi, chowdhury2017wihacs, zhu2017notifi} also leverage COTS Wi-Fi devices for device-free human activity recognition. However, these works are limited in their robustness when applied to real-world dynamic and complex through-wall scenarios.
Later studies addressed the challenges of activity recognition in complex environments.
For instance, Wang et al.~\cite{wang2018device} introduced a multi-domain feature extraction framework by incorporating spatial structural information, significantly improving sensing accuracy in through-wall and NLoS scenarios. 
TW-See~\cite{wu2019tw} focused on human activity recognition specifically in through-wall scenarios, leveraging an opposite robust PCA (Or-PCA) approach to enhance signal processing and achieving an high accuracy with commodity Wi-Fi devices. 
Sun et al.~\cite{sun2021through} extended the scope further by using Wi-Fi passive radar with iterative adaptive processing, employing an SDR receiver equipped with wideband antennas. This setup enabled detection of both major movements (e.g., walking) and fine-grained activities (e.g., typing, breathing) even through concrete walls.
To further enhance the understanding of Wi-Fi signal propagation in through-wall scenarios, Zhang et al.~\cite{zhang2023understanding} proposed a theoretical refraction-aware Fresnel model, providing new insights into how walls influence Wi-Fi sensing and improving the accuracy of through-wall sensing systems.

Unlike the above-mentioned works focused on legitimate functional applications via Wi-Fi, Lu et al.~\cite{lu2022actlistener} explicitly proposed the malicious potential of RF sensing by investigating covert user activity monitoring through omnidirectional Wi-Fi signals.  
The study proposes ActListener, an attack method that requires no physical access to devices or prior knowledge of activity recognition models. By detecting variations in Wi-Fi signals to infer user behavior and leveraging signal modeling and generative model-based calibration, ActListener transforms eavesdropped signals into those received by legitimate devices for activity recognition.
In addition to analyzing body movements directly through RF signals, Fafoutis et al.~\cite{fafoutis2017privacy} found that physical activity information from wireless wearable devices can leak through the Bluetooth Low Energy (BLE) wireless channel, even if the data is encrypted. 
An adversary can infer users' activities by analyzing variations in the RSS of the BLE signal.

\paragraph{\textbf{Reconstruction-level Eavesdropping}}
Early works~\cite{yang2007design, ralston2010real} attempted to use UWB radar combined with techniques such as synthetic aperture radar (SAR) or MIMO phased array radar for through-wall target reconstruction and imaging. 
However, these approaches were limited by technical factors such as low resolution and insufficient processing power, resulting in the generation of coarse-grained heatmaps rather than high-resolution imaging or precise target profiles. 
However, these studies laid the foundation for subsequent advancements in through-wall reconstruction and imaging technology with higher accuracy and real-time capabilities.
Adib et al.~\cite{adin2015capture} proposed RF-Capture, a system leveraging 5.46~GHz FMCW radar signals to capture and reconstruct 3D human shapes and movements through walls. By analyzing RF signal interactions with the human body, RF-Capture outperforms earlier approaches by generating high-resolution target heatmap profiles.
Recently,
WiCamera~\cite{xu2024wicamera} introduces a novel Wi-Fi imaging prototype that leverages vortex electromagnetic waves (VEMWs) and commodity 3×3 MIMO devices to reconstruct silhouette heatmap of stationary human targets through orbital angular momentum-based wavefront imaging and GAN-powered refinement.

Building on coarse heatmap representation, recent research has advanced to more structured outputs, such as \textit{human skeleton reconstruction}. Zhao et al.\cite{zhao2018through} utilized Wi-Fi signals to estimate human poses through walls and occlusions, innovatively integrating visual inputs with RF signals via cross-modal learning. This approach enables the system to predict human 2D poses even in fully occluded scenarios, demonstrating RF signals' potential for pose estimation in obstructed environments. 
Similar Wi-Fi-based 2D skeleton reconstruction is also implemented in research~\cite{wang2019person}.
In contrast, RF-Pose3D~\cite{zhao2018rf} reconstructs full 3D skeletons, tracking 14 key points (e.g., head, shoulders, knees) in real-time. 
RF-Pose3D excels in multi-person and unseen environments, showcasing its potential for real-time, high-resolution 3D skeletal reconstruction. Recent works~\cite{jiang2020towards, ren2022gopose} extend these capabilities using commercial Wi-Fi hardware for 3D skeletal reconstruction.

To advance beyond skeletal reconstruction, recent efforts have focused on \textbf{3D human mesh reconstruction} leveraging RF signals, which provide detailed representations of both body pose and shape. Unlike skeleton-based methods, mesh reconstruction captures full surface geometry, enabling richer applications such as motion analysis and healthcare while posing greater privacy risks due to its fine-grained detail. 
For instance, 
RF-Avatar~\cite{zhao2019through} reconstructs dynamic 3D human meshes in real-time using custom Wi-Fi devices and time-series RF signal analysis. Similarly, Wi-Mesh~\cite{wang2022wi} employs parametric Skinned Multi-Person Linear (SMPL) model~\cite{loper2015smpl} and commodity Wi-Fi devices to simultaneously infer body pose and shape, leveraging 2D AoA estimation and deep neural networks for high-fidelity mesh reconstruction.

In addition to Wi-Fi signals, recent works have also explored the capabilities of mmWave signal for human skeletal~\cite{Arindam2020mpose,shi2022mpose,wu2024mmhpe} and mesh reconstruction~\cite{xue2021mmmesh, chen2022mmbody,xue2022m4esh}. 
While mmWave signals offer superior resolution and robustness in complex environments, their penetration capability is limited to thinner obstacles or walls due to their high-frequency, short-wavelength characteristics.

\subsubsection{\textbf{Other Privacy Information Exposure}}
Beyond commonly discussed privacy threats, researchers have explored a range of other RF sensing-based attacks that target different forms of private information exposure.

\paragraph{\textbf{Through-wall Handwritten Content Eavesdropping}}
RadSee~\cite{zhang2025radsee} realizes the recognition of handwritten letters through walls without direct visual contact. 
This work utilizes 6~GHz FMCW radar, leveraging phase feature extraction to achieve millimeter-level hand movement detection, combined with high-gain patch antennas to enhance signal penetration and reduce environmental interference. This study reveals the security and privacy risks of wall-transparent handwriting detection, which poses a new challenge for personal information protection.

\paragraph{\textbf{Through-wall Screen Eavesdropping}}
WaveSpy \cite{li2020wavespy} introduces a novel side-channel attack that enables the inference of LCD screen content through walls, allowing for the extraction of sensitive information without requiring direct visual access. 
This method leverages a 24~GHz mmWave FMCW radar to detect the alignment patterns of liquid crystals, capture high-resolution liquid crystal state responses, and subsequently infer the displayed screen content.
Experimental results demonstrate that WaveSpy can accurately recognize various types of screen displays, including text editors, social media platforms, and online banking interfaces, while also inferring sensitive user inputs such as PIN codes, graphical unlock patterns, and alphanumeric passwords. Additionally, the system exhibits a limited capability in recovering on-screen text.
However, WaveSpy cannot reconstruct full-screen content, particularly images and videos, due to the spatial resolution constraints of mmWave signals and its dependence on deep learning models, which restrict inference to text and key inputs based on training data.

\paragraph{\textbf{Virtual Reality Privacy Leakage}}
mmSpyVR~\cite{mei2024mmspyvr} introduces a novel side-channel attack that exploits mmWave radar to penetrate obstacles and infer VR users' private activities, exposing a previously unknown privacy vulnerability in VR systems. This method enables attackers to extract sensitive information without physical or virtual access to VR devices by leveraging a 60~GHz mmWave FMCW radar to detect VR user motions, capture high-resolution radar cross-section (RCS) features, and infer VR interactions.
Experimental results demonstrate that mmSpyVR can accurately recognize VR applications (e.g., gaming, chatting, browsing, and shopping) and infer keystroke inputs with high accuracy.

\subsubsection{\textbf{Countermeasures}}
To defend against malicious eavesdroppers attempting to extract private user information through RF sensing, researchers have proposed two primary strategies: 1) Proactively detecting potential eavesdroppers by analyzing RF signals or using baiting techniques to identify hidden listening devices and take countermeasures, and 2) Adversarially disrupting eavesdroppers by leveraging deep learning, physical-layer perturbations, or external hardware to obscure the eavesdropper's ability to recover genuine behavioral information.

\begin{table*}[htbp]
    \centering
    \renewcommand{\arraystretch}{1.4} 
    \setlength{\tabcolsep}{5.3pt} 
    \caption{Comparison of Countermeasures Against RF Eavesdropping}
    \begin{tabular}{l l l l l}
        \toprule
        \textbf{Category} & \textbf{Defense Strategy} & \textbf{Example} & \textbf{Real-time} & \textbf{Complexity} \\
        \bottomrule
        \multirow{2}{*}{Active Detection} & RF Leakage Detection  & \cite{chaman2018ghostbuster} & Near Real-time & Moderate (MIMO setup) \\
        & EM Radiation Monitoring& \cite{shen2021earfisher} & Near Real-time & Low (EM sensors) \\
        \midrule
        \multirow{3}{*}{Adversarial Disruption} & Data-layer Deep Learning & \cite{zhou2019adversarial, liu2022behavior} & Offline Processing & Low (Software only) \\
        & Physical-layer Obfuscation & \cite{meng2023secur, deng2024css} & Real-time & Moderate (Wi-Fi control) \\
        & Hardware-based Perturbation & \cite{qiao2016phycloak, yao2018aegis, staat2022irshield} & Real-time & High (IRS, relays) \\
        \bottomrule
    \end{tabular}
    \label{tab:defense3}
\end{table*}

\paragraph{\textbf{Active Detection of Potential Eavesdroppers}}

This strategy aims to identify anomalies in the wireless environment, such as signal leakage or abnormal devices, to locate or recognize potential eavesdroppers—even those operating in a fully passive mode without actively transmitting signals. For example, Ghostbuster~\cite{chaman2018ghostbuster} employs a detection mechanism based on RF leakage from hidden eavesdroppers. Even when an eavesdropper remains completely passive, its local oscillator (LO) inevitably emits an extremely weak RF signal. Ghostbuster utilizes MIMO antenna technology to spatially separate the eavesdropper’s leakage signal from regular transmissions and applies a zero-forcing (ZF) algorithm to cancel dominant wireless signals, thereby isolating and highlighting the hidden eavesdropper's leakage. This approach achieves a maximum detection range of 5 meters. 
Notably, Phantom Eavesdropping~\cite{shao2019phantom} introduces a targeted evasion technique that defeats Ghostbuster by whitening the LO RF leakage through dynamic frequency shifting, thereby concealing the eavesdropper’s spectral signature.

In contrast, EarFisher~\cite{shen2021earfisher} introduces a different detection mechanism using a ``baiting and electromagnetic radiation (EMR) monitoring'' approach. By transmitting specially crafted ``bait packets'', it tricks eavesdroppers into caching data, which induces distinctive electromagnetic emissions from their memory buses. By detecting these EMR variations, EarFisher effectively differentiates ordinary devices from eavesdroppers. This technique extends the detection range up to 25 meters and supports through-wall detection, making it highly practical in real-world applications.

\paragraph{\textbf{Adversarially Disrupting Eavesdroppers}}

The core idea is to actively tamper with or obfuscate the RF signals, so that eavesdroppers can not obtain the real behavioral information, and at the same time, ensure the normal communication experience of legitimate users as much as possible. 
Based on the type of signal manipulation, adversarial disruption strategies can be categorized into data-layer adversarial deep learning, transmitter-side physical-layer obfuscation, and external hardware-based perturbation.
\begin{itemize}[leftmargin=*]
    \item \textit{Data-layer adversarial deep learning:} At the data layer, researchers leverage deep neural networks (DNNs) or adversarial generative models to distort or remove behaviorally distinguishable features from RF sensing data, effectively preventing the eavesdropper from recognizing sensitive activities. The key advantage of this approach is that it does not require any modifications to the physical layer hardware and can be applied before data is published or processed in the cloud. However, its primary limitation is that it is mainly effective for offline data processing and is less effective against real-time eavesdropping. For instance, Zhou et al.~\cite{zhou2019adversarial} proposes a deep learning-based adversarial deep network (ADG sub-network) specifically designed to perturb CSI, significantly reducing the accuracy of classifying sensitive behaviors, while ensuring that normal behaviors remain correctly classified. Similarly, Liu et al.~\cite{liu2022behavior} employs a Siamese Network with an identity classifier to enforce similarity among RF signals corresponding to different behaviors, effectively removing behavioral information while retaining identity information. This technique is applicable across multiple wireless sensing platforms, including Wi-Fi, RFID, and millimeter-wave (mmWave) systems.
    \item \textit{Physical-layer obfuscation:} In contrast to data-layer methods, physical-layer obfuscation embeds perturbations directly into the transmitted signals, disrupting CSI at the physical layer or device driver level to prevent eavesdroppers from recovering meaningful behavioral information. This technique typically introduces dynamic ``fake activity signatures'' and encrypted vectors into transmitted signals. Legitimate users—who possess decryption keys or predefined codebooks—can still retrieve the original CSI for regular operations, whereas eavesdroppers only observe distorted or obfuscated CSI. Compared to data-layer methods, this approach can effectively defend against real-time eavesdropping rather than merely post-processing collected data. For example, Secur-Fi~\cite{meng2023secur} proposes a Wi-Fi antenna switching scheme that dynamically alters the antenna selection pattern based on a predefined behavior codebook, ensuring that fake behavior signals closely resemble real human motion patterns—thus deceiving eavesdroppers while still allowing legitimate users to reconstruct authentic activity signals. Building on this concept, CCS~\cite{deng2024css} introduces scrambling vectors at the Wi-Fi physical layer, deliberately generating channel variations that mimic human motion to mislead eavesdroppers. Meanwhile, legitimate receivers can use decryption keys to retrieve the genuine CSI, maintaining normal communication quality.
    \item  \textit{External hardware-based perturbation:} A third category of obfuscation techniques involves external hardware-based adversarial perturbations, which manipulate wireless signals indirectly using separate physical devices such as IRS, full-duplex relays, or rotating antennas. These devices alter, reconstruct, or overlay RF signals in the environment, effectively preventing eavesdroppers from obtaining accurate CSI. The main advantage of this approach is that it does not require modifications to existing Wi-Fi access points (APs) or devices, making it highly compatible with different wireless communication protocols. However, its downside lies in the additional hardware costs and potential deployment complexity. For instance, PhyCloak~\cite{qiao2016phycloak}  utilizes full-duplex amplify-and-forward (A\&F) relays to generate dynamic multipath effects indoors, thereby disrupting an eavesdropper's ability to track stable Doppler shifts and phase variations, significantly reducing its capacity to recognize activities. Similarly, Aegis~\cite{yao2018aegis} perturbs signal amplitude, Doppler frequency shift, and phase to create an environment where external eavesdroppers struggle to extract human motion information while allowing legitimate users to continue normal communications. Lastly, IRShield~\cite{staat2022irshield} leverages IRS technology to dynamically reconfigure wireless propagation paths, ensuring that eavesdroppers cannot establish stable channel models to infer user behavior, thus preserving privacy effectively.
\end{itemize}

\subsection{Summary and Insights}
\subsubsection{\textbf{Summary}}
This section systematically reviews the privacy threats introduced by RF sensing systems across three critical domains: acoustic eavesdropping, keystroke inference, and privacy behavior exposure. 
By leveraging the wall-penetrating, device-free, and contactless characteristics of RF signals, attackers can covertly capture private user information—including speech, typed inputs, gestures, behaviors, and even 3D body shapes—without physical access or user awareness.

RF-based privacy attacks have evolved from limited, constrained-scope techniques to highly detailed and fine-grained reconstructions, driven by advances in signal resolution (e.g., mmWave radar) and learning algorithms (e.g., GANs, deep fusion models). Meanwhile, the attack surface has extended from individual behaviors (e.g., speech commands or PINs) to full-scene and multi-device privacy compromise, including screens, headphones, and VR interactions.

To counter these threats, the section also reviews defense strategies tailored to each task, including access control and encryption, input obfuscation, signal shielding, RF anomaly detection, and adversarial training. However, current countermeasures often face trade-offs between usability and effectiveness, and few are validated under real-time, multi-modal, or adversarial settings—highlighting the need for more robust, adaptive, and system-level privacy protection mechanisms.

\subsubsection{\textbf{Insights}}
RF sensing is undergoing a fundamental paradigm shift in its relationship with user privacy. 
Once celebrated for their non-visual, contactless, and device-free characteristics, RF sensing technologies were widely adopted in domains such as healthcare and smart homes under the assumption that they offered stronger privacy protection than vision-based schemes. However, with advancements in signal resolution, deep learning algorithms, and multi-modal fusion, the very properties that once safeguarded RF sensing—wall penetration, physical transparency, and silent operation—have become inherent advantages for covert surveillance. This duality underscores a growing structural tension between sensing capability and privacy boundaries.

Concurrently, the integration of diverse RF modalities—such as mmWave, UWB, RFID, and LoRa—has expanded the attack surface from visible anatomical features like lips and throats to more covert carriers including headphone and the surfaces of everyday objects. Sensing resolution has increased from the centimeter level to the millimeter and even micron level.
The incorporation of deep generative models, such as GANs and diffusion models, further enhances signal reconstruction quality, pushing RF sensing toward perceptual limits previously attainable only by human observation. 
This has given rise to a new invasion paradigm: “multi-source sampling + model-driven reconstruction.”

Particularly noteworthy is the convergence of large language models (LLMs) with RF sensing, which signals the advent of a new stage of semantic-level privacy intrusion. 
Traditional RF systems primarily captured low-level physical features such as motion types, skeletal postures, or object trajectories, which lacked semantic richness. 
However, LLMs excel at contextual reasoning, semantic inference, and intent recognition—enabling attackers to infer coherent, human-readable meaning from fragmented, low-fidelity RF signals. 
Examples include reconstructing full conversations from intermittent throat vibrations, inferring user intent from gesture and gait patterns, or aligning multi-source signals into a shared semantic space for cross-modal reasoning. As a result, future RF-based attacks may evolve from merely “capturing actions” to “understanding intentions”—ushering in a new frontier of RF sensing semantic inference threats.

In contrast to the accelerating offensive capabilities, existing defense mechanisms remain fragmented, task-specific, and largely decoupled from the underlying communication protocols. 
Most countermeasures focus on perturbation injection or behavioral obfuscation but lack system-level, cross-layer integration. Effective privacy protection will require the development of a unified, end-to-end defense architecture that integrates real-time adversarial mitigation, protocol-level access control, and robust co-design across sensing, communication, and computation layers to handle increasingly complex and multi-modal threats.

Equally pressing is the conceptual gap in current RF privacy research—the absence of a standardized threat modeling framework. 
As this section demonstrates, RF sensing-based intrusions often follow a progressive trajectory from presence detection to identity recognition and ultimately full behavior reconstruction. 
Yet no formal metric currently exists to quantify this layered exposure or evaluate degrees of privacy leakage. 
To address this gap, academia and industry must establish a systematic model that maps sensing capabilities to privacy risk levels, forming a foundation for technical safeguards, ethical governance, and regulatory design. Only then can RF sensing evolve toward a future that is both technically powerful and ethically aligned—a future where privacy is not compromised by capability.

\section{Leveraging RF Sensing for Security Applications}
\label{sec:security}
\subsection{RF Sensing-based Human Authentication and Identification}
RF sensing-based human authentication and identification are essentially fine-grained behavior recognition tasks, designed to distinguish individuals by capturing subtle, person-specific variations in motion dynamics and physiological traits. These tasks leverage a variety of RF modalities - such as Wi-Fi~\cite{han2022wiid, shi2021wifi,shah2017wi,shi2017smart,kong2020continuous,shi2020towards,kong2023toward,shah2018wi,kong2021multiauth,lin2023contactless,li2018studying,kong2019fingerpass,wang2022caution,yu2023wifi,yang2020user,kong2022push,wei2024multi,huang2022continuous,liu2019proactive,zhang2016wifi,gu2022secure,liu2014practical,wang2016gait,ou2022wiwalk,pokkunuru2018neuralwave,yan2025pushing,wang2021gait,zhang2019widigr,zhang2021gaitsense,xu2020attention,li2016using,zhang2020gate,li2020wihf,yang2025environment,deng2022gaitfi,zhang2022metaganfi,jiang2023wi,shahzad2018augmenting,zeng2016wiwho,fei2019multi,fan2022wivi,xiao2024pattern,yan2024freegait,zhang2021wi,ming2019humanfi,yang2023gait,zhang2020gaitid,bai2023wifi,lv2017device,korany2020multiple,song2021imfi,yao2022evosense,ding2020wihi,chen2024wi,cao2025real,liu2022wicrew,zheng2017device,martins2024wifi,korany2019xmodal,lv2017wii,nipu2018human,wu2022widff}, mmWave~\cite{wang2024simultaneous,yang2024openauth,xie2024mmpalm,wang2022heartprint,xie2024palm,jiang2024behaviors,wang2022your,yang2020mu,liu2023wavoid,li2020vocalprint,xu2022mask,dong2021secure,diederichs2017wireless,yang2023wave,han2024mmsign,gu2019mmsense,xu2024gestureprint,li2023passive,li2023mmhsv,hao2022mmsafe,hof2020face,ozturk2021gaitcube,shan2024identitykd,zhao2024mn,challa2021face,guo2024millimeter,ni2022gait,cao2022cross,wang2024rdgait}, UWB~\cite{mokhtari2017non,vecchio2019method,cao2024uwb,arra2019personalized,rana2022markerless,rana2019non}, LoRa~\cite{ge2023logait,yang2023xgait,yin2023gait}, and RFID~\cite{jiang2022rf,yang2024rf,chen2023sensing,chen2022rfpass,huang2019id,paranjpay2024multi,zhang2020unobtrusive,zhang2018rfree,huang2022rfid,saxena2011vibrate} - to sense signal perturbations caused by human movement and biometric signatures.

Although closely related, authentication and identification differ in their objectives and application contexts. 
Authentication verifies a claimed identity through one-to-one matching, commonly used in access control or personalized service scenarios, and emphasizes reliability and resistance to spoofing. 
In contrast, identification aims to determine a person’s identity from among many via one-to-many matching, often applied in surveillance, crowd monitoring, or multi-user interaction settings, requiring high scalability and generalization capability.


\begin{table*}[htp!]
\centering
\renewcommand{\arraystretch}{1.3}
\setlength{\tabcolsep}{5pt}
\begin{threeparttable}
\caption{\textbf{Comparison of RF Sensing-Based Human Authentication Methods}}
\label{tab:rf_auth}
\begin{tabular}{c c c c c c c}
\toprule
\textbf{Scheme} & \textbf{Year} & \textbf{RF Modality} & \textbf{Sensing Target} & \textbf{Continuous Auth.} & \textbf{Multi-user Support} &\textbf{Adaptability} \\
\midrule

SmartAuth~\cite{shi2017smart} & 2017 & Wi-Fi &  Activity  & \CIRCLE & \Circle & \RIGHTcircle\\
WiAU~\cite{lin2018wiau} & 2018 & Wi-Fi & Activity  & \Circle & \Circle & \Circle\\
FingerPass~\cite{kong2019fingerpass} & 2019 & Wi-Fi & Finger Gestures & \CIRCLE & \Circle & \RIGHTcircle\\
FreeAuth~\cite{kong2022push} & 2022 & Wi-Fi & Undefined Body Gestures &  \Circle & \Circle & \CIRCLE\\
Wi-Access~\cite{shah2018wi} & 2018 & Wi-Fi & Typing Gestures  & \Circle & \Circle & \Circle\\
WiPass~\cite{gu2022secure} & 2022 & Wi-Fi & Keystroke Dynamics & \Circle & \Circle & \Circle\\
CAUTION~\cite{wang2022caution} & 2022 & Wi-Fi & Gait  & \Circle & \Circle & \CIRCLE\\
Multi-WiIR~\cite{wei2024multi} & 2024 & Wi-Fi & Gait  & \Circle & \CIRCLE & \RIGHTcircle\\
MultiAuth~\cite{kong2021multiauth} & 2021 & Wi-Fi & Gait & \Circle & \CIRCLE & \RIGHTcircle\\
OpenAuth~\cite{yang2024openauth} & 2024 & mmWave & Body Silhouette \& Posture & \CIRCLE & \CIRCLE & \CIRCLE\\
mmSign~\cite{han2024mmsign} & 2024 & mmWave & Handwritten Signature & \Circle & \Circle & \CIRCLE\\
HeartPrint~\cite{wang2022heartprint} & 2022 & mmWave & Heartbeat & \CIRCLE & \CIRCLE & \Circle \\
M-Auth~~\cite{wang2022your} & 2022 & mmWave & Respiration & \CIRCLE & \CIRCLE & \Circle \\
MN-UIV~\cite{zhao2024mn} & 2024 & mmWave & Breathing \& Heartbeat \& Posture & \CIRCLE & \Circle & \RIGHTcircle \\
mmPalm~\cite{xie2024mmpalm} & 2024 & mmWave & Palm & \Circle & \Circle & \Circle \\
mmFace~\cite{xu2022mask} & 2022 & mmWave & Facial Structure & \Circle  & \Circle  & \Circle \\
mmFaceID~\cite{jiang2024behaviors} & 2024 & mmWave &  Dynamic Facial Activity & \Circle  & \Circle  & \RIGHTcircle\\
VocalPrint~\cite{li2020vocalprint} & 2020 & mmWave & Vocal Vibration  & \CIRCLE & \Circle & \Circle\\
Dong et al.~\cite{dong2021secure}  & 2021 & mmWave & Vocal Vibration \& Lip Motion & \Circle & \Circle & \Circle\\
Au-Id~\cite{huang2019id} & 2019 & RFID & Activity   & \Circle & \Circle & \CIRCLE \\
RFPass~\cite{chen2022rfpass} & 2022 & RFID & Gait  & \Circle & \CIRCLE & \Circle \\
RFace~\cite{xu2021rface} & 2021 & RFID & Facial Structure  & \Circle & \Circle & \Circle \\
Arra et al.~\cite{arra2019personalized}  & 2019 & UWB & Gait & \CIRCLE & \Circle &\Circle\\
\bottomrule
\end{tabular}
\begin{tablenotes}
    \item \Circle ~for Low / No, \RIGHTcircle  ~for Middle, \CIRCLE  ~for High / Yes.
\end{tablenotes}
\end{threeparttable}
\end{table*}

\subsubsection{\textbf{Authentication}}
RF sensing enables contactless and device-free user authentication by capturing behavioral, physiological, or structural traits from RF signal interactions. This section introduces representative authentication approaches based on Wi-Fi, mmWave, and RFID sensing.
Table~\ref{tab:rf_auth} summarizes the key differences of representative RF sensing-based authentication methods.

\paragraph{\textbf{Wi-Fi-based Methods}}
Wi-Fi sensing-based human authentication leverages human-induced signal fluctuations to enable device-free and non-intrusive identity verification.

One prominent line of research focuses on \textit{activity-based} authentication~\cite{lin2018wiau, shi2017smart, shi2020towards,shi2021wifi}, which exploits the fact that individuals tend to exhibit unique patterns in their daily activities. By capturing and analyzing these behavioral signatures through Wi-Fi, systems can verify identity in a passive manner. 

Among activity-based methods, \textit{gait patterns} stand out as a stable and distinctive biometric.
Early systems focus on single-user gait authentication, extracting individual walking signatures from CSI and achieving high accuracy in controlled settings.
For instance, CAUTION~\cite{wang2022caution} employs a few-shot learning framework combined with open-set recognition, allowing it to authenticate users with limited training data while detecting unknown intruders based on gait-induced CSI features
To extend gait-based authentication to multi-user scenarios, recent works~\cite{kong2021multiauth,wei2024multi} tackle the challenge of signal entanglement caused by overlapping movements. For example, MultiAuth~\cite{kong2021multiauth} introduces a multipath time-of-arrival (ToA) algorithm to reconstruct individual CSI streams in multi-user settings, followed by deep learning models to perform parallel gait-based identification and spoofing detection.

Beyond activities, \textit{gesture-based} authentication~\cite{kong2019fingerpass, kong2020continuous, kong2022push, gu2022secure} represents a finer-grained approach, leveraging distinctive motion dynamics to enhance identity resolution. 
For instance, FingerPass~\cite{kong2019fingerpass} utilizes CSI phase variations induced by finger gestures to support real-time, continuous user authentication in smart homes. More recently, FreeAuth~\cite{kong2022push} pushes the boundary further by enabling gesture-independent authentication—extracting user-invariant physiological features regardless of specific gesture types via adversarial learning.
In parallel, WiPass~\cite{gu2022secure} explores micro-gesture authentication by capturing keystroke-induced CSI fluctuations. It enhances NLoS signal components to better sense subtle finger motions and leverages a CNN-SVM pipeline to achieve PIN-free, device-free identity verification.
Similarly, KeySign~\cite{fu2023keysign} leverages keystroke-induced CSI patterns for user authentication using a dual-receiver CNN-based framework.

\paragraph{\textbf{mmWave-based Methods}}
mmWave sensing has emerged as a key enabler for user authentication, owing to its high spatial resolution and exceptional sensitivity to subtle human motions. 
Compared to Wi-Fi sensing, mmWave systems offer the capability to capture not only coarse body movements but also fine-grained dynamic variations, thereby enabling richer biometric representations for contactless authentication.

A broad class of \textit{motion-based} authentication approaches has also been developed, leveraging mmWave’s ability to sense user-specific movement patterns with high precision. These methods span from coarse daily activities to fine-grained gestures such as in-air handwriting. For instance, OpenAuth~\cite{yang2024openauth} constructs posture-normalized human silhouettes and dynamic movement sequences, allowing continuous authentication under natural daily activities while ensuring user privacy in open-world environments.
To support more fine-grained identity verification, recent work has extended motion-based authentication to capture personalized \textit{handwriting behaviors}~\cite{li2023mmhsv,han2024mmsign}. For instance, mmSign~\cite{han2024mmsign} is an mmWave-based handwritten signature verification system that captures dynamic signing features such as velocity for secure document authentication. It leverages a meta-learning framework to adapt to new users with limited data, improving practicality and resistance to forgery.

Beyond behavioral traits, attention has gradually shifted to intrinsic physiological features with greater uniqueness and spoof resistance. Among them, \textit{heartbeat} and \textit{respiration} are involuntary rhythms that provide stable and difficult-to-mimic biometric signals. 
For instance, HeartPrint~\cite{wang2022heartprint} proposes a multi-user authentication system based on heartbeat-induced skin micro-vibrations, utilizing frequency-domain features derived from mmWave reflections for high-precision authentication. 
Similarly, M-Auth~\cite{wang2022your} leverages the morphological characteristics of respiratory waveforms, combined with dynamic beam steering and energy-based signal discrimination, to enable reliable multi-user authentication even at close proximity.
Furthermore, MN-UIV~\cite{zhao2024mn} integrates static physiological signals (e.g., breathing and heartbeat) and dynamic posture features (e.g., DRAI) for user identity verification, and introduces multimodal neural network architectures to achieve feature fusion and representation learning.

In addition to physiological traits, structural biometrics such as \textit{palmprints} and \textit{facial geometry} have also been explored for mmWave-based authentication.
mmPalm~\cite{xie2024mmpalm} proposes a palm-based authentication system that extracts geometric and textural features from reflected mmWave signals using a commercial device. It enhances robustness through virtual antenna synthesis, data augmentation, and domain adaptation, achieving high accuracy and strong spoof resistance.
Similarly, facial-structure-based authentication~\cite{hof2020face,xu2022mask,hof2020face} has shown great promise due to mmWave’s ability to sense 3D facial contours in a privacy-preserving manner. Hof et al.~\cite{hof2020face} first demonstrated the feasibility of capturing facial contours using mmWave radar, employing 802.11ad/y chips and autoencoder-based networks to achieve high-resolution face verification.
Further, mmFace~\cite{xu2022mask} employs SAR imaging and cross-modal virtual registration to synthesize mmWave facial templates from photos without requiring on-site data collection, enabling live face authentication even under mask-wearing conditions and defending against 2D image and 3D mask attacks.
Building on this, mmFaceID~\cite{jiang2024behaviors} introduces dynamic facial activities—such as subtle muscular movements during speech—as key biometric signals. It reconstructs facial expression parameters from mmWave reflections, enabling high-precision identity verification with built-in liveness detection.

Moreover, mmWave can capture \textit{vocal-induced micro-vibrations} around the throat or lips during speech, enabling secure voice-related identification while mitigating replay attack risks. For example, VocalPrint~\cite{li2020vocalprint} directly senses vocal fold vibrations via mmWave radar and extracts spectral and temporal features from reflected skin signals, forming a robust voice authentication framework that effectively resists mimicry, spoofing, and environmental interference.
Dong et al.~\cite{dong2021secure} integrate mmWave-sensed vocal cord vibrations (VCV) and lip movements (LM) to enable multi-modal voice-based authentication, achieving high verification accuracy and strong anti-spoofing capability in smart home environments.

\paragraph{\textbf{Other RF Modalities-based Methods}}

Recent works have explored RF sensing beyond Wi-Fi and mmWave for human authentication, focusing primarily on \textit{RFID} and \textit{UWB}. 

RFID-based user authentication leverages backscattered RF signals to extract user-specific physical and behavioral traits in a passive, device-free manner. Benefiting from its low cost, battery-free operation, and flexible deployment, RFID sensing has emerged as a promising modality for fine-grained biometric authentication.

One direction focuses on behavioral authentication from \textit{daily motion sequences}. Au-Id~\cite{huang2019id} captures user-indicating micro-motions—such as knocking—using phase and RSSI signals from a spatially arranged RFID tag array. It fuses these multi-modal signals via a CNN-LSTM architecture to extract user-specific representations, and incorporates transfer learning and one-class SVMs to support scalable, personalized authentication.
To achieve authentication invariant to environmental changes, RFPass~\cite{chen2022rfpass} exploits Doppler shift features for \textit{gait-based} identity recognition. It introduces the MDSS algorithm to isolate user-specific signal paths from multipath interference and applies a CNN-RNN model to extract robust spatial-temporal gait profiles, enabling authentication across diverse walking conditions.

In addition to behavioral traits, RFace~\cite{xu2021rface} explores physiological biometrics using \textit{facial features}. It utilizes a 7$\times$7 RFID tag array to capture 3D facial geometry and subsurface biomaterial signatures through RSS and phase differences. The system employs a distance-deflection disturbance suppression algorithm to ensure stability under pose variations, providing privacy-preserving face-based authentication without revealing visual facial information.

Unlike previous RFID sensing methods that infer identity from ambient signal variations, some works focus on \textit{RFID tag-side authentication}. 
For instance, Vibrate-to-Unlock~\cite{saxena2011vibrate} enables users to unlock passive RFID tags via phone vibrations encoding a PIN, adding a physical layer of control to tag access. In contrast, multi-model systems~\cite{paranjpay2024multi} integrate RFID as a secondary factor alongside vision-based gait recognition, forming a modular multi-factor authentication framework. These approaches highlight alternative roles of RFID beyond sensing—emphasizing tag protection and system-level integration rather than identity inference via RF signals.

In the domain of UWB-based sensing, wearable authentication systems have also been explored. For example, Arra et al.~\cite{arra2019personalized} propose a wearable UWB authentication framework that extracts gait patterns from inter-device distances measured across the body, using one-class classification for personalized verification. 
More recently, UWB-Auth~\cite{cao2024uwb} introduces a two-factor authentication platform that leverages UWB ranging and AoA measurements for real-time location verification, combined with user-owned tokens such as motion rings or fingerprint-enabled devices. This approach enhances security by incorporating physical proximity into the authentication process and defending against phishing and relay attacks

\begin{table*}[htp!]
\centering
\renewcommand{\arraystretch}{1.3}
\setlength{\tabcolsep}{3.3pt}
\begin{threeparttable}
\caption{\textbf{Comparison of Wi-Fi-Based Person Identification Methods}}
\label{tab:rf_id}
\begin{tabular}{c c c c c c c}
\toprule
\textbf{Scheme} & \textbf{Year} & \textbf{RF Modality} & \textbf{Sensing Target} & \textbf{Group Scale} & \textbf{Multi-user Support} & \textbf{Adaptability} \\
\midrule
WiWho~\cite{zeng2016wiwho} & 2016 & Wi-Fi & Gait & 2–6 & \Circle & \Circle \\
WiDIGR~\cite{zhang2019widigr} & 2019 & Wi-Fi & Gait & 3–6 & \Circle & \CIRCLE \\
GaitSense~\cite{zhang2021gaitsense} & 2021 & Wi-Fi  & Gait & 2–11 & \Circle & \CIRCLE \\
WiWalk~\cite{ou2022wiwalk} & 2022 & Wi-Fi & Gait & 3–5 & \CIRCLE~2& \Circle \\
XModal-ID~\cite{korany2019xmodal} & 2019 & Video $\rightarrow$ Wi-Fi & Human Mesh & 2-8 & \Circle & \CIRCLE \\
XGait~\cite{yang2023xgait} & 2023 & IMU $\rightarrow$ Wi-Fi / mmWave / LoRa & Gait & 5-15 & \Circle & \CIRCLE \\
GaitFi~\cite{deng2022gaitfi} & 2022 & Wi-Fi + Video & Gait & 12 & \Circle & \RIGHTcircle \\
WiHF~\cite{li2020wihf} & 2020 & Wi-Fi & Gesture & 6 & \Circle & \CIRCLE \\
BreathID~\cite{guo2022breathid} & 2022 & Wi-Fi & Respiration & 11 & \Circle & \Circle \\
GRi-Fi~\cite{wang2021gait} & 2022 & Wi-Fi & Gait \& Respiration & 10 & \Circle & \RIGHTcircle \\
WiPIN~\cite{wang2019wipin} & 2019 & Wi-Fi &  Biological Features & 2-30 & \Circle & \Circle \\
MU-ID~\cite{yang2020mu} & 2020 & mmWave & Gait  & 10 & \CIRCLE~2-4& \Circle \\
RDGait~\cite{wang2024rdgait} & 2024 & mmWave & Gait  & 125 & \CIRCLE~2-4 & \CIRCLE \\
mID~\cite{zhao2019mid} & 2019 & mmWave & Gait \& Body Shape  & 12 & \CIRCLE~2 & \RIGHTcircle \\
mmSense~\cite{gu2019mmsense} & 2019 & mmWave & Body Contour \& Vital Signs & 5 & \CIRCLE~5 & \RIGHTcircle \\
MN-UIV~\cite{zhao2024mn} & 2024 & mmWave  & Posture + Vital Signs & 3-11 & \Circle & \RIGHTcircle \\
GesturePrint~\cite{xu2024gestureprint} & 2024 & mmWave & Gesture & 17 & \Circle & \Circle \\
PointFace~\cite{zhong2023face} & 2023 & mmWave  & Facial Structure  & 9 & \Circle & \Circle \\
WavoID~\cite{liu2023wavoid} & 2023 & mmWave + Audio & Vocal Vibration \& Microphone Voice & 100 & \Circle & \RIGHTcircle \\
RFree-ID~\cite{zhang2018rfree} & 2018 & RFID & Gait  & 2-30 & \Circle & \Circle \\
RFPass~\cite{chen2023sensing} & 2023 & RFID & Gait  & 15 & \Circle & \CIRCLE \\
Rana et al.~\cite{rana2019non} & 2019 & UWB & Gait  & 15 & \Circle & \Circle \\
LoGait~\cite{ge2023logait} & 2023 & LoRa & Gait & 6 & \Circle & \Circle \\

\bottomrule
\end{tabular}
\begin{tablenotes}
    \item \Circle ~for Low / No, \RIGHTcircle  ~for Middle, \CIRCLE  ~for High / Yes.
\end{tablenotes}
\end{threeparttable}
\end{table*}

\subsubsection{\textbf{Identification}}
Unlike authentication, identification seeks to determine who a person is from a group. 
This one-to-many task requires higher discriminability and scalability. RF sensing enables such identification by capturing user-specific traits through Wi-Fi, mmWave, RFID, UWB, and LoRa signals.
Table~\ref{tab:rf_id} summarizes representative RF sensing-based identification methods.

\paragraph{\textbf{Wi-Fi-based Methods}}
Wi-Fi-based human identification focuses on recognizing individuals from their unique movement dynamics or inherent body characteristics. 

Early \textit{gait-based} identification methods~\cite{zeng2016wiwho,zhang2016wifi,wang2016gait} primarily focused on controlled environments with fixed walking paths and single walking directions, aiming to validate the presence of individual-specific patterns embedded in Wi-Fi signal. 
For instance, WiWho~\cite{zeng2016wiwho} proposed a CSI-based identification framework that extracts statistical gait features such as step period and stride length for person recognition. 
To enhance system usability and generalizability, subsequent works~\cite{zhang2019widigr,zhang2021gaitsense,zhang2022metaganfi,yan2025pushing} sought to overcome the fixed-path assumption and improve robustness to walking direction, speed variation, environmental changes, and non-gait behaviors. For instance, WiDIGR~\cite{zhang2019widigr} was the first to support direction-independent recognition by constructing a two-dimensional Fresnel zone, combining walking direction estimation and CSI spectrum reconstruction to unify diverse gait signals into a common representation space. 
Moreover, WiWalk~\cite{ou2022wiwalk} targeted practical multi-user scenarios, proposing a Deep Clustering-based approach to separate overlapped gait signals for simultaneous dual-user identification. 

To improve robustness across different environments and sensing modalities, recent research has explored \textit{cross-modal fusion} strategies.  
XModal-ID~\cite{korany2019xmodal} pioneered a Wi-Fi-video cross-modal identification framework by synthesizing CSI spectrograms from 3D video-based human meshes, enabling cross-modal identity verification even through walls. IMFi~\cite{song2021imfi} combined IMU and Wi-Fi sensing modalities, using IMU-derived personalized gait patterns to support environment-agnostic recognition without retraining. XGait~\cite{yang2023xgait} further extended cross-modality to a generative paradigm, translating IMU data into RF spectrograms via a transformer-based generator, enabling model-free deployment without any RF data collection.
Recent works have also introduced fine-grained multimodal collaborative systems to enhance real-time recognition performance. GaitFi~\cite{deng2022gaitfi} proposed a vision–Wi-Fi fused gait identification framework using a dual-branch architecture and joint loss optimization, achieving high-precision device-free identification across lighting and environmental conditions. RFCam~\cite{chen2022rfcam} focused on person-device association, fusing video and CSI data to map individuals in camera frames to their corresponding device MAC addresses through an uncertainty-aware multimodal model that integrates AoA, distance, activity, and visual context for accurate identification.

Beyond gait patterns, some studies~\cite{zheng2017device,li2020wihf} have explored more flexible \textit{activity features} for identity recognition. WiID~\cite{zheng2017device} introduced a user behavior-based model that encodes both explicit actions and implicit features to model action sequences for identity recognition.
In addition, WiHF~\cite{li2020wihf} proposed a real-time system that jointly performs gesture recognition and user identification by extracting domain-independent motion change patterns from CSI. 

In parallel, other works~\cite{guo2022breathid,wang2021gait,wang2019wipin,han2022wiid} have investigated physiological information (i.e., respiration patterns or biological features) for identification. 
For instance, BreathID~\cite{guo2022breathid} used Wi-Fi sensing to capture users’ resting respiration patterns, extracting stable features such as breathing frequency and amplitude, and applied WMD-DTW for static user identification. Building on this, GRi-Fi~\cite{wang2021gait} combined gait and respiratory signals, performing CSI segmentation and multimodal feature fusion to enable robust user identification.
Furthermore, WiPIN~\cite{wang2019wipin} proposed an bio-electromagnetic information-based identification system. This method introduces a novel perspective that Wi-Fi signals, when propagating through the human body, are modulated by individual-specific physical characteristics such as body shape, fat rate, and muscle composition, leading to distinguishable signal distortions. Han et al.~\cite{han2022wiid} further explored the modeling of bio-electromagnetic signals during intermittent motion states. By designing a motion sensitivity vector, the system automatically segments short-term stationary periods from continuous CSI data and extracts stable physiological features for identity recognition without requiring the user to maintain a specific posture.

\paragraph{\textbf{mmWave-based Methods}}

MmWave-based methods typically captures dynamic gait information of individuals during walking by extracting micro-Doppler features or constructing sparse point clouds. 

Compared with Wi-Fi-based sensing solutions, mmWave radar offers higher spatial and velocity resolutions, allowing for finer discrimination among \textit{multiple individuals} in complex environments. As a result, it demonstrates significant advantages in multi-person identification tasks.
Among the existing research, the most prominent category of methods relies on Range-Doppler Heatmaps to extract intra-gait velocity variation patterns~\cite{pegoraro2020multiperson,yang2020mu,li2023passive,ni2022gait,wang2024rdgait}. 
Representative works such as MU-ID~\cite{yang2020mu} and MCGait~\cite{li2023passive} build spatiotemporal features based on lower-limb motion or micro-Doppler spectrograms, and achieve high-accuracy identification of multiple individuals via segmentation-based separation. 
The recently proposed RDGait~\cite{wang2024rdgait} further incorporates an environment-independent ghost removal algorithm and attention-based temporal modeling, maintaining robust performance under various angles, elevations, and walking behaviors. These systems commonly employ CNNs and RNNs to learn identity-specific gait representations.
In addition to heatmap-based approaches, other studies have explored the use of sparse point clouds generated by mmWave radar for multi-person identification. For instance, mID~\cite{zhao2019mid} performs point cloud clustering and trajectory construction, and utilizes a bidirectional LSTM network for user identification. Compared with Doppler heatmaps, point cloud methods provide higher spatial interpretability but face challenges in temporal alignment and trajectory association.

In contrast to gait-centric approaches, mmSense~\cite{gu2019mmsense} adopts a distinct modeling strategy that combines static \textit{body contour features with vital signs such as respiration and heartbeat}. By leveraging the high directionality and low penetration capability of mmWave, it constructs environment-specific fingerprint maps to identify multiple users.
Similarly, MN-UIV~\cite{zhao2024mn} fuses dynamic posture and vital signs using dual-radar input for home user identification.
GesturePrint~\cite{xu2024gestureprint} captures individual motion styles from mmWave \textit{gesture} point clouds and identifies users using an attention-based network.

Unlike above motion-based approaches that rely on dynamic behavioral patterns, \textit{facial structure}-based methods~\cite{challa2021face,zhong2023face} extract anatomical structures from mmWave radar signals. 
For instance,
RFaceNet~\cite{challa2021face} utilizes 60~GHz mmWave radar imaging and applies a convolutional auto-encoder with a random forest classifier for face recognition. 
PointFace~\cite{zhong2023face} directly employs point clouds captured by a low-cost 77~GHz radar and leverages a lightweight PointNet-based network for privacy-preserving face recognition at the edge.

Beyond single-mmWave modality solutions, recent studies have proposed multi-modal mmWave-based identification systems that combine mmWave with other sensing modalities such as voice, vision, and physiological data, aiming to improve robustness, adaptability, and security.
For instance, 
WavoID~\cite{liu2023wavoid} integrates mmWave-sensed vocal vibrations and microphone voice signals to enhance voice-based authentication and resist spoofing.
Other studies explore cross-modal identity matching by linking mmWave with vision-based inputs. For example, Cao et al.~\cite{cao2022cross} proposed to leverage gait signatures from mmWave and RGB-D cameras to associate identities across camera-free and camera-allowed zones.
Mission~\cite{liu2024mission} aligns mmWave point clouds with RGB images to enable cross-space person re-identification.
In addition, recent works~\cite{yang2023xgait,fan2024video2mmpoint} explore cross-modal signal generation, translating IMU or video data into mmWave representations to reduce reliance on RF data collection and enhance generalization across environments.



\paragraph{\textbf{Other RF Modalities-based Methods}}
Beyond Wi-Fi and mmWave, other RF technologies such as RFID, UWB, and LoRa have also been explored for person identification, offering complementary trade-offs in deployment cost, energy efficiency, and spatial coverage.

RFID-based methods benefit from their low cost, passive operation, and ease of deployment, and are typically implemented using spatial arrays of commodity tags. One prominent line of work~\cite{chen2023sensing,zhang2018rfree,yang2024rf,jiang2022rf} focuses on extracting user-specific gait signatures by analyzing variations in phase and RSSI as individuals walk through RFID-tagged environments. For example, RFree-ID~\cite{zhang2018rfree} emphasizes robust gait segmentation under continuous walking conditions, while RFPass~\cite{chen2023sensing} introduces Doppler shift modeling and multipath DoA filtering to achieve environment-independent recognition. 
In parallel, another direction~\cite{feng2021rf,zhao2025pedestrian} enhances identity separability by combining dynamic gait features with static physical traits, such as body shape or spatial posture, using attention-based neural architectures.

UWB-based systems share a similar focus on gait dynamics but leverage ultra-wideband’s high temporal resolution and multipath resilience.Non-wearable systems~\cite{rana2022markerless,rana2019non,mokhtari2017non} utilize UWB radar to extract fine-grained gait signatures or 3D motion features, often focusing on passive, markerless recognition in smart home or healthcare settings.  
In contrast, wearable approaches~\cite{vecchio2019method} estimate gait-induced inter-device distances from body-worn UWB nodes and classify individuals based on spatial motion patterns, enabling identification among multiple users

LoRa-based methods offer a low-power, long-range alternative by capturing gait-induced signal variations in challenging environments such as underground coal mines~\cite{yin2023gait,ge2023logait}. While limited to single-user settings, they demonstrate the feasibility of identification using low-bandwidth signals across extended distances.

\subsection{RF Sensing-based Secret Key Generation}
RF sensing-based key generation can be regarded as a security-oriented sensing task, where the goal is to perceive and exploit physical-layer characteristics of the wireless environment—such as CSI and RSSI—to establish shared cryptographic keys between devices. Unlike conventional RF sensing tasks that focus on inferring human activities or identities, key generation centers on sensing channel consistency and physical co-presence, effectively transforming physical-layer signal dynamics into a secure communication primitive.
RF sensing-based key generation has proven particularly useful for IoT device pairing, wearable communications, and contactless authentication, offering a non-intrusive, lightweight, and environment-adaptive security solution.

\begin{figure}
\centering
\subfigure[Channel Reciprocity-based Approaches]{
\begin{minipage}[t]{0.71\linewidth}
\centering
\includegraphics[width=\linewidth]{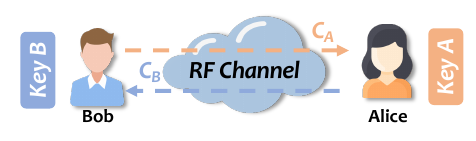}
\label{fig:recKey}
\end{minipage}
}

\subfigure[Proximity-based Approaches]{
\begin{minipage}[t]{0.7\linewidth}
\centering
\includegraphics[width=\linewidth]{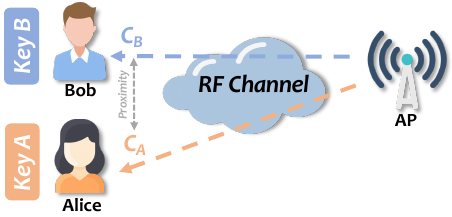}
\label{fig:proxKey}
\end{minipage}
}
\centering
\caption{RF sensing-based Key Generation Methods.}
\label{fig:keyGen}
\end{figure}

Depending on the spatial configuration of the communicating parties (Alice and Bob) and the type of channel information sensed, RF sensing-based key generation schemes can be broadly categorized into two paradigms: channel reciprocity-based methods and proximity-based methods, as shown in Fig.~\ref{fig:keyGen}.
In the channel reciprocity-based scheme, Alice and Bob generate the same key by leveraging their shared RF channel, ensuring the key remains secure from potential eavesdroppers. In contrast, the proximity-based scheme relies on Alice and Bob being physically close, allowing them to receive signals from a common third party to establish a shared key.

\textbf{Principle.} RF sensing-based secret key generation relies on three core principles: \textit{temporal variation}, \textit{channel reciprocity}, and \textit{spatial decorrelation}. These principles enable secure and efficient key generation and are explained below:

\begin{itemize}[leftmargin=*]
    \item \textit{Temporal Variation.} The wireless channel experiences dynamic effects like reflection, refraction, and scattering, especially in mobile environments. These effects result in unpredictable and random variations over time, which can be harnessed to generate cryptographic keys.
    \item \textit{Channel Reciprocity.} When uplink and downlink transmissions share the same carrier frequency, the channel properties observed by both ends (e.g., Alice and Bob) are reciprocal. This reciprocity ensures that Alice and Bob can obtain highly correlated channel measurements, enabling them to generate identical keys.
    \item \textit{Spatial Decorrelation.} According to communication theory, when an eavesdropper (Eve) is located more than half a wavelength away from the legitimate users, she experiences uncorrelated channel effects. This spatial property ensures that the keys generated by Alice and Bob cannot be deduced by Eve, thereby guaranteeing their security.
\end{itemize}

The principles of RF sensing-based key generation have been extensively modeled, validated, and applied across various wireless technologies, such as Wi-Fi and ultrawideband (UWB) systems.

\textbf{Protocol.} A typical protocol for RF secret key generation consists of four main steps: \textit{channel probing}, \textit{quantization}, \textit{information reconciliation}, and \textit{privacy amplification}.

\begin{itemize}[leftmargin=*]
    \item Channel Probing. This step requires bidirectional signal exchange between Alice and Bob.
Alice first transmits a signal to Bob, who measures the channel information using parameters such as RSSI, Channel Impulse Response (CIR), or Channel Frequency Response (CFR). Bob then replies to Alice, who measures the same parameter, completing a pair of measurements. This process is repeated until sufficient measurements (\(X_A\) and \(X_B\)) are collected. The choice of parameters depends on the wireless technology; for example, RSSI is widely available across standards, while CIR and CFR are specific to wideband systems like UWB or IEEE 802.11.

   \item \textit{Quantization.} To generate a binary cryptographic key, the collected analog measurements (\(X_u\)) are quantized into a binary sequence (\(K_u\)). Two common methods are typically used: (1) threshold-based quantization using mean and standard deviation to determine boundaries, and (2) CDF-based quantization leveraging the cumulative distribution function for multi-bit encoding.

    \item \textit{Information Reconciliation.} Due to channel variations and noise, there may be mismatches between the keys generated by Alice (\(K_A\)) and Bob (\(K_B\)). The mismatch rate, or Key Disagreement Rate (KDR), is given by:
    \begin{equation}
    \text{KDR} = \frac{1}{l_K} \sum_{i=1}^{l_K} |K_A(i) - K_B(i)|,
    \end{equation}
    where $l_K$ is the key length.
    Error correction codes (ECCs), such as Secure Sketch algorithms, are used to reconcile mismatched keys. For example, Alice sends a syndrome (\(s\)) derived from her key, allowing Bob to correct his key based on their shared ECC.

    \item \textit{Privacy Amplification.} To mitigate the risk of information leakage during reconciliation, privacy amplification is applied. Techniques such as extractors, universal hashing functions, or cryptographic hash functions are used to refine the key. After this step, Alice and Bob obtain a secure, identical key that can be used for symmetric encryption, such as AES-128, to protect subsequent communications.
\end{itemize}

\begin{table*}[t!]
\centering
\renewcommand{\arraystretch}{1.4} 
\setlength{\tabcolsep}{5.3pt} 

\caption{\textbf{Comparison of RF Sensing-based Secret Key Generation Schemes}}
\begin{tabular}{c c c c c c c c}
\toprule
\textbf{Reference} & \textbf{Year} & \textbf{RF Signal Type} & \textbf{Approach} & \textbf{Target Scene} & \textbf{KGR (bit/s)} & \textbf{KAR (\%)} & \textbf{Randomness} \\
\bottomrule
Mathur et al.~\cite{mathur2008radio} & 2008 & Wi-Fi CIR & Channel Reciprocity-based & Indoor & $\sim$1 & \textgreater80 & \ding{51} \\

Patwari et al.~\cite{patwari2009high} & 2010 & ZigBee RSSI & Channel Reciprocity-based & Indoor & 10-22 & 97.8-99.46 & \ding{51} \\

Zeng et al.~\cite{zeng2010exploiting} & 2010 & Wi-Fi RSSI & Channel Reciprocity-based & Indoor & 10 & 90 & \ding{55} \\

Premnath et al.~\cite{premnath2014secret} & 2014 & Bluetooth RSSI & Channel Reciprocity-based & Indoor & - & \textgreater79 & \ding{51} \\

Xi et al.~\cite{xi2016instant} & 2016 & Wi-Fi CSI & Proximity-based & Indoor & 90-120 & 96.5-98 & \ding{51} \\

Xu et al.~\cite{xu2018lora} & 2018 & LoRa RSSI & Channel Reciprocity-based & Indoor/Outdoor & 18-31 & 98-100 & \ding{51} \\

Ruotsalainen et al.~\cite{ruotsalainen2019experimental} & 2020 & LoRa RSSI & Channel Reciprocity-based & Indoor/Outdoor & - & 71-85 & \ding{51} \\

Gao et al.~\cite{gao2021novel} & 2021 & LoRa RSSI & Channel Reciprocity-based & Indoor/Outdoor & 13.8 & - & \ding{51} \\
\bottomrule
\end{tabular}
\label{tab:key_gen}
\end{table*}

\subsubsection{Channel Reciprocity-based Approaches}
Channel reciprocity-based key generation has attracted extensive research interest and has been implemented across various wireless technologies, such as ZigBee, Wi-Fi, LoRa, and so on. In such systems, Alice and Bob leverage the reciprocal characteristics of the wireless channel between them to generate shared secret keys.

ZigBee, based on IEEE 802.15.4, operates at the 2.4~GHz band and is widely used in wireless sensor networks. Aono et al.~\cite{aono2005wireless} first introduced a practical ZigBee-based key generation protocol, utilizing an electronically steerable parasitic array radiator (ESPAR) antenna to induce channel fluctuations. Patwari et al.~\cite{patwari2009high} proposed the high-rate uncorrelated bit extraction (HRUBE) framework, achieving a bit rate of 22~bit/s with a key disagreement rate (KDR) of 2.2\%. Ali et al.~\cite{ali2013eliminating} explored key generation in body area networks, demonstrating feasibility across dynamic and static environments. More recently, Li et al.~\cite{li2017secret} presented an RSSI trajectory-based key generation system enhanced with Bloom filters and Karhunen--Loeve Transform (KLT), achieving robust performance and generating a 128-bit key within 1~second.

Wi-Fi, based on the IEEE 802.11 family of standards, has become a dominant platform for key generation research due to its ubiquitous presence and support for fine-grained CSI. Early works, such as Mathur et al.~\cite{mathur2008radio}, relied on RSSI and CIR peaks, achieving good key agreement without reconciliation. However, the advent of CSI-based methods significantly improved key generation performance. Liu et al.~\cite{liu2013fast} proposed the first practical CSI-based system, achieving 60--90~bits per packet. Follow-up studies explored CSI-based key generation mechanisms from different perspectives~\cite{xi2014keep,zhang2019design,zhang2016efficient,zhang2016experimental,zhao2012efficient}. 
For instance, some studies focus on the perspectives of multi-user environments~\cite{zhang2019design} and CSI correlations~\cite{xi2014keep} to improve key randomness and agreement.

LoRa, a long-range, low-power IoT technology, has seen limited yet promising key generation studies. Ruotsalainen et al.~\cite{ruotsalainen2019experimental} evaluated the effects of spreading factors and bandwidths on key generation performance, showing feasibility even under static conditions. Xu et al.~\cite{xu2018lora} designed a complete protocol, achieving 18--31~bit/s across stationary and mobile scenarios with a compressive sensing-based reconciliation method. Zhang et al.~\cite{zhang2018channel} introduced a differential value-based approach, improving randomness by quantizing keys based on received power trends in large-scale environments. Other studies explored key generation schemes for LoRa networks from different perspectives~\cite{yang2024scenario, yang2022vehicle, yang2023chirpkey, gao2021novel}.

Despite Bluetooth's widespread use in smartphones and wearable devices, key generation research is limited. Premnath et al.~\cite{premnath2014secret} demonstrated the first system, leveraging Bluetooth's frequency hopping to ensure resilience under Wi-Fi interference.
The advent of 5G introduces advanced techniques, including massive MIMO, mmWave communications, and full duplex, which provide new opportunities for channel reciprocity-based key generation. 
For instance, Jiao et al.~\cite{jiao2018physical} exploited mmWave MIMO’s AoA and AoD properties, achieving a high bit agreement ratio under low SNR. 
Chen et al.~\cite{chen2020beam} proposed pilot reuse and beam domain techniques to reduce overhead in multi-user massive MIMO systems. Further studies have explored full duplex probing~\cite{vogt2018secret} to enhance key generation rate while mitigating eavesdropping risks.

\subsubsection{Proximity-based Approaches}
Another research direction exploits the co-location property of mobile devices to generate keys. As illustrated in Fig.~\ref{fig:proxKey}, their radio signals will be highly similar when two devices are physically co-located. Below, we review several representative works that leverage the co-location property of IoT devices.

Amigo~\cite{varshavsky2007amigo} is one of the earliest systems that utilized a shared radio environment to authenticate mobile devices without explicit user involvement. Amigo adopted the Diffie–Hellman (D-H) protocol to establish keys, followed by a commitment scheme to defend against man-in-the-middle (MITM) attacks. The similarity of radio signals measured by two devices was used to verify physical proximity. The authors demonstrated that Amigo is resilient to MITM, eavesdropping, and spoofing attacks.
Mathur et al.~\cite{mathur2011proximate} proposed \textit{Proximate}, a system that generates secret keys for mobile devices in close proximity by measuring their wireless radio signals. Proximate followed the traditional key generation pipeline: quantization, reconciliation, and privacy amplification. Unlike Amigo, it removed reliance on the Diffie–Hellman protocol. However, its bit generation rate was relatively low, achieving only 1--3.5~bits/s.
To address the low bit rate issue, Xi et al.~\cite{xi2016instant} introduced The Dancing Signals (TDS), a CSI-based authentication and key generation system for co-located devices. TDS proposed a novel approach where keys were generated randomly and encoded using CSI features. Only devices with highly similar CSI measurements could decode the key, enabling agreement on a shared key. By decoupling key generation from CSI values, TDS achieved a significantly improved bit rate, reaching hundreds of bits per second. Moreover, TDS was extended to support group-based key generation for multiple devices.
Some proximity-based systems rely on a different observation: when a nearby sender moves very close to one of the antennas on a receiver, the receiver observes a significant RSSI variation. In contrast, if the sender is farther away, the RSSI difference between the two antennas remains small. Representative works based on this observation include Neighbor~\cite{cai2011good}, Wanda~\cite{pierson2016wanda}, and Move2Auth~\cite{zhang2017proximity}. 
Good Neighbor was the first scheme to pair devices using this principle. Wanda extended Good Neighbor by incorporating channel reciprocity to generate secret keys. Move2Auth adapted these ideas to enable smartphones to authenticate nearby IoT devices.

While co-location-based approaches provide an efficient mechanism for proximity-based authentication, they suffer from a key limitation: the distance between two legitimate devices must be very short. For example, Proximate~\cite{mathur2011proximate} requires a distance of 1.25~cm, TDS~\cite{xi2016instant} requires 5~cm, and Move2Auth~\cite{zhang2017proximity} works within 20~cm. Such requirements reduce the practicality of these methods, as modern wireless transceivers are embedded in mobile devices, making it difficult to position antennas within such close proximity in real-world scenarios.

A summary of representative key generation approaches is provided in Tab.~\ref{tab:key_gen}.




\subsection{RF Sensing-based Intrusion Detection}

RF sensing has emerged as a promising complementary modality for intrusion detection~\cite{wang2019wi,jin2018whole,ding2018robust,bao2019wisafe,lin2020revisiting,devoti2020pasid,ni2022gait,cai2020foreign,mallikarjun2020intruder}. Compared with traditional monitoring systems that rely on vision or infrared sensing~\cite{chowdhry2015smart}, RF sensing-based method offers enhanced privacy protection and demonstrates stronger robustness and adaptability in challenging environments such as poor lighting and occlusion. 

Among various RF modalities, \textbf{Wi-Fi}-based intrusion detection has become the most widely explored approach, owing to its reliance on existing network infrastructure, low deployment cost, and convenient access to signal data. For instance, APID~\cite{tian2018wifi}introduces a statistical hypothesis testing mechanism based on the coefficient of variation of CSI amplitude, achieving adaptive intrusion detection without the need for offline training. EID-T~\cite{zhu2021environmental} further incorporates a self-organizing map (SOM) neural network for CSI feature extraction and employs a multi-antenna joint decision mechanism to enhance detection accuracy and system stability. In addition, PetFree~\cite{lin2020revisiting} proposes an effective interference height (EIH) model that estimates the vertical distance of a moving target relative to the Wi-Fi link to distinguish humans from pets, thereby significantly reducing false alarms caused by non-human objects. Overall, Wi-Fi-based systems emphasize lightweight deployment and high adaptability, making them well-suited for typical indoor environments such as homes, offices, and commercial spaces.

In contrast, \textbf{mmWave} offers significant advantages in terms of spatial resolution and target discrimination, making it more suitable for complex environments or scenarios with stringent security requirements. 
For instance, PASID~\cite{devoti2020pasid}leverages beam alignment procedures in mmWave communication systems to detect subtle variations in beam power profiles, enabling high-precision passive intrusion detection without interfering with the communication process. Building upon this, MGait~\cite{ni2022gait} utilizes micro-Doppler signatures extracted from mmWave radar and applies an open-set identification network to perform individual recognition and intruder rejection in multi-person scenarios. 
In domain-specific applications, Cai et al.~\cite{cai2020foreign} target safety monitoring at railway crossings, employing FMCW mmWave radar along with MTI and CFAR algorithms to accurately detect foreign objects and estimate motion trajectories. Compared to Wi-Fi systems, mmWave-based solutions provide superior performance in spatial modeling and multi-target discrimination, and are more suitable for deployment in high-security, structurally complex environments.

For large-scale, long-distance, and energy-constrained scenarios, low-power wide-area network (LPWAN) technologies such as \textbf{LoRa} serve as a valuable complement to Wi-Fi and mmWave systems. The intrusion detection system~\cite{mallikarjun2020intruder} combines a PIR sensor with a GPS module to trigger and localize intrusion events, and transmits the alerts via LoRa to a remote TTN server. This design enables low-cost, long-range boundary protection in areas lacking power supply and network connectivity, such as remote industrial zones or forest perimeters. Although LoRa-based systems may not match the spatial resolution or semantic detection capabilities of Wi-Fi or mmWave solutions, they provide indispensable value in energy-sensitive and infrastructure-limited deployments.

In summary, RF sensing has established a multi-modal, hierarchical, and scenario-adaptive framework for intrusion detection. Wi-Fi-based approaches are ideal for lightweight, indoor applications due to their deployment flexibility and cost-effectiveness; mmWave systems offer high-resolution sensing and semantic-level discrimination for more demanding security scenarios; and LoRa-based solutions fill the gap in large-scale, low-power deployments. 

\subsection{RF Sensing-based Surveillance Video Enhancement}
RF sensing offers new opportunities to enhance existing surveillance video systems beyond the limits of traditional cameras. Recent advances focus on two key directions: detecting visual forgeries and reconstructing video from RF signals.

\subsubsection{\textbf{Surveillance Video Forgery Detection}}
Traditional video surveillance systems are increasingly exposed to severe security threats, such as frame replacement, intra-frame editing, and Deepfake-based impersonation. 
These attacks can often bypass both human perception and conventional vision-based algorithms, thereby undermining the usability, forensic reliability, and legal validity of surveillance footage. 
In contrast, RF signals—due to their inherent resistance to forgery and strong correlation with human activities—offer a trustworthy alternative for sensing the physical world. Leveraging RF signals as an auxiliary modality provides a promising direction for establishing cross-modal forgery detection frameworks that enhance the authenticity verification of video content.

To this end, many studies have explored the integration of RF sensing as a complementary data source to reinforce the reliability of surveillance video. 
An early effort in this direction is SurFi~\cite{lakshmanan2019surfi}, which detects surveillance camera looping attacks by correlating human activity patterns extracted from video feeds and Wi-Fi CSI signals. Rather than relying on semantic features like body pose, SurFi compares low-level temporal and frequency attributes across modalities, demonstrating the viability of RF-based consistency checks for real-time forgery detection.
Building upon this idea of cross-modal consistency, Secure-Pose~\cite{huang2021towards} takes a step further by incorporating structured semantic features into the detection pipeline. 
It extracts human pose representations—Joint Heat Maps (JHMs) and Part Affinity Fields (PAFs)—from both video frames and Wi-Fi CSI, and detects inconsistencies between the two modalities to identify frame-level and object-level tampering. 
Secure-Pose not only enables forgery detection but also supports localization of tampered regions, such as inserted or removed human subjects, thereby demonstrating the feasibility of cross-modal semantic divergence as an effective signal of forgery.
Building upon this foundation, the WiSil~\cite{fang2023nowhere} simplifies the semantic representation by replacing complex human pose estimation with silhouette reconstruction. It utilizes CSI-derived wavefront features and employs the U-Net architecture to generate silhouette maps from Wi-Fi signals. 
Compared with Secure-Pose, WiSil maintains high detection accuracy while improving model generalizability and deployability by avoiding the need for fine-grained pose annotations. 
Follow-up work~\cite{liu2024real} employs lightweight model compression techniques for real-time performance on edge devices, and introduce forgery trace localization through silhouette discrepancy maps, improving the interpretability and responsiveness of the system.

These works collectively highlight the unique advantages of RF sensing as a non-intrusive and hard-to-fake modality in verifying the authenticity of surveillance video. They pave the way toward the development of cross-modal, multi-channel, and trustworthy intelligent surveillance systems.

\subsubsection{\textbf{Surveillance Video Reconstruction}}
In addition to verifying video content authenticity, recent studies have begun to explore the use of RF sensing technologies to recover or reconstruct surveillance video, particularly in scenarios where camera footage becomes unavailable due to obstruction, physical damage, or malicious attacks.

Early attempts to reconstruct visual information from RF signals predominantly focused on Wi-Fi CSI.
Wi2Vi~\cite{kefayati2020wi2vi} proposed a deep learning framework that maps Wi-Fi CSI to video frames. The system incorporates a CSI encoder, a cross-domain translator, and an image decoder to generate coarse-grained grayscale surveillance video from CSI data collected using commercial COTS Wi-Fi devices. Although the reconstruction quality was limited in detail, Wi2Vi was the first to demonstrate that semantic information embedded in RF signals can be translated into the visual domain via deep learning, offering an important proof-of-concept.
Building on this foundation, CSI2Video~\cite{li2022recovering} introduced a two-stage synthesis framework. The system first extracts human pose representations—such as Joint Heat Maps (JHMs) and Part Affinity Fields (PAFs)—from Wi-Fi CSI, and then fuses them with pre-captured visual appearance data to generate more realistic RGB video frames. Emphasizing real-time performance and low deployment cost, CSI2Video operates entirely using standard IEEE 802.11n Wi-Fi devices, enhancing its practicality in real-world deployments.

While Wi-Fi-based methods demonstrated the feasibility of RF-to-video translation, they are inherently constrained by limited spatial resolution and multipath interference.
To address these limitations, M$^2$Vision~\cite{han2024seeing} leverages COTS mmWave radar for surveillance video reconstruction. Compared with Wi-Fi, mmWave offers finer spatial and motion resolution. M$^2$Vision introduces a dual-stage denoising algorithm and a virtual antenna-enhanced heatmap generation method to extract detailed profile and motion heatmaps from mmWave signals. These heatmaps are then fused with prior knowledge of the target’s appearance and environmental context via a deep multi-modal generative network, resulting in high-fidelity video reconstruction.

The above advances underscore the growing practical value and research potential of RF sensing-assisted visual reconstruction in intelligent surveillance systems.

\subsection{Summary and Insights}
\subsubsection{\textbf{Summary}}

This section systematically reviews how RF sensing technologies are leveraged to enhance security applications across two major categories: \textit{1) identity-centric security mechanisms}, and \textit{2) environment-centric intelligent surveillance}. 
By exploiting the unique propagation characteristics of RF signals, RF sensing enables contactless, device-free, and context-aware security solutions well-suited for smart environments and IoT ecosystems.

In the first category, RF sensing empowers identity-centric security mechanisms including human authentication~\cite{han2024mmsign,xu2022mask,li2021vocalprint}, identification~\cite{yang2020mu,mokhtari2017non,zhang2018rfree}, and physical-layer key generation~\cite{yang2023chirpkey,ju2023random,gao2021novel}. 
RF-based authentication and identification methods span multiple RF modalities including Wi-Fi, mmWave, UWB, LoRa, and RFID. These systems recognize individuals based on biometric and behavioral signatures such as gait, gesture, respiration, and heartbeat. 
Wi-Fi offers ubiquitous and cost-effective solutions using CSI-based activity recognition, while mmWave enables high-resolution sensing for micro-movement-based biometrics. 
Complementary technologies (e.g., RFID, UWB) provide long-range and low-power alternatives, extending RF sensing-based authentication to diverse real-world settings.

Meanwhile, RF-based secret key generation can be viewed as a sensing task that focuses on extracting entropy from the physical RF channel. By perceiving fine-grained channel dynamics—such as temporal fluctuations, channel reciprocity, and spatial decorrelation—devices can securely derive shared cryptographic keys without relying on pre-shared secrets or centralized infrastructure. 
Channel reciprocity and physical proximity serve as the foundation for key generation across technologies such as ZigBee~\cite{patwari2009high}, Wi-Fi~\cite{xi2016instant}, LoRa~\cite{ruotsalainen2019experimental}, and Bluetooth~\cite{premnath2014secret}. However, applying these methods in real-world scenarios remains challenging due to signal instability in dynamic environments, hardware inconsistencies across devices, and strict proximity requirements.

In the second category, RF sensing facilitates intelligent intrusion detection~\cite{wang2019wi,ding2018robust} as part of a broader secure sensing framework. 
Wi-Fi-based systems identify unauthorized presence by analyzing CSI variance, enabling non-invasive, low-cost indoor monitoring. 
mmWave radar contributes higher spatial resolution and target discrimination for complex environments, while LoRa offers long-range coverage for energy-constrained settings.

Beyond physical intrusion detection, RF signals also reinforce the authenticity and continuity of surveillance video. 
Cross-modal forgery detection frameworks~\cite{huang2021towards,fang2023nowhere} compare RF-derived and vision-derived human representations to reveal tampering such as frame replacement or identity spoofing. 
Meanwhile, RF-to-vision reconstruction systems~\cite{kefayati2020wi2vi,han2024seeing} utilize RF signals to recover human silhouettes or semantic video when visual streams are unavailable due to occlusion, failure, or attack. 
These studies demonstrate that RF sensing can serve as a trustworthy, hard-to-spoof modality for visual security enhancement.

These developments underscore the future potential of RF sensing technology for secure, passive and resilient identity verification, communication protection and environment awareness in the future ubiquitous computing environments.

\subsubsection{\textbf{Insights}}
RF sensing is transitioning from a simple data acquisition method to a core component of security mechanisms. 
Current research suggests that RF sensing holds several unique advantages in security-critical scenarios. 

First, compared to vision or infrared-based sensing, RF signals offer better privacy protection, strong penetration capability, and higher robustness in complex conditions such as occlusion or low-light environments. 
Second, RF signals are highly sensitive to human presence and motion, naturally encoding individual differences and behavioral continuity, making them well-suited for identity recognition and behavioral analysis. 
Third, the physical-layer consistency and non-forgeability of RF signals make them ideal for enhancing the trustworthiness and robustness of multimodal security systems, particularly as a complement to visual modalities.

Looking ahead, RF sensing is expected to shift from passive sensing to proactive defense, moving beyond mere identity recognition and state monitoring toward behavioral intent inference, dynamic trust assessment, and adversarial manipulation detection. This evolution will lead to the development of a security-aware sensing layer capable of autonomous risk judgment and response.

\textit{First, RF sensing will progress from basic activity recognition to context-aware intent reasoning, enabling more intelligent security policy adaptation.} 
Most existing systems focus on recognizing actions, but in many high-stakes security scenarios, this is not sufficient. For instance, when a person lingers near an ATM, are they waiting in line or preparing for malicious activity? When someone approaches a restricted area at night, is it an accident or deliberate probing? These questions require integrating RF signals with temporal, spatial, and historical behavioral context to infer intent and assess risk. Future work may explore the fusion of RF features with abnormal trajectory modeling, behavioral history retrieval, and semantic scene labeling to construct intent-aware threat graphs, enabling early-stage risk mitigation.

\textit{Second, RF sensing will play a pivotal role in dynamic spatial trust modeling and real-time access control.} 
Traditional systems often assume that once a user is authenticated, their subsequent actions are trustworthy. However, in many real-world environments, trust should depend not just on identity, but also on whether a user’s behavior aligns with spatial and contextual expectations. 
For instance, RF sensing can enable continuous tracking of user location and behavior patterns, allowing systems to assess whether a person remains within authorized areas or deviates from expected routes. This forms the foundation of a three-dimensional trust mechanism based on space–behavior–identity correlations, supporting real-time permission adjustment and policy enforcement. Such capabilities will help overcome the limitations of static authentication and move toward risk-adaptive and behavior-aware security control.

\textit{Finally, the RF modality will emerge as both a new attack surface and a defensive anchor in multimodal security systems.} 
While the integration of RF with visual modalities is becoming increasingly common, it also opens new attack vectors. Adversaries may forge RF signals to impersonate legitimate users, synchronize RF and visual cues to create Deepfake-style “cross-modal deception,” or manipulate EM reflections to interfere with sensing. 
RF is no longer just an information source—it is a potential target. 
Conversely, due to their physical-layer consistency and non-spoofability, RF signals can also serve as trustworthy references for detecting cross-modal inconsistencies and falsified content. 
Future research should explore adversarial modeling and joint validation across modalities, such as RF–vision consistency checkers and multimodal forgery provenance frameworks, thereby establishing a trust-centric, RF-grounded ecosystem for secure sensing and computation.

RF sensing is evolving into a security-critical modality that supports intent inference, dynamic trust modeling, and multimodal adversarial defense. Its physical-layer robustness and contextual awareness position it as a foundational layer for secure and intelligent systems. Future research should focus on making RF sensing more adaptive, trustworthy, and deeply integrated into real-world decision-making pipelines.

\section{Challenges and Future Directions}
\label{sec:challenge}
\subsection{Core Challenges in RF Sensing Security and Privacy}
RF sensing security and privacy involves a dual perspective of attack and defense, both of which constrain each other and jointly drive the development of this field. 
From both attack and defense perspectives, the following core challenges currently exist.

\subsubsection{\textbf{Attack Perspective}}
Based on the analysis and insights derived from the preceding sections, we identify several fundamental challenges that attackers typically face.

\paragraph{Attackers Attempting to Compromise RF Sensing Systems}
First, attackers attempting to compromise RF sensing systems face several inherent difficulties. 

\begin{itemize}[leftmargin=*]
\item \textit{Complex Signal Characteristics.}
RF sensing systems utilize complex, high-dimensional features (e.g., CSI~\cite{liu2014practical,zeng2019farsense,wang2017phasebeat}, micro-Doppler features~\cite{yang2020mu,qian2018widar2,han2024mmsign}) that contain significant redundancy and high sensitivity to minor variations. Precisely and consistently manipulating these signals is thus inherently challenging for attackers.

\item \textit{Black-box Models.}  
Most RF sensing models operate in a black-box manner~\cite{liu2024time,nallabolu2023emulation,li2024practical}, limiting attackers' knowledge of internal structures and parameters. This severely restricts targeted adversarial attacks, forcing attackers to rely on less effective transfer-based or generic approaches.

\item \textit{Physical-Domain Constraints.}  
Attacks executed in the physical domain face practical issues, including hardware imperfections~\cite{chen2023metawave}, multipath interference~\cite{ali2015keystroke}, and signal attenuation~\cite{jiang2020towards}. These factors significantly reduce attack reliability and reproducibility in real-world scenarios.

\item \textit{Limited Attack Generalizability.}  
RF sensing features are highly environment-dependent~\cite{wang2022placement}, meaning attacks designed in controlled settings often fail in dynamic, real-world conditions. Attackers must constantly adapt to environmental changes, substantially increasing attack complexity and cost.
\end{itemize}

\paragraph{Attackers Attempting to Steal Sensitive Information}
Second, attackers who attempt to steal sensitive user information face the following core difficulties:
\begin{itemize}[leftmargin=*]
    \item \textit{Covert Signal Acquisition.}
    Successfully invading privacy via RF sensing requires covertly capturing and interpreting environmental RF signals. However, secretly deploying receiving hardware or continuously monitoring signals without detection poses significant practical challenges, as any such attempt risks exposure by security measures~\cite{chaman2018ghostbuster}.
        
    \item \textit{Signal Quality and Stability.}  
    Although RF signals can penetrate physical obstacles, they often suffer severe attenuation, multipath interference, and distortion~\cite{seybold2005introduction}, especially in uncontrolled settings. These impairments reduce signal quality, limiting attackers' ability to reliably extract sensitive information in realistic, long-range, or through-wall scenarios.
    
    \item \textit{Poor Generalizability.}  
    RF-based privacy invasion attacks rely heavily on environment-specific signal features. Variations across different settings necessitate large, diverse datasets to train robust models. However, gathering comprehensive data from multiple real-world scenarios remains challenging, resulting in limited generalizability~\cite{wang2025survey} and significantly reduced attack success rates.
    
    \item \textit{Semantic Complexity.} Recent advances have enabled attackers to infer simple semantic information, such as basic activity~\cite{lu2022actlistener} or short input sequences like PINs~\cite{hu2023password}. 
    However, moving beyond such coarse-grained or short-form classification toward detailed semantic reconstruction—such as recovering complete typed sentences, or inferring context-aware intent—remains a significant challenge.

\end{itemize}

\subsubsection{\textbf{Defense Perspective}}
From a defensive perspective, securing RF sensing systems involves two core objectives: (1) ensuring the system’s resilience against adversarial manipulation, and (2) preventing its misuse for privacy invasion. 
Based on the analysis and discussions in previous sections, current defensive strategies face several fundamental and common challenges in addressing these two objectives.

\paragraph{Insufficient Robustness and Generalization of Defensive Mechanisms}

A primary obstacle in safeguarding RF sensing integrity lies in the \textit{limited robustness and generalization} of existing defense techniques. 
Current defensive mechanisms are typically developed under controlled laboratory conditions, lacking effectiveness against the complex, dynamic conditions and unknown attack strategies encountered in real-world scenarios. 
Furthermore, the increasing reliance on deep learning~\cite{ahmad2024wifi} further compounds this vulnerability, exposing systems to a wide range of AI-specific threats. 
Consequently, defensive mechanisms must simultaneously contend with both environmental perturbations and sophisticated adversarial manipulations—conditions under which many current solutions fail to maintain effectiveness.
In addition, existing systems~\cite{wang2022caution,li2021vocalprint} decouple identity authentication from ongoing behavioral monitoring, lacking dynamic trust modeling that captures space–behavior–identity consistency. This gap undermines the ability to continuously validate user intent or detect anomalous behavior after initial access.

\paragraph{Privacy-Protection Trade-offs in Practical Deployments}

Similarly, defense strategies aimed at preventing the misuse of RF sensing for privacy invasion confront significant difficulties in \textbf{practical deployments}. 
Privacy-preserving measures such as signal obfuscation, CSI encryption~\cite{qiao2016phycloak,deng2024css,meng2023secur}, while reducing the risk of information leakage, often compromise critical system functionalities. 
Consequently, a challenging trade-off emerges between the strength of privacy protection and overall system performance. 
Meanwhile, existing defense approaches largely rely on passive strategies, lacking the capability to monitor, detect, and proactively respond to sophisticated eavesdropping attacks in real-time, further limiting their effectiveness.

\paragraph{Absence of Standardization and Regulatory Guidance}

A more fundamental challenge is the \textit{absence of unified standards and regulatory frameworks} for RF sensing security. The lack of clear industry guidelines and standardized best practices results in uneven security across different RF sensing systems, creating vulnerabilities that attackers can exploit. Therefore, establishing comprehensive security standards and regulatory oversight constitutes an essential foundational challenge to be addressed in future research.

\subsection{Emerging Threat Trends in RF Sensing}

\subsubsection{\textbf{New Attack Mediums and Vectors}}
With the evolution of 6G Integrated Sensing and Communication (ISAC)~\cite{dong2022sensing}, RF signals are no longer limited to data transmission but also serve environmental sensing functions. Technologies such as mmWave~\cite{hong2021role}, terahertz (THz) waves~\cite{jiang2024terahertz}, and massive MIMO~\cite{jeon2021mimo} enhance spatial perception capabilities, but they also introduce new security and privacy vulnerabilities.

Unlike traditional sensing signals, 5G/6G signals feature stronger channel stability, higher bandwidth, and more advanced beamforming and multi-antenna technologies. These characteristics allow attackers to exploit existing infrastructure for more covert and long-range RF sensing attacks without the need to deploy additional signal sources.
For example, attackers can monitor environmental changes by passively analyzing the CSI of 5G/6G signals, thereby inferring the target's location, gait, and even heartbeat information in a contactless scenario, thereby leaking privacy. 
Furthermore, as ISAC integrate sensing functions directly into cellular communication frameworks~\cite{yu2022location}, the network itself becomes vulnerable to sensing-based manipulation. Attackers could tamper with beamforming directions or utilize IRS to redirect signal paths, inducing target localization errors or spoofing the sensing system's interpretation of the environment. 

This transition from dedicated sensing devices to ubiquitous, infrastructure-level sensing dramatically expands the threat surface. In the future, as sensing capabilities become embedded in standard communication protocols, the very networks that enable connectivity may also serve as platforms—or victims—of sophisticated RF sensing-based privacy invasions.

\subsubsection{\textbf{Advanced Attack Techniques}}
In addition to exploiting emerging infrastructure, future attackers may also manipulate sensing processes through physical interference, multimodal signal exploitation, and AI-driven techniques.

\paragraph{Environmental Manipulation Attack-}
RF sensing relies heavily on the physical properties of the environment, such as reflection, attenuation, and scattering~\cite{seybold2005introduction}. This dependency makes it susceptible to adversarial manipulation, where an attacker can alter environmental parameters—such as temperature, humidity, or material reflectivity—to deceive the sensing system and mislead its detection mechanisms.
For instance, at different temperatures, the molecular arrangement of certain materials may change, leading to variations in mmWave scattering characteristics~\cite{chen2020thermowave}. 
By exploiting this phenomenon, an attacker can modify the properties of the target or its surrounding environment to disrupt RF sensing accuracy and manipulate perception outcomes.
Moreover, recent studies have highlighted the potential of IRS in RF sensing applications~\cite{shao2022target}. Attackers can leverage this technology to dynamically adjust the reflectivity of surface materials, thereby altering RF signal propagation paths and preventing the sensing system from accurately interpreting the environment.

\paragraph{Cross-Domain Physical Attacks}
Cross-domain physical attacks refer to adversarial techniques that utilize non-RF signals—such as acoustic waves or mechanical vibrations—to interfere with or deceive RF sensing systems. Rather than directly manipulating the RF signal itself, this attack targets the signal reception or interpretation processes through physical means, thereby introducing sensing errors and undermining system stability and accuracy. It often exploits the inherent physical characteristics of RF devices or the target object, with resonance effects commonly employed to amplify the impact.

For instance, in RF sensing systems, signal quality plays a critical role in ensuring the accuracy and robustness of sensing tasks. As the antenna is a key component in RF signal acquisition~\cite{zeng2021exploring, lan2021metasense, zeng2019farsense}, attackers can emit acoustic waves at specific frequencies to induce resonant modes in the antenna structure, thereby generating high levels of noise that disrupt normal signal collection and prevent accurate interpretation of the sensed information.

Moreover, many RF sensing systems are designed to extract weak vibration patterns from the target, such as mmWave-based speech recognition~\cite{hu2023mmecho,xu2024mmear,wang2022mmphone} that detects micro-vibrations of the target. In such scenarios, attackers may induce spurious or carefully crafted vibration patterns on the target using precisely modulated mechanical or acoustic excitations. These deceptive inputs can lead the sensing system to misidentify or misclassify the observed phenomena, effectively spoofing the perception outcome.

\paragraph{Multimodal Collaborative Attacks}
Future research can leverage different RF signal modalities, such as Wi-Fi CSI and mmWave radar, to conduct coordinated attacks that maximize their respective advantages, enabling more efficient and covert intrusions. 
For instance, LoRa signal~\cite{zhang2020exploring} excels in long-range target detection, while mmWave signal~\cite{blandino2023ieee} provides higher spatial resolution. 
Exploiting these characteristics, attackers can dynamically adapt their strategies based on different scenarios, integrating multiple sensing modalities to achieve adaptive multimodal RF sensing privacy invasions. 
For instance, in non-through-wall environments, mmWave signals can be used for fine-grained sensing, whereas in through-wall scenarios, attackers can utilize prior knowledge obtained from mmWave sensing and combine it with LoRa signals to penetrate barriers, overcoming the penetration limitations of mmWave radar and making the sensing attack more precise and stealthy.

\paragraph{AI-Driven Attacks}
Many attack strategies in RF sensing originate from computer vision, such as adversarial attacks~\cite{li2024practical} and backdoor attacks~\cite{zhao2023backdoor}. 
Given this connection, future RF sensing security research can draw inspiration from computer vision attack techniques to explore novel threats. For example, membership inference attacks~\cite{shokri2017membership} could compromise data privacy by allowing attackers to determine whether a specific data sample was used in training, while model extraction attacks~\cite{juuti2019prada} enable adversaries to reconstruct a model's parameters, posing a risk to intellectual property and proprietary algorithms.
Given the deep integration of deep learning with RF sensing, these attacks could be maliciously applied to sensing in the future.

Moreover, with the rapid proliferation of large-scale AI models~\cite{gozalo2023chatgpt}, RF sensing systems are beginning to adopt LLMs and multimodal foundation models~\cite{li2024llmcount}, bringing new security and privacy challenges. One critical concern is memory retention and data leakage: recent studies have shown that LLMs may memorize fragments of their training data~\cite{hartmann2023sok}, allowing attackers to extract sensitive information through carefully crafted prompts. If future RF sensing models—or multimodal AI systems incorporating RF data—are trained on private RF signal datasets, they may likewise store and unintentionally expose sensitive behavioral patterns or biometric signatures.

More importantly, the integration of foundation models is expected to drive RF sensing beyond low-level signal classification (e.g., activity recognition) toward context-aware, high-level semantic inference, such as intent prediction, emotional state estimation, or behavioral profiling. While such advances could significantly enhance the intelligence of RF sensing systems, they also amplify the risk of semantic-level privacy violations, as attackers may leverage fine-grained signal cues to infer private user intentions or mental states without explicit consent. Furthermore, the use of automated AI-driven tools~\cite{xu2024autoattacker} may further scale these attacks, enabling more efficient and adaptive RF-based privacy invasions.

\subsection{Advanced Defense Strategies}
\subsubsection{\textbf{Technological Approaches}}
RF sensing security faces multiple challenges, including dynamic environmental changes, adversarial perturbations, and machine learning vulnerabilities. 
To enhance overall system security and robustness, a multi-layered security defense framework is essential. This framework can be categorized into three key layers: 
\textit{physical layer security}, which prevents signal tampering and eavesdropping through signal control and security protocols; 
\textit{algorithmic defense}, which enhances RF sensing resilience against interference using multimodal data, robust AI training, and model validation; 
and \textit{privacy protection mechanisms}, which mitigate data leakage risks and enhance privacy through differential privacy, federated learning, and other techniques.

\begin{itemize}[leftmargin=*]
    \item In the physical layer security, various techniques can be employed to safeguard RF sensing. Secure channel design (e.g., frequency hopping and spread spectrum techniques)~\cite{gunther2020modeling} can mitigate malicious interference, while beamforming and directional communication~\cite{lin2017physical} help minimize signal leakage and prevent eavesdropping. In addition, to ensure the integrity and authenticity of RF sensing signals, signal fingerprinting and Physically Unclonable Functions (PUFs)~\cite{yadav2022review, gope2018lightweight} can be used to verify the source of received signals, preventing adversaries from tampering with or spoofing sensing data.

    \item In the algorithmic defense layer, RF sensing systems can leverage adversarial training~\cite{bai2021recent} by introducing adversarial examples during model training to enhance resistance against malicious attacks. Model validation~\cite{raschka2018model} techniques can monitor model parameter variations to prevent data poisoning or backdoor attacks. Furthermore, RF sensing can adopt the multimodal sensor fusion approach~\cite{bijelic2020seeing,lahat2015multimodal} from computer vision, integrating data from Wi-Fi and mmWave to improve system resilience in complex environments.
    
    \item In the privacy protection layer, differentially private~\cite{abadi2016deep} can be implemented by introducing noise or limiting CSI granularity, reducing the risk of attackers inferring personal information from RF sensing data. 
    In addition, federated learning~\cite{li2020review} for RF Models allows multiple RF devices to train models locally without transmitting raw data, thus minimizing the likelihood of data leaks. Lastly, adversarial channel simulation~\cite{zhang2023generative} can be introduced to systematically assess the security and robustness of RF sensing systems under different channel conditions and interference environments, ensuring stability and attack resistance in real-world applications.
\end{itemize}

However, one of the fundamental challenges in RF sensing security and privacy is implementing robust protection mechanisms without compromising system performance or usability~\cite{qu2024privacy}. 
Applications such as health monitoring and autonomous driving require low latency and high responsiveness to function effectively. 
While strong encryption, authentication, and anomaly detection mechanisms are critical for RF sensing security, they can introduce computational overhead and communication delays, making them impractical for real-time applications. 
Moreover, many RF sensing systems rely on existing communication signals (e.g., Wi-Fi, LoRa, ZigBee), which were not originally designed for secure sensing. As a result, integrating complex encryption or adversarial defense mechanisms into RF sensing pipelines can further exacerbate latency issues and disrupt critical applications.

To address this challenge, future research must explore lightweight security mechanisms~\cite{goyal2016lightweight} that can defend against attacks while preserving sensing accuracy and responsiveness.

\subsubsection{\textbf{Standardization and Compliance}}
While technical defenses provide immediate protection against RF sensing attacks, a long-term security strategy requires standardized guidelines and regulations to enforce best practices across the industry.
Unlike traditional IT systems, RF sensing technologies (such as device-free Wi-Fi sensing or home radar sensors) currently lack comprehensive security standards and regulatory oversight. 
While IEEE 802.11bf~\cite{ropitault2024ieee} has introduced a framework for Wi-Fi Sensing, enabling Wi-Fi signals to perform environmental sensing and target detection, the standard does not yet address security and privacy protection nor provide guidelines for mitigating attacks. 
In addition, RFID security standards such as ISO/IEC 29167~\cite{iso29167} and ISO/IEC 18000~\cite{iso18000} define encryption and authentication mechanisms for RFID systems, but they primarily apply to passive RFID systems and do not cover device-free RF sensing technologies such as Wi-Fi, mmWave, or 5G/6G-based sensing. 
Similarly, IEEE 802.15.4 (ZigBee, LoRa)~\cite{9144691} provides security protocols for IoT wireless communication but lacks specific provisions for RF sensing privacy protection.

As RF sensing becomes ubiquitous in IoT and smart environments, there is an increasing need for industry standards and government regulations to define security protocols, data processing policies, and user consent requirements. Currently, the European General Data Protection Regulation (GDPR)~\cite{gdpr2016} has introduced privacy protection requirements for RF data storage and processing, but it has not yet established specific provisions for RF sensing data. Similarly, the U.S. IoT Cybersecurity Improvement Act~\cite{iot_cybersecurity_act2020} primarily focuses on IoT device security but lacks regulatory oversight on RF sensing privacy risks.
In the future, regulatory bodies should leverage existing security standards—such as the encryption mechanisms in ISO/IEC 29167—and integrate them with sensing technology frameworks like IEEE 802.11bf to establish security guidelines for Wi-Fi-, mmWave-, and 5G/6G-based sensing systems. 
For example, regulators may define retention periods for raw RF data, which may contain sensitive biometric signatures, and mandate that RF devices adopt a ``Security by Design'' approach to ensure minimal data collection and strict access controls. 
Additionally, RF sensing systems may be required to support CSI Encryption or implement privacy-preserving data processing techniques based on Differential Privacy to mitigate risks associated with data leakage and unauthorized tracking.

\section{Conclusion}
\label{sec:conclusion}
RF sensing has demonstrated immense potential across diverse applications but also brings forth significant security and privacy challenges. 
In this survey, we systematically reviewed the threat landscape, ranging from model integrity attacks to covert surveillance. 
We categorized these threats into intrinsic system vulnerabilities and malicious sensing misuse, and analyzed how they manifest across various tasks such as activity recognition, gesture control, localization, and autonomous driving. 
In response, we surveyed a range of defense strategies—spanning physical-layer protections, signal-level obfuscation, and model-level robustness—that aim to mitigate these risks. We also discussed how RF sensing can serve as a security enabler in tasks like user authentication, intrusion detection, and video forgery analysis. Looking forward, the integration of RF sensing with 6G and large-scale AI models presents new challenges and opportunities, calling for cross-layer defense frameworks and standardized privacy risk models to ensure its secure and ethical deployment.

\bibliographystyle{IEEEtran}
\bibliography{ref}

\end{document}